\documentclass[aps,twocolumn,superscriptaddress,nofootinbib,longbibliography,notitlepage,floatfix]{revtex4-2}

\usepackage{minitoc}
\usepackage{graphicx,outlines,subfigure}
\usepackage{physics,graphicx,diagbox,markdown}
\usepackage{tabularx}
\usepackage{titletoc}
\usepackage{amsmath,mathtools}
\usepackage{amstext,cuted}
\usepackage{amssymb}
\usepackage{xfrac}
\usepackage[colorlinks,citecolor=magenta]{hyperref}
\usepackage{graphicx}
\usepackage{amsmath}
\usepackage{amstext}
\usepackage{amssymb}
\usepackage{longtable,booktabs}
\usepackage{hyperref}
\usepackage{url}
\usepackage{subfigure}
\usepackage{dsfont}
\usepackage{booktabs}
\usepackage{amsbsy}
\usepackage{dcolumn}
\usepackage{amsthm}
\usepackage{times}
\usepackage{bm}
\usepackage{esint}
\usepackage{multirow}
\usepackage{hyperref}
\usepackage{cleveref}
\usepackage{amsmath}
\usepackage{amssymb}
\usepackage{mathrsfs}
\usepackage{amsbsy}
\usepackage{dcolumn}
\usepackage{bm}
\usepackage{multirow}
\usepackage{color}
\usepackage{extarrows}
\usepackage{datetime}
\usepackage{comment}
\usepackage[super]{nth}
\usepackage{tikz}
\usepackage{makecell}
\usepackage{aligned-overset}
\usetikzlibrary{arrows.meta}
\hypersetup{
  colorlinks=true,
  linkcolor=blue,
  filecolor=blue,
  urlcolor=blue,
}
\usepackage{scalerel}
\usepackage{tikz}
\usetikzlibrary{svg.path}

\definecolor{orcidlogocol}{HTML}{A6CE39}
\tikzset{
  orcidlogo/.pic={
    \fill[orcidlogocol] svg{M256,128c0,70.7-57.3,128-128,128C57.3,256,0,198.7,0,128C0,57.3,57.3,0,128,0C198.7,0,256,57.3,256,128z};
    \fill[white] svg{M86.3,186.2H70.9V79.1h15.4v48.4V186.2z}
                 svg{M108.9,79.1h41.6c39.6,0,57,28.3,57,53.6c0,27.5-21.5,53.6-56.8,53.6h-41.8V79.1z M124.3,172.4h24.5c34.9,0,42.9-26.5,42.9-39.7c0-21.5-13.7-39.7-43.7-39.7h-23.7V172.4z}
                 svg{M88.7,56.8c0,5.5-4.5,10.1-10.1,10.1c-5.6,0-10.1-4.6-10.1-10.1c0-5.6,4.5-10.1,10.1-10.1C84.2,46.7,88.7,51.3,88.7,56.8z};
  }
}

\newcommand\orcidicon[1]{\href{https://orcid.org/#1}{\mbox{\scalerel*{
\begin{tikzpicture}[yscale=-1,transform shape]
\pic{orcidlogo};
\end{tikzpicture}
}{|}}}}
\tikzset{bold/.style={color=blue, line width=2pt}}
\tikzset{redop/.style={circle,fill=red}}
\tikzset{blueop/.style={circle,fill=blue}}

\usepackage{datetime}
\usepackage{comment}
\usepackage{lineno}
\graphicspath{{Pictures/}}
\usepackage{xcolor}

\makeatletter

\newcommand{\Rmnum}[1]{\expandafter\@slowromancap\romannumeral #1@}
\makeatother

\begin{document}
\title{Fractonic superfluids. \Rmnum{3}. Hybridizing higher moments}
\author{Han-Xie Wang\orcidicon{0009-0002-6414-0830}}
\thanks{These authors contributed equally to this work.}

	\affiliation{Guangdong Provincial Key Laboratory of Magnetoelectric Physics and Devices, State Key Laboratory of Optoelectronic Materials and Technologies,
and School of Physics, Sun Yat-sen University, Guangzhou, 510275, China}
 \author{Shuai A. Chen}\thanks{These authors contributed equally to this work.}
\affiliation{Max Planck Institute for the Physics of Complex Systems, N\"{o}thnitzer Stra{\ss}e 38, Dresden 01187, Germany}
 \author{Peng Ye\orcidicon{0000-0002-6251-677X}}
\email{yepeng5@mail.sysu.edu.cn}
	\affiliation{Guangdong Provincial Key Laboratory of Magnetoelectric Physics and Devices, State Key Laboratory of Optoelectronic Materials and Technologies,
and School of Physics, Sun Yat-sen University, Guangzhou, 510275, China}

\date{\today}

\begin{abstract}

Fractonic superfluids are featured by the interplay of spontaneously broken charge symmetry and mobility constraints on single-particle kinematics due to the conservation of higher moments, such as dipoles, angular charge moments, and quadrupoles.  Building on prior studies by Yuan \textit{et al.} [\href{https://doi.org/10.1103/PhysRevResearch.2.023267}{Phys. Rev. Res. 2, 023267 (2020)}] and Chen \textit{et al.} [\href{https://doi.org/10.1103/PhysRevResearch.3.013226}{Phys. Rev. Res.  3, 013226 (2021)}], we study a class of fractonic superfluids, termed \textit{hybrid fractonic superfluids} (HFS), in which bosons of multiple species interact while moment hybridization is conserved. We explore the consequences of hybridization via  two model series: \textit{Model Series A}, conserving total moments of the same order across species, and \textit{Model Series B}, conserving total moments of different orders. In Model Series A, we analyze dipole moment hybridization and extend the discussion to higher-order moments, examining the ground state, Goldstone modes, correlation functions, and so on. We compute the minimal spatial dimensions, where the total charge symmetry begins to get partially broken  via particle-hole condensation, leading to true off-diagonal long-range order. In Model Series B, we focus on HFS with hybrid dipole-quadrupole conservation. For both series, we introduce Bose-Hubbard-type lattice models that reduce to either of both series  in the weak Hubbard interaction regime. We perform a mean-field analysis on the global phase diagram and discuss  experimental realizations in strongly tilted optical lattices via a third-order perturbation theory. This work, alongside prior studies, completes a trilogy on fractonic superfluids, uncovering symmetry-breaking physics emerging from higher moment conservation, leaving various promising studies for future investigation.

\end{abstract}
\maketitle

\section{Introduction}

In recent years, fracton topological order has been gaining increasing attention.  
Exactly solvable lattice models of fracton topological order are typically described by fracton codes, such as the three-dimensional X-cube model~\cite{fractonorder1,fractonorder2,fractonorder3,fracton16,PhysRevLett.129.230502,fracton3,fracton2,fracton1,fracton19,fracton,fracton4,fracton17,fracton18,fracton51,fracton52,fracton27,fracton53,fracton54,fracton15,PhysRevB.110.205108}.  
Topological excitations in fracton codes are commonly referred to as fractonic excitations, including fractons, subdimensional particles (e.g., lineons and planeons), and more intricate spatially extended excitations that are subject to mobility restriction~\cite{fracton18,fracton51,fracton52}.  
In contrast to conventional topological orders, such as the toric code model, where topological excitations can be freely moved to any location via string operators, fractonic excitations in fracton topological order exhibit mobility restrictions. For instance, a single fracton cannot move freely along any directions, while a single lineon can only move along a straight line. Due to these mobility restrictions, fracton codes are believed to provide robust fault-tolerant quantum memory~\cite{fractonorder2,fractonorder3,fracton16,PhysRevLett.129.230502}. The exploration of fracton topological order has also stimulated studies on subsystem symmetry-protected topological order (SPT)~\cite{SSPT9,SSPT11,SSPT3,SSPT10,SSPT12,SSPT5,SSPT6,SSPT4,SSPT7,SSPT8,SSPT1,SSPT2}, substantially enlarging the territory of SPT phases.

Another active research direction involves treating fractons as constituent particles of many-body systems, where mobility restrictions arise from the conservation of higher-order moments~\cite{fracton58,fracton59,Fractongauge,fracton55,fracton56,fracton57,fracton60,fracton20,fracton26,diffusionofhigher-moment,Fractonicsuperfluids1,Fractonicsuperfluids2,NSofFractonicsuperfluids,fracton23,Fractonicsuperfluidsdefect,reviewofFractonicsuperfluids,fracton5,2023JHEP...09..184J,fracton33,fracton46,fracton8,2024JHEP...07..197J,fracton38,fracton9,fracton11,fracton44,fracton28,fracton7,fracton29,fracton42,fracton25,fracton10,fracton12,fracton49,fracton50,fracton21,fracton14,fracton22,fracton30,Lifshitzduality,HMWT,DBHM,DBHM2,fracton31,fracton35,fracton45,fracton6,fracton24,fracton34,fracton36,fracton37,fracton39,fracton40,fracton41,fracton32,fracton43,fracton48,fracton47,PhysRevLett.132.071603}.  Such systems are referred to as \textit{fractonic many-body systems}. 
It is well known that conventional total charge conservation is associated with a global $U(1)$ symmetry. Similarly, these new conserved quantities formed by higher moments are associated with new $U(1)$ symmetries. Upon gauging, the resulting gauge degrees of freedom are symmetric tensors; therefore, these new symmetries are sometimes alternatively  referred to as ``higher-rank symmetries''. To sum up, this line of research has evolved into a platform for exploring new possibility of symmetries, stimulating    interests and ideas from dynamics, hydrodynamics, gravity, and other domains \cite{fracton5,fracton7,fracton8,fracton9,fracton11,fracton28,fracton29,fracton33,fracton38,fracton42,fracton44,fracton46,fracton10,fracton12,fracton25,fracton49,fracton50,2024JHEP...07..197J,2023JHEP...09..184J}.

On the other hand, spontaneous symmetry breaking (SSB) is a fundamental concept spanning a broad range of research areas in theoretical physics, from the Standard Model of particle physics to the Ginzburg-Landau theory of second-order phase transitions. In condensed matter physics, superfluids serve as canonical examples of systems with SSB, where the full charge $U(1)$ symmetry group is spontaneously broken, and off-diagonal long-range order (ODLRO) is established. A celebrated phenomenon in this context is the Kosterlitz-Thouless topological phase transition, which occurs at finite temperature in two dimensions (2D). 
The study of SSB, particularly in the context of continuous global symmetries, has long been associated with important topics such as Goldstone mode counting, the Mermin-Wagner theorem concerning the stability of ODLRO at finite temperatures, and the infrared divergences induced by quantum fluctuations at zero temperature. In recent years, growing interest in generalized symmetries~\cite{Gaiotto:2014kfa,Mcgreevy_generalizedsym} has led to significant advances across various subfields of theoretical physics. 
Within the framework of generalized symmetries, long-range entangled states, which do not fall under the conventional classification of SSB phases, are now interpreted as SSB phases of higher(-form) symmetries. As the fracton physics is evolving rapidly, it is timely and important to investigate SSB physics in fracton systems. More specifically, for   fractonic many-body quantum systems, with their exotic conserved quantities and corresponding new symmetries, key questions arise: \textit{What are the physical consequences of the SSB of such new symmetries? What are the resulting phases of matter?}
Motivated by these questions, a series of exotic SSB phases of matter, termed \textit{fractonic superfluids}, were proposed in Refs.~\cite{Fractonicsuperfluids1} and have been explored in details in Refs.~\cite{Fractonicsuperfluids1,Fractonicsuperfluids2,NSofFractonicsuperfluids,fracton23,Fractonicsuperfluidsdefect,reviewofFractonicsuperfluids}.
Interestingly, symmetry breaking in the presence of dipole conservation has become  a rapidly developing topic in the literature of  theoretical physics, e.g.,  Refs.~\cite{2023JHEP...09..184J,fracton33,fracton46,fracton8,2024JHEP...07..197J,PhysRevLett.132.071603} where the notions of fracton superfluids and dipole superfluids are introduced.

In fractonic superfluids, the constituent bosons that are condensed are either completely immobile~\cite{Fractonicsuperfluids1} or movable only along fixed directions~\cite{Fractonicsuperfluids2,NSofFractonicsuperfluids,fracton23,Fractonicsuperfluidsdefect}. 
Due to the similar restriction of mobility, these particles are also referred to as fractons and lineons respectively by borrowing the terminology in fracton topological order and fracton codes.
For simplicity, dipole moments, which are first-order moments, are often used in concrete model studies. The mobility restriction on fractons can be realized by imposing conservation of dipole moments, \(\hat{Q}^{(i)} = \int d^{d}\mathbf{x} \, \hat{\rho} x_{i}\)~\cite{Fractonicsuperfluids1}. Additionally, the conservation of angular charge moments (abbreviated as ``angular  moments''), \(\hat{Q}_{a,b} = \int d^{d}\mathbf{x} \, ( \hat{\rho}_{a}x_{b} - \hat{\rho}_{b}x_{a} )\), was explored in Refs.~\cite{Fractonicsuperfluids2,NSofFractonicsuperfluids,fracton23,Fractonicsuperfluidsdefect} to capture the partial restriction of mobility. Here, \(\hat{\rho}\) represents the density operator, and the indices \(a, b\) label the species of bosons.
 The total particle number conservation (i.e., ``charge'' conservation) is implicitly assumed when a higher-moment conservation law is considered.

In Refs.~\cite{Fractonicsuperfluids1,Fractonicsuperfluids2}, minimal models of fractonic superfluids are constructed and analyzed across various aspects, including symmetry transformations, Noether currents, ground-state wavefunctions, Gross-Pitaevskii equations, hydrodynamics, and vortex configurations. Through low-energy effective field theories, it has been shown that conservation of higher moments tends to induce strong quantum fluctuations, making the establishment of  ODLRO more challenging under higher-moment conservation. Specifically, in systems with charge and dipole moment conservation, charge symmetry breaking occurs only in spatial dimensions $d\geq3$ at zero temperature.
Recently, Ref.~\cite{HMWT} demonstrated more generally that charge symmetry breaking is forbidden at finite temperature even in four dimensions as long as dipole conservation is imposed. In Ref.~\cite{NSofFractonicsuperfluids}, continuity equations and Navier-Stokes-like equations were derived from the Euler-Lagrange equations of fractonic superfluids. Critical velocity fields and density currents were obtained, leading to a Landau-like criterion. 
Ref.~\cite{fracton23} examined the emergence of higher-rank symmetries and angular moment conservation at low energies through renormalization group analysis, providing insights into their experimental realization. Meanwhile, Ref.~\cite{Fractonicsuperfluidsdefect} investigated finite-temperature phase transitions in fractonic superfluids, showing that the topological vortices (symmetry defects) proposed in Ref.~\cite{Fractonicsuperfluids2} form a hierarchical structure of bound states, resulting in hierarchical confinement-deconfinement transitions upon increasing temperature.

Bose-Hubbard-type lattice models incorporating higher-moment conservation are also of significant interest, as they extend the conventional Bose-Hubbard model, which realizes the superfluid phase in the weak interaction regime. Dipolar Bose-Hubbard lattice models have been constructed and studied extensively in various spatial dimensions, yielding profound physical insights (e.g., Refs.~\cite{DBHM,DBHM2,fracton31,fracton35,fracton44,fracton45}). In the small-$U$ limit, these lattice models at long wavelengths reduce to the   many-boson model of fractonic superfluids discussed in Ref.~\cite{Fractonicsuperfluids1}.

 In the fractonic superfluid systems and associated Bose-Hubbard lattice models discussed above, the formulation of conserved quantities plays a crucial role in shaping the low-energy physics. Different conserved quantities can lead to entirely distinct outcomes in the interplay between SSB and fracton physics, influencing phenomena such as vortex configurations, the Mermin-Wagner theorem, finite-temperature phase diagrams, and hydrodynamics. Therefore, it is worthwhile to explore new conserved quantities that substantially extend beyond straightforward generalizations of previously studied cases.

Following the research on fractonic superfluids~\cite{Fractonicsuperfluids1,Fractonicsuperfluids2}, in this paper, we continue to consider spontaneous symmetry breaking and other aspects in higher-moment-conserving systems.
We introduce the concept of \textit{hybrid fractonic superfluids} (HFS) by imposing the conservation of \textit{hybrid} moments. HFS represent a class of higher-moment symmetry-breaking phases characterized by hybrid moments, which are constructed from the moments of bosons belonging to different species. Therefore, HFS models inherently involve at least two distinct species of bosons. We systematically investigate HFS and their realizations in Bose-Hubbard lattice models.
Specifically, we focus on two representative series of HFS (referred to as \textit{Model Series A} and \textit{Model Series B}), distinguished by the mechanisms of moment hybridization. Notably, previous studies~\cite{Fractonicsuperfluids2,NSofFractonicsuperfluids,fracton23,Fractonicsuperfluidsdefect} on angular moments can also be literally  regarded as a special subclass of fractonic superfluids with hybrid moments and could be categorized as \textit{Model Series C} if needed. Given the profound physical implications uncovered in these prior studies—which likely represent only the tip of the iceberg of hybridization—it is imperative to systematically explore the broader impacts of hybridization, a key motivation for the present work.   This article, together with Refs.~\cite{Fractonicsuperfluids1,Fractonicsuperfluids2}, forms a trilogy of fractonic superfluids. From the trilogy as well as the extended studies in Refs.~\cite{NSofFractonicsuperfluids,fracton23,Fractonicsuperfluidsdefect,reviewofFractonicsuperfluids}, we attempt to uncover SSB physics of new symmetries brought by new conservation laws in quantum many-body systems.

Leaving technical details and coherent discucssions in the main text, here we summarize key facts of Model Series A, Model Series B, and the Bose-Hubbard-type lattice model respectively.
\begin{itemize}
\item  In Model Series A, we consider the hybridization of moments of the \textit{same} order moments constituted  by bosons of \textit{different} species. 
Arbitary models in Model Series A can be represented by a tuple \([d,N,m,m']\), where \(d\) is the spatial dimension, \(N\) is the order of the moments participating in the hybridization, \(m\) is the total number of species of bosons, and \(m^\prime\) denotes the number of species of bosons involved in the hybridization.
   We specifically study the \([d,1,2,2]\) model of Model Series A in which total dipole moments are conserved. 
In this model, we examine the ground states, Goldstone modes, Noether currents, and correlation functions. 
We find that one of the dispersion relations is linear.
In two spatial dimensions, 
 the correlation function \(\langle\hat{\Phi}_{1}^{\dagger}(\mathbf{x})\hat{\Phi}_{2}(\mathbf{x})\hat{\Phi}_{2}^{\dagger}(\mathbf{0})\hat{\Phi}_{1}(\mathbf{0})\rangle\) saturates to a constant at long distances. 
As a result, the system undergoes the spontaneous breaking of \textit{relative} charge symmetry (denoted as $U(1)_{-,C}$) that is associated with the conservation of  particle number \textit{difference} of different species of bosons, thus supporting true ODLRO.  In other words, the spontaneous breaking of relative charge symmetry in \([d,1,2,2]\) model is triggered by nonzero order parameter 
$\langle\hat{\Phi}_1(\mathbf{x})\hat{\Phi}_2^\dagger(\mathbf{x})\rangle
$.
These   results are also extended to arbitrary \([d,N,m,m^\prime]\) models, and partial spontaneous breaking of charge symmetry occurs in spatial dimensions $d=N+1$.

\item In Model Series B, we consider the hybridization of moments of \textit{different} order constituted  by bosons of \textit{different} species.
 Except for the order requirement, the other settings are the same as those in Model Series A. 
 We specifically study an example with conservation of hybridization of dipole moments of species $\hat{\rho}_1=\hat{\Phi}_1^\dagger\hat{\Phi}_1$ and quadrupole moments of species $\hat{\rho}_2=\hat{\Phi}_2^\dagger\hat{\Phi}_2$, and also find one of the dispersion relations is linear.
In two spatial dimensions, the correlation function $\langle\hat{\Phi}_{2}^{\dagger}(\mathbf{x})\hat{\Phi}_{2}(\mathbf{0})\rangle$ decays in a power law, and the correlation function $\langle\hat{\Phi}_{1}^{\dagger}(\mathbf{x})\hat{\Phi}_{1}(\mathbf{0})\rangle$ saturates to a constant at long distances.
Therefore, the spontaneous breaking of charge symmetry of bosons of species $1$ occurs, and the system has a true ODLRO.
This spontaneous breaking in Model Series B is characterized by the order parameter $\langle\hat{\Phi}_1(\mathbf{x})\rangle$. 
A summary of comparison between different model realizations with two species of bosons is given in Table \ref{break}. 
\begin{table*}[htbp]
\centering
\caption{Comparison of patterns of charge symmetry breaking in four different model realizations with two species of bosons in two spatial dimensions at $T=0$. The specific analysis of the \([d,1,2,2]\) model is given in Sec.~\ref{EFTofM1}, and the hybrid dipole-quadrupole moment conserving model is discussed in Sec.~\ref{EFTofM2}. $U(1)_{1,C}$ and $U(1)_{2,C}$ denote $U(1)$ symmetries associated with the particle number (charge) conservation of first and second species of bosons respectively. $U(1)_{-,C}$ denote the \textit{relative} charge symmetry  associated with the particle number \textit{difference} of the two species of bosons. }
\renewcommand{\arraystretch}{1.6}  
\resizebox{\textwidth}{!}{
\begin{tabular}{l c c c c} 
\hline
\hline
\makecell[l]{Superfluid Models\\ with Two Species of Bosons} & 
\makecell[c]{Two Decoupled\\ Conventional Superfluids} & 
\makecell[c]{Two Decoupled Fractonic Superfluids \cite{Fractonicsuperfluids1}\\ with Separate Dipole Conservations} & 
\makecell[c]{\([d,1,2,2]\) Model\\ in Model Series A} & 
\makecell[c]{Hybrid Dipole-Quadrupole Conserving\\ Model in Model Series B} \\
\hline
Higher Moment Conservation & N/A & Dipole Moments of Two Species & Hybrid Dipole Moments & Hybrid Dipole-Quadrupole Moments \\
 
Broken Charge Symmetry & 
\makecell{$U(1)_{1,C} \otimes U(1)_{2,C}$} & 
\makecell{N/A} & 
\makecell{$U(1)_{-,C}$} & 
\makecell{$U(1)_{1,C}$} \\
 
ODLRO & 
\makecell{\checkmark} & 
\makecell{$\times$} & 
\makecell{$\checkmark$} & 
\makecell{$\checkmark$} \\
 
Order Parameter & 
\makecell{$\langle \hat{\Phi}_1(\mathbf{x}) \rangle$, $\langle \hat{\Phi}_2(\mathbf{x}) \rangle$} & 
\makecell{N/A} & 
\makecell{$\langle\hat{\Phi}_1(\mathbf{x})\hat{\Phi}_2^\dagger(\mathbf{x})\rangle$} & 
\makecell{$\langle\hat{\Phi}_1(\mathbf{x})\rangle$} \\
\hline
\hline
\end{tabular}}
\label{break}
\end{table*}

\item  As the third piece of the trilogy of fractonic superfluids, Model Series A and Model Series B presented in this paper incorporate hybridization of   moments and  significantly enrich the physics of fractonic superfluids. For instance, the hybridization influences the value of critical dimensions for ODLRO of charge operators, see Sec.~\ref{SecCorrelationM1} for more details. While spontaneous breaking of charge symmetry in fractonic superfluids with dipole conservation~\cite{Fractonicsuperfluids1}  provides prototypical  case study of the generalized Mermin-Wagner theorem~\cite{fracton30,HMWT},   the hybrid moment conservation presented in this work provides a new platform for generalized Mermin-Wagner theorems.

\item We further construct Bose-Hubbard-type lattice models  whose small Hubbard interaction energy ($U$) limit recovers  Model Series A and Model Series B at long wavelengths. To obtain the global phase diagram of lattice models, we perform a mean-field approximation for \([d,1,2,2]\) lattice model (i.e., the  Hubbard model corresponding to the HFS labeled by \([d,1,2,2]\)).
The mean-field results indicate that the introduction of hybridization brings exotic intermediate phases between Mott insulator (MI) phase and HFS phase. It will be  valuable to study these newly-introduced Bose-Hubbard models by using large-scale numerical simulations, such as quantum Monte Carlo and tensor network. Finally, 
we also discuss the realization of \([d,1,2,2]\) lattice model in   \textit{strongly tilted optical lattices} by means of a third-order perturbation theory. A summary of comparison between Bose-Hubbard-type models with different conservation laws is given in Table \ref{BHMtable}.
\begin{table}[htbp]
  \caption{Comparison of three different Bose-Hubbard-type models. This table illustrates the conserved quanities and order parameters of these lattice models. The order parameters  are applied to characterize  exotic intermediate phases between MI (Mott insulating phase) and HFS (hybrid fractonic superfluid).}
  \renewcommand{\arraystretch}{1.6}  
\resizebox{\columnwidth}{!}{
  \begin{tabular}{l c c}
      \hline
      \hline
      Lattice model & Conserved quanities & Order parameter \\
      \hline
      Bose Hubbard Model [Ref.~\cite{Zhai_2021_BHM}]& \makecell[c]{Charge} & $\langle\hat{b}_\mathbf{i}\rangle$\\
      Dipolar Bose Hubbard Model [Ref.~\cite{DBHM}]& \makecell[c]{Charge \& Dipole moment} & $\langle\hat{b}_\mathbf{i}\rangle$ \& $\langle \hat{b}_{\mathbf{i}+\hat{x}_i}^\dagger\hat{b}_{\mathbf{i}}\rangle$\\
      \makecell[l]{Lattice Model Series A [Eq.~(\ref{LM1})]} & \makecell[c]{Charge \& Hybrid \\dipole-dipole moment} & $\langle\hat{b}_{a,\mathbf{i}}\rangle$ \& $\langle \hat{b}_{a,\mathbf{i}+\hat{x}_i}^\dagger\hat{b}_{a,\mathbf{i}}\rangle$\& $\langle \hat{b}_{1,\mathbf{i}}^\dagger\hat{b}_{2,\mathbf{i}}\rangle$\\
      \hline
      \hline
  \end{tabular}}
  \label{BHMtable}
\end{table}

\end{itemize}

The remaining part of this paper is organized as follows. In Sec.~\ref{Conserved quantities and the model Hamiltonian}, previously studied fractonic superfluids are first reviewed in brief. Then the idea of hybridization is introduced in order to construct two series of  minimal models (Model Series A and B) in the continuum space.  The two series of models are studied respectively in Sec.~\ref{EFTofM1} and \ref{EFTofM2}. In Sec.~\ref{Lattice model},   Bose-Hubbard-type lattice models are constructed.   Sec.~\ref{Summary and Outlook} provides a summary of this paper and offers additional perspectives for future research.

\section{Construction of Hamiltonian of hybrid moments}

\label{Conserved quantities and the model Hamiltonian} In this section, we start with a review of fractonic superfluids without hybridization~\cite{Fractonicsuperfluids1}. Then, by introducing two or more species of bosons in the system, we subsequently develop two distinct series of models with   hybridization. During the procedure of model construction, we introduce the definition of conserved quantities and associated symmetry transformations on boson operators.

\subsection{Review of fractonic superfluids}

Initially, the concept of fractonic superfluids was  introduced in   single-component boson systems where    multipole moments are conserved.
We set $\hat{\Phi}^\dagger(\mathbf{x})$ and $\hat{\Phi}(\mathbf{x})$ to be bosonic creation and annihilation operators and $\hat{\rho}(\mathbf{x})=\hat{\Phi}^\dagger(\mathbf{x})\hat{\Phi}(\mathbf{x})$ is a density operator. 
The creation and annihilation operators satisfy the standard commutation relations $[\hat{\Phi}(\mathbf{x}),\hat{\Phi}^\dagger(\mathbf{y})]=\delta^d(\mathbf{x}-\mathbf{y})$, where $\mathbf{x}=(x_1,\cdots,x_d)$ is a spatial coordinate and we neglect it in the following for symbol convenience.
We formulate the charges related to the multipole moments systematically as 
\begin{align}
\hat{Q}_{n,d}^{(i)}=\int d^{d}\mathbf{x}\hat{\rho}\mathcal{M}_{n,d}^{(i)}
\,,
\end{align}
where $\mathcal{M}_{n,d}^{(i)}$ denotes the monomials obtained by multiplying the spatial rectangular coordinates with the coefficients equal to $1$.
The notation $n$ stands for the number of multiplied spatial rectangular coordinates in $\mathcal{M}_{n,d}^{(i)}$, and $d$ stands for spatial dimensions.
Each $\hat{Q}_{n,d}^{(i)}$ denotes one component of the $n$-order moment in $d$ spatial dimensions.
The notation $i$ denotes a index of the monomial for $n$th-order moments in the lexicographic ordering\footnote{In the lexicographic ordering, the arrangement of monomials obtained by multiplying the same number of $x$'s follows the rule: for two monomials, the order of $x_1$ is firstly compared, with the monomial with the greater order of $x_1$ coming first. If two monomials have the same order of $x_1$, compare the order  of $x_2$, and so on.
For instance, in two spatial dimensions, we can sort the monomials of quadrupole moments as $\mathcal{M}_{2,2}^{(1)}=x_{1}^{2}$, $\mathcal{M}_{2,2}^{(2)}=x_{1}x_{2}$, and 
$\mathcal{M}_{2,2}^{(3)}=x_{2}^{2}$.
In three spatial dimensions, we can sort the monomials of quadrupole moments as $\mathcal{M}_{2,3}^{(1)}=x_{1}^{2}$, $\mathcal{M}_{2,3}^{(2)}=x_{1}x_{2}$, $\mathcal{M}_{2,3}^{(3)}=x_{1}x_{3}$, $\mathcal{M}_{2,3}^{(4)}=x_{2}^{2}$, $\mathcal{M}_{2,3}^{(5)}=x_{2}x_{3}$, and $\mathcal{M}_{2,3}^{(6)}=x_{3}^{2}$.}, and the index $i$ is listed from $1$ to $C_{d-1+n}^{n}=\frac{(d-1+n)!}{(d-1)!n!}$.

 In addition, the conservation of charge $\hat{Q}=\int d^d\mathbf{x}\hat{\rho}$ is also required.
Given the maximal order of the monomials as $N$, we can construct a symmetry group $\mathcal{G}$ with an element 
\begin{align}
U=\exp\bigg[-i(\lambda\hat{Q}+\sum_{n=1}^{N}\sum_{i=1}^{C_{d-1+n}^{n}}\lambda_{n,d}^{(i)}\hat{Q}_{n,d}^{(i)})\bigg]
\end{align} which leads to the following transformation on boson operators:
\begin{align}
\hat{\Phi}^{\prime}=U\hat{\Phi}U^{\dagger}=\hat{\Phi}e^{i(\lambda+\sum_{n}^{N}\sum_{i}^{C_{d-1+n}^{n}}\lambda_{n,d}^{(i)}\mathcal{M}_{n,d}^{(i)})}\,.
\end{align}
 Here, $\lambda, \lambda_{n,d}^{(i)}\in\mathbb{R}$ are group parameters. Hereafter, 
this  symmetry will be alternatively called a higher-rank symmetry since the symmetry leads to  higher-rank tensor gauge fields upon gauging.  The kinetic term $\mathcal{H}_{0}$ is no longer Gaussian, e.g. the Hamiltonian constructed in the Appendix of Ref.~\cite{Fractonicsuperfluids1}: 
\begin{align}
\mathcal{H}_{0}= & \sum_{i_{1},\cdots,i_{N+1}}^{d}K^{i_{1}\cdots i_{N+1}} (\hat{\Phi}^{\dagger})^{N+1} (\nabla_{i_{1}\cdots i_{N+1}}\log\hat{\Phi}^{\dagger})\nonumber\\
 & \hat{\Phi}^{N+1}(\nabla_{i_{1}\cdots i_{N+1}}\log\hat{\Phi}),
\label{Htype0}
\end{align}
where $\nabla_{i_{1}\cdots i_{n}}=\partial_{i_{1}}\cdots\partial_{i_{n}}$ and symmetric coefficients $K^{i_{1}\cdots i_{N+1}}$. 
The operator $\log \hat{\Phi}$ is a very useful notation to simplify the Hamiltonian, see Appendix of Ref.~\cite{Fractonicsuperfluids1}.
For convenience, we set the coefficients $K^{i_{1}\cdots i_{N+1}}>0$. We can apply the potential term $V(\hat{\Phi}^{\dagger},\hat{\Phi})=-\mu\hat{\Phi}^{\dagger}\hat{\Phi}+\frac{g}{2}\hat{\Phi}^{\dagger}\hat{\Phi}^{\dagger}\hat{\Phi}\hat{\Phi}$, where $\mu$ is the chemical potential and $g>0$ describes onsite repulsive interaction. A fractonic superfluid phase can be reached for a positive chemical potential.

To get an understanding on the fractonic superfluid phase, we can look at the following two examples. When the dipole moments are conserved, the kinetic term takes the form as
\begin{align}
\mathcal{H}_{0}= & \sum_{i,j}^{d}K^{ij}(\hat{\Phi}^{\dagger}\partial_{i}\partial_{j}\hat{\Phi}^{\dagger}-\partial_{i}\hat{\Phi}^{\dagger}\partial_{j}\hat{\Phi}^{\dagger})\nonumber\\
 & (\hat{\Phi}\partial_{i}\partial_{j}\hat{\Phi}-\partial_{i}\hat{\Phi}\partial_{j}\hat{\Phi}).
\label{dipolefracton}
\end{align}
The Hamiltonian is more complicated when the quadrupole moments are conserved: 
\begin{align}
&\mathcal{H}_{0}=\sum_{i,j,k}^{d}K^{ijk}\big(\hat{\Phi}^{\dagger2}\partial_{i}\partial_{j}\partial_{k}\hat{\Phi}^{\dagger}-\hat{\Phi}^{\dagger}\partial_{i}\hat{\Phi}^{\dagger}\partial_{j}\partial_{k}\hat{\Phi}^{\dagger}\nonumber\\
&-\hat{\Phi}^{\dagger}\partial_{j}\hat{\Phi}^{\dagger}\partial_{i}\partial_{k}\hat{\Phi}^{\dagger}-\hat{\Phi}^{\dagger}\partial_{k}\hat{\Phi}^{\dagger}\partial_{i}\partial_{j}\hat{\Phi}^{\dagger}\nonumber\\
&+2\partial_{i}\hat{\Phi}^{\dagger}\partial_{j}\hat{\Phi}^{\dagger}\partial_{k}\hat{\Phi}^{\dagger}\big)\big(\hat{\Phi}^{2}\partial_{i}\partial_{j}\partial_{k}\hat{\Phi}-\hat{\Phi}\partial_{i}\hat{\Phi}\partial_{j}\partial_{k}\hat{\Phi}\nonumber\\
&-\hat{\Phi}\partial_{j}\hat{\Phi}\partial_{i}\partial_{k}\hat{\Phi}-\hat{\Phi}\partial_{k}\hat{\Phi}\partial_{i}\partial_{j}\hat{\Phi}+2\partial_{i}\hat{\Phi}\partial_{j}\hat{\Phi}\partial_{k}\hat{\Phi}\big).\label{quadrupolefracton}
\end{align}
For the two examples, the dynamics of a single particle are strongly restricted while a bound state becomes free when the cooperative motions are required to conserve multipole moments: dipole moments for Hamiltonian in Eq.~(\ref{dipolefracton}) and quadrupole moments for Eq.~(\ref{quadrupolefracton}). The corresponding fractonic superfluid phase then has one Goldstone mode with a high-order dispersion.
For a general $N$ as the largest order of the monomials, the effective theory describes the Goldstone mode with the dispersion relation $\omega\propto\left|\mathbf{k}\right|^{N+1}$.
Systems characterized by Goldstone mode with dispersion relation $\omega\propto\left|\mathbf{k}\right|^{N+1}$ exhibit ODLRO in spatial dimensions $d\geq N+2$, as shown in Ref.~\cite{Fractonicsuperfluids1}.
Therefore, this system breaks its charge symmetry in spatial dimensions $d\geq N+2$.
The spontaneous breaking of charge symmetry in fractonic superfluids manifests as a consequence of the generalized Mermin-Wagner theorem~\cite{fracton30,HMWT}.

\subsection{Construction of Model Series A denoted as $[d,N,m,m^\prime]$}

Next, we begin to construct models with hybridization of moments. We first focus on  Model Series A in which hybrid moments   are formed by moments of the same order.  The general construction looks very tedious. The readers who are more interested in concrete models can directly read from Eq.~(\ref{ddcon00}).

Consider a system with $m$ species of bosons, 
the operators $\hat{\Phi}_a^\dagger(\mathbf{x})$ and $\hat{\Phi}_a(\mathbf{x})$ create and annihilate a boson of species $a$ respectively, and satisfy the commutation relations $[\hat{\Phi}_a(\mathbf{x}),\hat{\Phi}_b^\dagger(\mathbf{y})]=\delta_{ab}\delta^d(\mathbf{x}-\mathbf{y})$. 
In such a system, we consider the following conserved quantities in Eqs.~(\ref{type1con1}, \ref{type1con2}, \ref{type1con3}, \ref{type1con4}):
\begin{itemize}
  \item The charges of bosons of species $a$ are conserved (the range of $a$ is $1\leq a\leq m$):
   \begin{align}
    \hat{Q}_{a}=\int d^{d}\mathbf{x}\hat{\rho}_{a}\,.\label{type1con1}
  \end{align} 
  \item The hybrid moments defined below are conserved. $N$ is the order of the moments which participate in the hybridization (the range of the label $i$ is $1\leq i\leq C_{d-1+N}^{N}$):
  \begin{align}
    \hat{Q}_{\text{mix}}^{(i)} = \int d^{d}\mathbf{x}  \sum_{a=1}^{m'} \hat{\rho}_{a}  \mathcal{M}_{N,d}^{(i)}\,.\label{type1con2}
  \end{align} 
  
  \item All the moments whose orders are lower than $N$ are conserved (the ranges of $a$, $n$, and $i$ are $1\leq a\leq m$,  $1\leq n\leq N-1$, and $1\leq i\leq C_{d-1+n}^{n}$ respectively): 
  \begin{align}
    \hat{Q}_{a,n,d}^{(i)}=\int d^{d}\mathbf{x}\hat{\rho}_{a}\mathcal{M}_{n,d}^{(i)}\,.\label{type1con3}
  \end{align}
  
  \item The moments of bosons of species $m'+1\leq a\leq m$ do not participate in the hybridization. Their $N$th-order moments are independently conserved (the ranges of $a$ and $i$ are $m'+1\leq a\leq m$, $1\leq i\leq C_{d-1+N}^{N}$):
  \begin{align}
    \hat{Q}_{a,N,d}^{(i)}=\int d^{d}\mathbf{x}\hat{\rho}_{a}\mathcal{M}_{N,d}^{(i)}\,.\label{type1con4}
  \end{align}
\end{itemize}

Compared with the separate conservation of the $N$th-order moments for each species of bosons, the number of conserved quantities in the present hybridization case decreases. That means freeing up some mobility of bosons.
We can see this better in the corresponding symmetry. 
We denote a group $\mathcal{G}$ generated by the conserved quantities [Eqs.~(\ref{type1con1}, \ref{type1con2}, \ref{type1con3}, \ref{type1con4})], and an element of the group 
\begin{align}
U= & \exp\Bigg[-i\sum_{a=1}^{m}(\lambda_{a}\hat{Q}_{a}+\sum_{n}^{N-1}\sum_{i}^{C_{d-1+n}^{n}}\lambda_{a,n}^{(i)}\hat{Q}_{a,n,d}^{(i)})\nonumber\\
 & -i\sum_{i}^{C_{d-1+N}^{N}}\lambda_{\text{mix}}^{(i)}\hat{Q}_{\text{mix}}^{(i)}-i\sum_{a=m'+1}^{m}\sum_i^{C_{d-1+N}^N}\lambda_{a,N}^{(i)}\hat{Q}_{a,N,d}^{(i)}\Bigg]
\end{align}
leads to the following transformation on boson operators: 
\begin{align}
  &\hat{\Phi}_{a}^{\prime}=U\hat{\Phi}_{a}U^{\dagger}\nonumber\\
  =&\hat{\Phi}_{a}\exp\bigg\{i\Big[\lambda_{a}+\sum_{n}^{N-1}\sum_{i}^{C_{d-1+n}^{n}}\lambda_{a,n}^{(i)}\mathcal{M}_{n,d}^{(i)}+\sum_{b=1}^{m'}\delta_{ab}\nonumber\\
  &\sum_{i}^{C_{d-1+N}^{N}}\lambda_{\text{mix}}^{(i)}\mathcal{M}_{N,d}^{(i)}+\sum_{b=m'+1}^{m}\delta_{ab}\sum_{i}^{C_{d-1+N}^{N}}\lambda_{a,N}^{(i)}\mathcal{M}_{N,d}^{(i)}\Big]\bigg\}\,,
\end{align}
where $\lambda_a, \lambda_{a,n}^{(i)}, \lambda_{\text{mix}}^{(i)}\in\mathbb{R}$ are group parameters.
The group $\mathcal{G}$ can be written as the direct product of $U(1)$ groups,
\begin{align}
  \mathcal{G}=\bigg(\bigotimes_a^m U(1)_{a,C}\bigg)\otimes\bigg(\bigotimes_a^m\bigotimes_{n,i}U(1)_{a,n,i}\bigg)\nonumber\\
  \otimes\bigg(\bigotimes_i U(1)_{\text{mix-A},i}\bigg)\otimes\bigg(\bigotimes_{a=m'+1}^{m}\bigotimes_i U(1)_{a,N,i}\bigg)\,
\end{align}
where   group elements of  $U(1)_{a,C}$, $U(1)_{a,n,i}$, and $U(1)_{\text{mix-A},i}$ are respectively labeled by  $\lambda_a$, $\lambda_{a,n}^{(i)}$, and $\lambda_{\text{mix}}^{(i)}$.
The notation $\bigotimes_{\cdots} U(1)_{\cdots}$ here denotes the direct product of multiple corresponding $U(1)$ groups.

A Hamiltonian satisfying this conservation law needs to commute with the conserved quantities [Eqs.~(\ref{type1con1}, \ref{type1con2}, \ref{type1con3}, \ref{type1con4})].
The minimal Hamiltonian can be written as: 
\begin{align}
&\mathcal{H}_{0}\!=\! \sum_{a}^{m}\!\sum_{i_{1},\cdots,i_{N+1}}^{d}\!K_{a}^{i_{1}\cdots i_{N+1}}\!(\hat{\Phi}_{a}^{\dagger})^{N+1}\!(\nabla_{i_{1},\cdots,i_{N+1}}\log\hat{\Phi}_{a}^{\dagger})\nonumber\\
&\hat{\Phi}_{a}^{N+1}(\nabla_{i_{1},\cdots,i_{N+1}}\log\hat{\Phi}_{a})+\sum_{a\neq b}^{m'}\sum_{i_{1},\cdots,i_{N}}^{d}\Gamma_{a,b}^{i_{1}\cdots i_{N}}(\hat{\Phi}_{a}^{\dagger})^{N}\nonumber\\
&(\hat{\Phi}_{b}^{\dagger})^{N}[(\nabla_{i_{1},\cdots,i_{N}}\log\hat{\Phi}_{a}^{\dagger})-(\nabla_{i_{1},\cdots,i_{N}}\log\hat{\Phi}_{b}^{\dagger}) ]\nonumber\\
 & \hat{\Phi}_{a}^{N}\hat{\Phi}_{b}^{N} [(\nabla_{i_{1},\cdots,i_{N}}\log\hat{\Phi}_{a})-(\nabla_{i_{1},\cdots,i_{N}}\log\hat{\Phi}_{b}) ].
\label{Htype1}
\end{align}
For convenience, we also set $K_{a}^{i_{1}\cdots i_{N+1}}>0$, $\Gamma_{a,b}^{i_{1}\cdots i_{N}}>0$.
In addition, $\Gamma_{a,b}^{i_{1}\cdots i_{N}}=\Gamma_{b,a}^{i_{1}\cdots i_{N}}$ is needed.

In the construction above, the spatial dimensions $d$, the order $N$, the number of boson species involved in hybridization $m^\prime$, and the total number of boson species $m$ are tunable for Model Series A.
A tuple $[d,N,m,m^\prime]$ can be used to uniquely denote one example of Model Series A.

With the above preparation, we  begin to construct the simplest example of Model Series A, denoted as \([d,1,2,2]\) model.
\([d,1,2,2]\) model can be regarded as a system in which the total dipole moments are conserved along with the conservation of the charges of each species,
\begin{align}
  \hat{Q}_{a}=&\int d^{d}\mathbf{x}\hat{\rho}_{a}\qquad (a=1,2)\,,\label{ddcon00}\\
  \hat{Q}_{\text{mix}}^{(i)}=&\int d^{d}\mathbf{x}(\hat{\rho}_{1}+\hat{\rho}_{2})x_{i}\quad (i=1,\cdots,d)\,.
\label{ddcon}
\end{align}
The total symmetry group $\mathcal{G}$ can be written as 
\begin{align}
  \mathcal{G}=(U(1)_{1,C}\otimes U(1)_{2,C})\otimes(\bigotimes_i U(1)_{\text{mix-A},i} )\,.
\end{align}
An element of the group $\mathcal{G}$ is given by
\begin{align}
U=\exp\big[-i\big(\lambda_{1}\hat{Q}_{1}+\lambda_{2}\hat{Q}_{2}+\sum_{i}^{d}\lambda_{\text{mix}}^{(i)}\hat{Q}_{\text{mix}}^{(i)}\big)\big]\,.
\end{align}
And the transformation on field operators can be written as
\begin{align}
\hat{\Phi}_{a}^{\prime}=U\hat{\Phi}_{a}U^{\dagger}=\hat{\Phi}_{a}\exp\big[i\big(\lambda_{a}+\sum_{i}^{d}\lambda_{\text{mix}}^{(i)}x_{i}\big)\big]\,.
\end{align}
To satisfy this conservation law, the minimal Hamiltonian of \([d,1,2,2]\) model with  an energy potential term is 
\begin{align}
\mathcal{H}= & \sum_{a}^{2}\sum_{i,j}^{d}K_{a}^{ij} (\hat{\Phi}_{a}^{\dagger}\partial_{i}\partial_{j}\hat{\Phi}_{a}^{\dagger}-\partial_{i}\hat{\Phi}_{a}^{\dagger}\partial_{j}\hat{\Phi}_{a}^{\dagger} )\nonumber\\
 &  \cdot(\hat{\Phi}_{a}\partial_{i}\partial_{j}\hat{\Phi}_{a}-\partial_{i}\hat{\Phi}_{a}\partial_{j}\hat{\Phi}_{a} )\nonumber\\
 & +\sum_{a\neq b}^{2}\sum_{i}^{d}\Gamma_{a,b}^{i} (\hat{\Phi}_{a}^{\dagger}\partial_{i}\hat{\Phi}_{b}^{\dagger}-\hat{\Phi}_{b}^{\dagger}\partial_{i}\hat{\Phi}_{a}^{\dagger} )\nonumber\\
 & 
\cdot (\hat{\Phi}_{a}\partial_{i}\hat{\Phi}_{b}-\hat{\Phi}_{b}\partial_{i}\hat{\Phi}_{a} )+V (\hat{\Phi}^{\dagger},\hat{\Phi} )\,,
\label{Hmdd}
\end{align}
where $V (\hat{\Phi}^{\dagger},\hat{\Phi} )=\sum_{a}^{2}-\mu_{a}\hat{\Phi}_{a}^{\dagger}\hat{\Phi}_{a}+\frac{g}{2}\hat{\Phi}_{a}^{\dagger}\hat{\Phi}_{a}^{\dagger}\hat{\Phi}_{a}\hat{\Phi}_{a}$, and the coefficients satisfy $K_{a}^{ij}>0$, $\Gamma_{a,b}^{i}>0$, $K_{a}^{ij}=K_{a}^{ji}$, $\Gamma_{a,b}^{i}=\Gamma_{b,a}^{i}$, and $g>0$. We can also derive the corresponding conserved quantities and Noether currents. For the conserved charge $\hat{Q}_{a}= \int d^{d}\mathbf{x}\hat{\Phi}_{a}^{\dagger}\hat{\Phi}_{a}=\int d^{d}\mathbf{x}\hat{\rho}_{a}$,  the current is given by ($i=1,\cdots,d$):
\begin{align}
\hat{J}_{i}^{a}& = i\sum_{j}^{d}K_{a}^{ij}\partial_{j} [\hat{\Phi}_{a}^{\dagger 2}(\hat{\Phi}_{a}\partial_{i}\partial_{j}\hat{\Phi}_{a}-\partial_{i}\hat{\Phi}_{a}\partial_{j}\hat{\Phi}_{a})-\text{h.c.}]\nonumber\\
 & +2i\sum_{b\neq a}^{2}\Gamma_{a,b}^{i} [ (\hat{\rho}_{a}\hat{\Phi}_{b}\partial_{i}\hat{\Phi}_{b}^{\dagger}-\hat{\rho}_{b}\hat{\Phi}_{a}\partial_{i}\hat{\Phi}_{a}^{\dagger} )-\text{h.c.} ]\,.\label{Ncurrent1}
\end{align}
For the hybrid moments $\hat{Q}_{\text{mix}}^{(j)}=\int d^{d}\mathbf{x}(\hat{\rho}_{1}x_{j}+\hat{\rho}_{2}x_{j})=\int d^{d}\mathbf{x}\hat{\rho}_{\text{mix}}^{(j)}$,  the current is given by ($i=1,\cdots,d$):
\begin{align}
\hat{\mathcal{D}}_{i}^{(j)}= & x_{j}\hat{J}_{i}^{1}+x_{j}\hat{J}_{i}^{2}-i\sum_{a}^{2}K_{a}^{ij}\nonumber\\
 & [\hat{\Phi}_{a}^{\dagger 2} (\hat{\Phi}_{a}\partial_{i}\partial_{j}\hat{\Phi}_{a}-\partial_{i}\hat{\Phi}_{a}\partial_{j}\hat{\Phi}_{a} )-\text{h.c.} ]\,.\label{Ncurrent2}
\end{align}

The $\Gamma_{a,b}^{i}$ term is the emerging term from the hybridization of moments. 
This Hamiltonian density [Eq.~(\ref{Hmdd})] is the simplest example of HFS in the Model Series A. The effective field theory of this Hamiltonian will be studied in Sec. \ref{EFTofM1}.

\subsection{Construction of Model Series B}

In the construction of Model Series B, in a system with $m$ species of bosons, we consider the hybridization of moments of different orders from different species. 
Although general formulas in Model Series B  are expected to be more complex than Model Series A,  let us  present conserved quantities  below for completeness (a concrete example will be provided shortly): 
\begin{itemize}
  \item The charges of bosons of species $a$ are conserved (the range of $a$ is $1\leq a\leq m$):  
  \begin{align}
    \hat{Q}_{a}=  \int d^{d}\mathbf{x}\hat{\rho}_{a}\,,\label{type2con1}
  \end{align}
  
  \item The hybrid moments defined below are conserved:  
   \begin{align}
    \hat{Q}_{\text{mix}}^{(i_{1})\cdots(i_{m^\prime})}=  \int d^{d}\mathbf{x}\sum_{a}^{m^\prime}\hat{\rho}_{a}\mathcal{M}_{N_{a},d}^{(i_{a})}\ell^{N_{m^\prime}-N_{a}}\,.\label{type2con2}
  \end{align}
 Here, \(m^\prime\) denotes the number of species of bosons involved in the hybridization.
 We use $N_a$ to denote the orders of the moments of the species $a$ which participate in the hybridization. 
  For briefness, we let $N_{a}$ increase as $a$ increase, i.e. $N_{1}<N_{2}\cdots<N_{m^\prime}$.
  $\ell$ with a dimension of length denotes the coefficient by which moments of different orders are converted to each other. The introduction of $\ell$ ensures that the quantities in $\hat{Q}_{\text{mix}}^{(i_{1})\cdots(i_{m^\prime})}$ have the same dimensions. 
  For each $\hat{Q}_{\text{mix}}^{(i_{1})\cdots(i_{m^\prime})}$, the index $i_a$ can arbitary taken in the range $1\leq i_a\leq C_{d-1+N_a}^{N_a}$, but cannot be repeated with other $\hat{Q}_{\text{mix}}^{(i_{1})\cdots(i_{m^\prime})}$.
  The number of the conserved quantities $\hat{Q}_{\text{mix}}^{(i_{1})\cdots(i_{m^\prime})}$ is $C_{d-1+N_1}^{N_1}$. 
  \item All the moments of species $a$ whose orders are lower than $N_a$ are conserved (the ranges of $a$, $n$, and $i$ are $1\leq a\leq m^\prime$,  $1\leq n\leq N_a-1$, and $1\leq i\leq C_{d-1+n}^{n}$ respectively):
   \begin{align}
    \hat{Q}_{a,n,d}^{(i)}= \int d^{d}\mathbf{x}\hat{\rho}_{a}\mathcal{M}_{n,d}^{(i)}\,.\label{type2con3}
  \end{align}
  \item The components of $N_a$th-order moments which do not participate in the hybridization are conserved:
  \begin{align}
    \hat{Q}_{a,N_a,d}^{(i_a)}=  \int d^{d}\mathbf{x}\hat{\rho}_{a}\mathcal{M}_{N_a,d}^{(i_a)}\,.\label{type2con4}
  \end{align}
   Due to the fact that the numbers of components of different order moments are not equal in two spatial dimensions and above, some components of the $N_{a}$th-order moments do not participate in the hybridization, and they are required to be conserved in Model Series B.
  The index $i_a$ takes all $1\leq i_a\leq C_{d-1+N_a}^{N_a}$ that are not taken in Eq.~(\ref{type2con2}). The range of $a$ is $1\leq a\leq m^\prime$.
\end{itemize}

The above conserved quantities look apparently more tedious than Model Series A. In the following, we directly move to concrete model construction. In order to present a  concrete model as simple as possible, we assume $m^\prime=m$ and  fix the coefficient $m$ and $N_{a}$, e.g, $m=2$, $N_{1}=1$, and $N_{2}=2$. 
In a system with two species of bosons, we mix dipole moments of bosons of species 1 and quadrupole moments of bosons of species 2. For convenience, the conserved quantities are given by: 
\begin{align}
\hat{Q}_{a}= & \int d^{d}\mathbf{x}\hat{\rho}_{a}\quad(a=1,2)\,,\label{dqcon1}\\
\hat{Q}_{2}^{(i)}= & \int d^{d}\mathbf{x}\hat{\rho}_{2}x_{i}\quad(i=1,\cdots,d)\,,\label{dqcon2}\\
\hat{Q}_{\text{mix}}^{i,ii}= & \int d^{d}\mathbf{x}(\hat{\rho}_{1}x_{i}\ell+\hat{\rho}_{2}x_{i}^{2})\quad(i=1,\cdots,d)\,,\label{dqcon3}\\
\hat{Q}_{2}^{ij}= & \int d^{d}\mathbf{x}\hat{\rho}_{2}x_{i}x_{j}\quad(i,j=1,\cdots,d,\quad i\neq j)\,.
\label{dqcon4}
\end{align}
We denote a group $\mathcal{G}$ generated by the conserved quantities [Eqs.~(\ref{dqcon1}, \ref{dqcon2}, \ref{dqcon3}, \ref{dqcon4})], and an element of the group 
\begin{align}
U= & \exp\big[-i\big(\lambda_{1}\hat{Q}_{1}+\lambda_{2}\hat{Q}_{2}+\sum_{i}^{d}\lambda_{2}^{(i)}\hat{Q}_{2}^{i}\nonumber\\
 & +\sum_{i}^{d}\lambda_{\text{mix}}^{(i)}\hat{Q}_{\text{mix}}^{i,ii}+\sum_{i\neq j}^{d}\lambda_{2}^{(ij)}\hat{Q}_{2}^{ij}\big)\big]
\end{align}
leads to the following transformations on field operators: 
\begin{align}
\hat{\Phi}_{1}^{\prime}=  U\hat{\Phi}_{1}U^{\dagger}=&\hat{\Phi}_{1}\exp\Big[i\big(\lambda_{1}+\sum_{i}^{d}\lambda_{\text{mix}}^{(i)}x_{i}\ell\big)\Big],\\
\hat{\Phi}_{2}^{\prime}=  U\hat{\Phi}_{2}U^{\dagger}=&\hat{\Phi}_{2}\exp\Big[i\big(\lambda_{2}+\sum_{i}^{d}\lambda_{2}^{(i)}x_{i} +\sum_{i}^{d}\lambda_{\text{mix}}^{(i)}x_{i}^{2}\nonumber\\
 &+\sum_{i\neq j}^{d}\lambda_{2}^{(ij)}x_{i}x_{j}\big)\Big]\,,
\end{align}
where $\lambda_1, \lambda_2, \lambda_{2}^{(i)}, \lambda_{\text{mix}}^{(i)}, \lambda_2^{(ij)}\in\mathbb{R}$ are group parameters.
The group $\mathcal{G}$ can be written as the direct product of $U(1)$ groups,
\begin{align}
  \mathcal{G}=(\bigotimes_a^2U(1)_{a,C})\otimes(\bigotimes_i U(1)_{2,i}) 
  \otimes(\bigotimes_i U(1)_{\text{mix-B},i})\nonumber\\\otimes(\bigotimes_{i\neq j} U(1)_{2,\{i,j\}})\,,
\end{align}
where group elements of  $U(1)_{a,C}$, $U(1)_{2,i}$, $U(1)_{\text{mix-B},i}$, and $U(1)_{2,\{i,j\}}$ are respectively labeled by   $\lambda_a$, $\lambda_{2}^{(i)}$, $\lambda_{\text{mix}}^{(i)}$, $\lambda_2^{(ij)}$.
 
 A Hamiltonian satisfying this conservation law needs to commute with conserved quantities [Eqs.~(\ref{dqcon1}, \ref{dqcon2}, \ref{dqcon3}, \ref{dqcon4})].
We can write down the minimal Hamiltonian of Model Series B with the potential terms,
  \begin{align}
&    \mathcal{H}= \nonumber\\
  & \!  \sum_{i,j}^{d}\!K_{1}^{ij}\!(\hat{\Phi}_{1}^{\dagger}\partial_{i}\partial_{j}\hat{\Phi}_{1}^{\dagger}-\partial_{i}\hat{\Phi}_{1}^{\dagger}\partial_{j}\hat{\Phi}_{1}^{\dagger})\!\!(\hat{\Phi}_{1}\partial_{i}\partial_{j}\hat{\Phi}_{1}-\partial_{i}\hat{\Phi}_{1}\partial_{j}\hat{\Phi}_{1})\nonumber\\
     &\! +\!\!\sum_{i,j,k}^{d}\!K_{2}^{\!ijk\!}\!\bigg(\hat{\Phi}_{2}^{\dagger2}\partial_{i}\partial_{j}\partial_{k}\hat{\Phi}_{2}^{\dagger}\!-\!\hat{\Phi}_{2}^{\dagger}\partial_{i}\hat{\Phi}_{2}^{\dagger}\partial_{j}\partial_{k}\hat{\Phi}_{2}^{\dagger}\!-\!\hat{\Phi}_{2}^{\dagger}\partial_{j}\hat{\Phi}_{2}^{\dagger}\partial_{i}\partial_{k}\hat{\Phi}_{2}^{\dagger}\nonumber\\
&     -\hat{\Phi}_{2}^{\dagger}\partial_{k}\hat{\Phi}_{2}^{\dagger}\partial_{i}\partial_{j}\hat{\Phi}_{2}^{\dagger}+2\partial_{i}\hat{\Phi}_{2}^{\dagger}\partial_{j}\hat{\Phi}_{2}^{\dagger}\partial_{k}\hat{\Phi}_{2}^{\dagger}\bigg)\nonumber\\
     & \cdot\bigg(\hat{\Phi}_{2}^{2}\partial_{i}\partial_{j}\partial_{k}\hat{\Phi}_{2}-\hat{\Phi}_{2}\partial_{i}\hat{\Phi}_{2}\partial_{j}\partial_{k}\hat{\Phi}_{2}-\hat{\Phi}_{2}\partial_{j}\hat{\Phi}_{2}\partial_{i}\partial_{k}\hat{\Phi}_{2}\nonumber\\
     &-\hat{\Phi}_{2}\partial_{k}\hat{\Phi}_{2}\partial_{i}\partial_{j}\hat{\Phi}_{2}+2\partial_{i}\hat{\Phi}_{2}\partial_{j}\hat{\Phi}_{2}\partial_{k}\hat{\Phi}_{2}\bigg)\nonumber\\
     & +\sum_{i}^{d}\Gamma_{1,2}^{i}(\ell\hat{\Phi}_{1}^{\dagger}\hat{\Phi}_{2}^{\dagger}\partial_{i}^{2}\hat{\Phi}_{2}^{\dagger}-\ell\hat{\Phi}_{1}^{\dagger}\partial_{i}\hat{\Phi}_{2}^{\dagger}\partial_{i}\hat{\Phi}_{2}^{\dagger}-2\partial_{i}\hat{\Phi}_{1}^{\dagger}\hat{\Phi}_{2}^{\dagger2} )\nonumber\\
     &\cdot(\ell\hat{\Phi}_{1}\hat{\Phi}_{2}\partial_{i}^{2}\hat{\Phi}_{2}\!-\!\ell\hat{\Phi}_{1}\partial_{i}\hat{\Phi}_{2}\partial_{i}\hat{\Phi}_{2}\!-\!2\partial_{i}\hat{\Phi}_{1}\hat{\Phi}_{2}^{2})\!+\!V\! (\hat{\Phi}^{\dagger},\hat{\Phi})\,.
    \label{Hmdq}
\end{align}
For convenience, the coefficients need to satisfy: $K_{1}^{ij}=K_{1}^{ij}>0$, $K_{2}^{ijk}=K_{2}^{ikj}=K_{2}^{jik}=K_{2}^{jki}=K_{2}^{kij}=K_{2}^{kji}>0$, $\Gamma_{1,2}^{i}>0$, and $g>0$. The $\Gamma_{1,2}^{i}$ terms are the emerging terms from the hybridization of moments. 
In the same way as Model Series A, we can also obtain Noether currents of Model Series B. 
The $\Gamma_{1,2}^{i}$ terms liberate part of the boson mobility restriction. 
The effective field theory of this Hamiltonian is performed in Sec. \ref{EFTofM2}.

\section{Hybrid fractonic superfluids: Model Series A}

\label{EFTofM1} In the preceding section, we have constructed the minimal Hamiltonians with conservation of hybrid moments. In the present and next sections, we subsequently study physical properties of Model Series A and Model Series B.  In this section, we study physical properties of Model Series A. We start our analysis with the \([d,1,2,2]\) model and then directly generalize to arbitary \([d,N,m,m^\prime]\) models.

\subsection{Classical ground state wavefunctions}

Classically, the energy density $\mathcal{E}$ of the \([d,1,2,2]\) model in Eq.~(\ref{Hmdd}) has the form as: 
\begin{align}
\mathcal{E}= & \sum_{a}^{2}\sum_{i,j}^{d}K_{a}^{ij}\left|\phi_{a}\partial_{i}\partial_{j}\phi_{a}-\partial_{i}\phi_{a}\partial_{j}\phi_{a}\right|^{2}\nonumber\\
 & +\sum_{a\neq b}^{2}\sum_{i}^{d}\Gamma_{a,b}^{i}\left|\phi_{a}\partial_{i}\phi_{b}-\phi_{b}\partial_{i}\phi_{a}\right|^{2}\nonumber\\
 & +\sum_{a}^{2}-\mu_{a}\left|\phi_{a}\right|^{2}+\frac{g}{2}\left|\phi_{a}\right|^{4}\,
\end{align}
which is obtained by replacing the non-Hermitian operators $\Phi_a$ in Eq.~(\ref{Hmdd}) by their coherent-state eigenvalue $\phi_a\in\mathbb{C}$. 
Considering the chemical potential $\mu_{1}>0$, $\mu_{2}>0$, at $\left|\phi_{a}\right|=\sqrt{\rho_{a0}}=\sqrt{\frac{\mu_{a}}{g}}$, the potential reaches the minimal value. Then, the energy density reduces to:  
 $ \mathcal{E}=  \sum_{a}^{2}\sum_{i,j}^{d}K_{a}^{ij}\rho_{a0}^{2}(\partial_{i}\partial_{j}\theta_{a})^{2}  
  +\sum_{a\neq b}^{2}\sum_{i}^{d}\Gamma_{a,b}^{i}\rho_{10}\rho_{20}(\partial_{i}\theta_{a}-\partial_{i}\theta_{b})^{2}-\sum_{a}^{2}\frac{\mu_{a}^{2}}{2g}\,.
 $ As the coefficients $K_{a}^{ij}>0$ and $\Gamma_{a,b}^{i}>0$, 
to make these terms reach the minimal value, we need $\partial_{i}\partial_{j}\theta_{a}=0$ and $\partial_{i}\theta_{a}=\partial_{i}\theta_{b}$ for any $a$ and $b$, which leads to  the classical ground state $|\text{GS}_{\bm{\beta}}^{\{\theta_{a0}\}}\rangle=\bigotimes_{\mathbf{x}}|\text{GS}_{\bm{\beta}}^{\{\theta_{a0}\}}\rangle_{\mathbf{x}}$. Here 
 $ |\text{GS}_{\bm{\beta}}^{\{\theta_{a0}\}}\rangle_{\mathbf{x}}=  \frac{1}{C}\prod_{a}^{2}\exp\left[\sqrt{\rho_{a0}}e^{i(\theta_{a0}+\sum_{i}^{d}\beta_{i}x_{i})}\hat{\Phi}_{a}^{\dagger}(\mathbf{x})\right]|0\rangle\,.
 $ 
$C$ is the normalization factor, and $|0\rangle$ is the quantum vacuum state. $\bm{\beta}=\{\beta_1,\cdots,\beta_d\}$ has the dimension of momentum. 
A significant feature of this ground state is the formation of ODLRO.
The  correlation functions are given by: 
 $ \langle \hat{\Phi}_{a}^{\dagger}(\mathbf{x})\hat{\Phi}_{b}(\mathbf{0})\rangle=\sqrt{\rho_{a0}\rho_{b0}}e^{i(\theta_{b0}-\theta_{a0}-\sum_{i}\beta_{i}x_{i})}$, 
  $\langle\hat{\Phi}_{a}^{\dagger}(\mathbf{x})\hat{\Phi}_{b}(\mathbf{x})\hat{\Phi}_{a}(\mathbf{0})\hat{\Phi}_{b}^{\dagger}(\mathbf{0})\rangle=\rho_{a0}\rho_{b0}$, 
  and $\langle \hat{\Phi}_{a}^{\dagger}(\mathbf{x})\hat{\Phi}_{b}^{\dagger}(\mathbf{x})\hat{\Phi}_{a}(\mathbf{0})\hat{\Phi}_{b}(\mathbf{0})\rangle=\rho_{a0}\rho_{b0}e^{-2i\sum_{i}\beta_{i}x_{i}}\,
 $. The zero in bold font $\mathbf{0}$ denotes the coordinate origin.  For the hybrid fractonic superfluid phase in the \([d,1,2,2]\) model, the order parameters are defined  as the expectation values on the ground state:  
 $    \langle\hat{\Phi}_1(\mathbf{x})\rangle=\sqrt{\rho_{10}}e^{i(\theta_{10}+\sum_i\beta_i x_i)}$,  
   $ \langle\hat{\Phi}_2(\mathbf{x})\rangle=\sqrt{\rho_{20}}e^{i(\theta_{20}+\sum_i\beta_i x_i)}$, and $ \langle\hat{\Phi}_1(\mathbf{x})\hat{\Phi}_2^\dagger(\mathbf{x})\rangle=\sqrt{\rho_{10}\rho_{20}}e^{i(\theta_{10}-\theta_{20})}$.
 After condensation, the Noether currents [Eq.~(\ref{Ncurrent1}, \ref{Ncurrent2})] become 
\begin{align}
\hat{\rho}_{a}= & \rho_{a0},\nonumber\\
\hat{J}_{i}^{a}= & -2K_{a}^{ij}\rho_{a0}^{2}\sum_{j}^{d}\partial_{i}\partial_{j}^{2}\hat{\theta}_{a}+4\Gamma_{1,2}^{i}\rho_{10}\rho_{20}(\partial_{i}\hat{\theta}_{1}+\partial_{i}\hat{\theta}_{2})\,,
\end{align}
and 
\begin{align}
\hat{\rho}_{\text{mix}}^{(j)}= & (\rho_{10}+\rho_{20})x_{j},\nonumber\\
\hat{\mathcal{D}}_{i}^{(j)}= & x_{j}\hat{J}_{i}^{1}+x_{j}\hat{J}_{i}^{2}+2\sum_{a}^{2}K_{a}^{ij}\rho_{a0}^{2}\partial_{i}\partial_{j}\hat{\theta}_{a}\,.
\end{align}

\subsection{Goldstone modes and quantum fluctuations}

\label{M1 Goldstone modes and quantum fluctuations} 
Quantum fluctuations tend to destroy true ODLRO. In this subsection, we introduce the effect of quantum fluctuations.
Considering quantum fluctuations, we have $\hat{\Phi}_{a}(\mathbf{x})=\sqrt{\rho_{a0}+\hat{f}_{a}(\mathbf{x})}e^{i\hat{\theta}_{a}(\mathbf{x})}$,
where $\hat{f}_{a}(\mathbf{x},t)\ll\rho_{a0}$ are the quantum fluctuations of the density fields, and $\hat{\theta}_a$ is the quantum fluctuations of the phase fields. The commutation relation between $\hat{\theta}_a$ and $\hat{f}_a$ is given by $[\hat{\theta}_a(\mathbf{x}),\hat{f}_b(\mathbf{y})]=-i\delta^{(d)}(\mathbf{x}-\mathbf{y})\delta_{ab}$. Therefore, the canonical momentum operator conjugate to $\hat{\theta}_a$ is given by $\hat{\pi}_a=-\hat{f}_a$,    satisfying   $[\hat{\theta}_{a}(\mathbf{x}),\hat{\pi}_{b}(\mathbf{y})]= i\delta_{ab}\delta^{d}(\mathbf{x}-\mathbf{y}).$
To the second order, we can derive the effective Hamiltonian: 
\begin{align}
\mathcal{H}= & \sum_{a}^{2}\bigg\{\frac{g}{2}\hat{\pi}_{a}^{2}+\sum_{i,j}^{d}K_{a}^{ij}\bigg[\rho_{a0}^{2}(\partial_{i}\partial_{j}\hat{\theta}_{a})^{2}+\frac{(\partial_{i}\partial_{j}\hat{\pi}_{a})^{2}}{4}\bigg]\nonumber\\
 & +\sum_{b\neq a}^{2}\sum_{i}^{d}\Gamma_{a,b}^{i}\bigg[\frac{(\rho_{a0}\partial_{i}\hat{\pi}_{b}-\rho_{b0}\partial_{i}\hat{\pi}_{a})^{2}}{4\rho_{a0}\rho_{b0}}\nonumber\\
 & +\rho_{a0}\rho_{b0}(\partial_{i}\hat{\theta}_{b}-\partial_{i}\hat{\theta}_{a})^{2}\bigg]\bigg\}\,.\label{effHM1}
\end{align}
The time evolutions of $\hat{\theta}_a$ and $\hat{\pi}_a$ are given by
\begin{align}
    \dot{\hat{\theta}}_a=&-i\left[\hat{\theta}_a,\int d^d\mathbf{x}\mathcal{H}\right]=g\hat{\pi}_{a}+\frac{1}{2}\sum_{i,j}^{d}K_{a}^{ij}\partial_{i}^{2}\partial_{j}^{2}\hat{\pi}_{a}\nonumber\\
    &-\sum_{b\neq a}^{2}\sum_{i}^{d}\Gamma_{a,b}^{i}(\frac{\rho_{b0}}{\rho_{a0}}\partial_{i}^{2}\hat{\pi}_{a}-\partial_{i}^{2}\hat{\pi}_{b})\,,\label{timeevo1}\\
  \dot{\hat{\pi}}_a=&-i\left[\hat{\pi}_a,\int d^d\mathbf{x}\mathcal{H}\right]=-2\sum_{i,j}^{d}K_{a}^{ij}\rho_{a0}^2\partial_i^2\partial_j^2\hat{\theta}_a\nonumber\\
  &+2\sum_{b\neq a}^2\sum_i^2\Gamma_{a,b}^i\rho_{a0}\rho_{b0}(\partial_i^2\hat{\theta}_a-\partial_i^2\hat{\theta}_b)\,.
\end{align}
Considering the isotropic case $K_{a}^{ij}=\frac{\kappa}{2}$, $\Gamma_{a,b}^{i}=\frac{\Gamma}{2}$, we define  the coherence length  
\begin{align}
  \xi_c=\pi\sqrt{\frac{2\Gamma}{g}}\,.\label{coh1}
\end{align}
In the long-wave length limit, i.e.,  $\left|\mathbf{k}\right|\ll2\pi\xi_{c}^{-1}$, we have $g\hat{\pi}_a \gg-\sum_{b\neq a}^{2}\sum_{i}^{d}\Gamma_{a,b}^{i}\partial_{i}^{2}\hat{\pi}_{a}$, and thus, with finite $\rho_{b0}/\rho_{a0}$, the last term in Eq.~(\ref{timeevo1}) is negligible. 
In addition, the term $\frac{1}{2}\sum_{i,j}^{d}K_{a}^{ij}\partial_{i}^{2}\partial_{j}^{2}\hat{\pi}_{a}$ with higher-order gradient is also negligible.
Thus, we can safely make the approximation: $\hat{\pi}_{a}=\dot{\hat{\theta}}_a/g$.
Plugging this approximation into Eq.~(\ref{effHM1}), and excluding the higher derivative terms of $\hat{\theta}_a$ and $\dot{\hat{\theta}}_a$, we can obtain the following effective Hamiltonian
\begin{align}
  \mathcal{H}= & \sum_{a}^{2}\Bigg[\frac{g}{2}\hat{\pi}_{a}^{2}+\frac{\kappa\rho_{a0}^{2}}{2}\sum_{i,j}^{d}(\partial_{i}\partial_{j}\hat{\theta}_{a})^{2}\nonumber\\
   & +\frac{\Gamma\rho_{a0}\rho_{b0}}{2}\sum_{b\neq a}^{2}\sum_{i}^{d}(\partial_{i}\hat{\theta}_{a}-\partial_{i}\hat{\theta}_{b})^{2}\Bigg]\,.
  \label{Hmdd1}
\end{align}
In the momentum space, the above Hamiltonian can be further expressed in the following compact form:
\begin{align}
H= & \frac{1}{2(2\pi)^{d}}\int d^{d}\mathbf{k}\psi^{\dagger} (\mathbf{k}) M (\mathbf{k}) \psi (\mathbf{k}) \,,
\label{Hmddmommain}
\end{align}
 where we set $\psi (\mathbf{k}) =\begin{bmatrix}\hat{\pi}_{1} (\mathbf{k})  & \hat{\pi}_{2} (\mathbf{k})  & \hat{\theta}_{1} (\mathbf{k})  & \hat{\theta}_{2} (\mathbf{k}) \end{bmatrix}^{T}$,
$\psi^{\dagger} (\mathbf{k}) =\begin{bmatrix}\hat{\pi}_{1} (-\mathbf{k})  & \hat{\pi}_{2} (-\mathbf{k})  & \hat{\theta}_{1} (-\mathbf{k})  & \hat{\theta}_{2} (-\mathbf{k}) \end{bmatrix}$,
$M (\mathbf{k}) =\text{diag}[M_{1} (\mathbf{k}) ,M_{2} (\mathbf{k}) ]$.
The matrices $M_{1} (\mathbf{k}) $, $M_{2} (\mathbf{k}) $ are given by $ M_{1} (\mathbf{k}) =\text{diag}(g,g)$ and 
\begin{align}
 & M_{2} (\mathbf{k}) =\nonumber\\
 & \begin{bmatrix}\kappa\rho_{10}^{2}\left|\mathbf{k}\right|^{4}+2\Gamma\rho_{10}\rho_{20}\left|\mathbf{k}\right|^{2} & -2\Gamma\rho_{10}\rho_{20}\left|\mathbf{k}\right|^{2}\\
-2\Gamma\rho_{10}\rho_{20}\left|\mathbf{k}\right|^{2} & \kappa\rho_{20}^{2}\left|\mathbf{k}\right|^{4}+2\Gamma\rho_{10}\rho_{20}\left|\mathbf{k}\right|^{2}\nonumber
\end{bmatrix}\,.
\end{align}
In the momentum space,   $\hat{\theta}_a (\mathbf{k}) $ and $\hat{\pi}_b(\mathbf{k}')$ satisfy the commutation relation: $[\hat{\theta}_{a} (\mathbf{k}) ,\hat{\pi}_{b}\ (\mathbf{k}') ]= i(2\pi)^{d}\delta_{ab}\delta^{(d)}(\mathbf{k}+\mathbf{k}^{\prime})$.
The derivation of the Hamiltonian in Eq.~(\ref{Hmddmommain}) is given in Appendix~\ref{Hmddmom}.

Next, we apply the Bogoliubov transformation to send the above Hamiltonian into a diagonal form. For this purpose, 
we introduce a transformation matrix $T (\mathbf{k}) $ which changes the basis from $\psi (\mathbf{k}) $ into  $\phi (\mathbf{k}) $ defined as: 
$\phi (\mathbf{k})=[\begin{matrix}\hat{\alpha} (\mathbf{k} ) & \hat{\beta} (\mathbf{k} ) & \hat{\alpha}^{\dagger} (-\mathbf{k} ) & \hat{\beta}^{\dagger} (-\mathbf{k} )\end{matrix}]^{T}=T (\mathbf{k} )\psi (\mathbf{k} )\, 
 $. Here   $\hat{\alpha} (\mathbf{k} )$ and $\hat{\beta} (\mathbf{k} )$ are annihilation operators of two independent quasiparticle modes,  satisfying the following commutation relations: $  [\hat{\alpha} (\mathbf{k}) ,\hat{\alpha}^{\dagger}\ (\mathbf{k}')  ]= [\hat{\beta} (\mathbf{k}) ,\hat{\beta}^{\dagger}\ (\mathbf{k}')  ]=(2\pi)^{d}\delta^{(d)}(\mathbf{k}-\mathbf{k}^{\prime}),  
   [\hat{\alpha} (\mathbf{k}) ,\hat{\alpha}\ (\mathbf{k}') ]= [\hat{\beta} (\mathbf{k}) ,\hat{\beta}\ (\mathbf{k}')  ]= [\hat{\alpha} (\mathbf{k}) ,\hat{\beta}\ (\mathbf{k}') ]
 = [\hat{\alpha} (\mathbf{k} ),\hat{\beta}^{\dagger}\ (\mathbf{k}')  ]= [\hat{\beta} (\mathbf{k}) ,\hat{\alpha}^{\dagger}\ (\mathbf{k}')  ]=0\,. $
The detailed expression of  $T (\mathbf{k}) $ is given in Appendix \ref{M1Tmatrix}. As a result,  the Hamiltonian  in Eq.~(\ref{Hmddmommain}) is transformed into 
\begin{align}
    H= & \frac{1}{2(2\pi)^{d}}\phi^{\dagger} (\mathbf{k}) \left[T^{-1} (\mathbf{k}) \right]^{\dagger}M (\mathbf{k}) T^{-1} (\mathbf{k}) \phi (\mathbf{k}) \,,\label{M1diagH}
\end{align}
where the matrix $ [T^{-1} (\mathbf{k})  ]^{\dagger}M (\mathbf{k}) T^{-1} (\mathbf{k}) =\text{diag} [D_1 (\mathbf{k}) ,D_2 (\mathbf{k}) ,D_3 (\mathbf{k}) ,D_4 (\mathbf{k})  ]$ is a diagonal matrix with the diagonal elements:
\begin{align}
    &D_{1} (\mathbf{k}) =D_{3} (\mathbf{k}) \nonumber\\
    = & \Bigg\{\frac{g}{2}\Big[\kappa(\rho_{10}^2+\rho_{20}^2)\left|\mathbf{k}\right|^4+4\Gamma\rho_{10}\rho_{20}\left|\mathbf{k}\right|^2\nonumber\\
    &-\sqrt{\kappa^2(\rho_{10}^2-\rho_{20}^2)\left|\mathbf{k}\right|^8+16\Gamma^2\rho_{10}^2\rho_{20}^2\left|\mathbf{k}\right|^4}\Big]\Bigg\}^{\frac{1}{2}},\\
    &D_{2} (\mathbf{k}) =D_{4} (\mathbf{k}) \nonumber\\
    = & \Bigg\{\frac{g}{2}\Big[\kappa(\rho_{10}^2+\rho_{20}^2)\left|\mathbf{k}\right|^4+4\Gamma\rho_{10}\rho_{20}\left|\mathbf{k}\right|^2\nonumber\\
    &+\sqrt{\kappa^2(\rho_{10}^2-\rho_{20}^2)\left|\mathbf{k}\right|^8+16\Gamma^2\rho_{10}^2\rho_{20}^2\left|\mathbf{k}\right|^4}\Big]\Bigg\}^{\frac{1}{2}}\,.
\end{align}
Then, we can obtain the following Hamiltonian formed  by the creation and annihilation operators of Goldstone modes,
\begin{align}
\!\!\!\!H= & \frac{1}{(2\pi)^{d}}\!\int \!d^{d}\mathbf{k}D_{1} (\mathbf{k}) \hat{\alpha}^{\dagger} (\mathbf{k}) \hat{\alpha} (\mathbf{k}) \!+\!D_{2} (\mathbf{k}) \hat{\beta}^{\dagger} (\mathbf{k}) \hat{\beta} (\mathbf{k}) ,
\label{Hmddfinmain}
\end{align}
where the normal ordering is applied. The derivation of this Hamiltonian is in Appendix \ref{Hmddfin}. The dispersion relations are 
\begin{align}
\omega_{1}= & D_{1} (\mathbf{k}) \approx\sqrt{\frac{g}{2}\kappa(\rho_{10}^{2}+\rho_{20}^{2})}\left|\mathbf{k}\right|^{2}\,,\\
\omega_{2}= & D_{2} (\mathbf{k}) \approx\sqrt{4g\Gamma\rho_{10}\rho_{20}}\left|\mathbf{k}\right|\,.
\label{M1dispersion}
\end{align}

\subsection{Correlation functions}
\label{SecCorrelationM1}

  \begin{table*}[htbp]
\centering
\caption{Correlation functions of the \([d,1,2,2]\) model at zero temperature. This table illustrates the broken charge symmetry of the \([d,1,2,2]\) model and the corresponding order parameters for \(d = 1\,,2\,,3\). The detailed analysis is provided in Sec.~\ref{SecCorrelationM1}. In Sec.~\ref{SecArbitaryM1}, we generalize these results to arbitrary \([d,N,m,m^\prime]\) models. The parameters \(c_1\), \(c_2\), \(\rho_{10}\), \(\rho_{20}\), and \(\xi_c\) are positive real constants, with \(\xi_c\) specifically denoting the coherence length.}
\renewcommand{\arraystretch}{1.6}  
\resizebox{\textwidth}{!}{
\begin{tabular}{l c c c c c c}
\hline
\hline
$d$ & $\langle \hat{\Phi}_{1}^{\dagger}(\mathbf{x}) \hat{\Phi}_{1}(\mathbf{0}) \rangle$ & 
     $\langle \hat{\Phi}_{1}^{\dagger}(\mathbf{x}) \hat{\Phi}_{2}(\mathbf{0}) \rangle$ &
     $\langle \hat{\Phi}_{1}^{\dagger}(\mathbf{x}) \hat{\Phi}_{2}(\mathbf{x}) \hat{\Phi}_{2}^{\dagger}(\mathbf{0}) \hat{\Phi}_{1}(\mathbf{0}) \rangle$ &
     $\langle \hat{\Phi}_{1}^{\dagger}(\mathbf{x}) \hat{\Phi}_{2}^{\dagger}(\mathbf{x}) \hat{\Phi}_{1}(\mathbf{0}) \hat{\Phi}_{2}(\mathbf{0}) \rangle$ &
     \makecell{Broken Charge Symmetry} &
     \makecell{Order Parameter} \\
\hline
$1$ & $\rho_{10} e^{-\frac{\left|\mathbf{x}\right|}{2c_{1}}} \left( \frac{\left|\mathbf{x}\right|}{\xi_{c}} \right)^{-\frac{1}{\pi c_{2}}}$ & 0 & 
    $\rho_{10} \rho_{20} \left( \frac{\left|\mathbf{x}\right|}{\xi_{c}} \right)^{-\frac{4}{\pi c_{2}}}$ & 
    $\rho_{10} \rho_{20} e^{-\frac{2}{c_1} \left|\mathbf{x}\right|}$ & N/A & N/A \\
 
 $2$ & $\rho_{10} \left( \frac{\left|\mathbf{x}\right|}{\xi_{c}} \right)^{-\frac{1}{2\pi c_{1}}} e^{-\frac{1}{2\pi c_{2} \xi_{c}}}$ & 
    $\sqrt{\rho_{10} \rho_{20}} \left( \frac{\left|\mathbf{x}\right|}{\xi_{c}} \right)^{-\frac{1}{2\pi c_{1}}} e^{-\frac{1}{2\pi c_{2} \xi_{c}}}$ & 
    $\rho_{10} \rho_{20} e^{-\frac{2}{\pi c_{2} \xi_{c}}}$ & 
    $\rho_{10} \rho_{20} \left( \frac{\left|\mathbf{x}\right|}{\xi_c} \right)^{-\frac{2}{\pi c_1}}$ & $U\left(1\right)_{-,C}$ & 
    $\langle \hat{\Phi}_1(\mathbf{x}) \hat{\Phi}_2^\dagger(\mathbf{x}) \rangle$ \\
 
 $3$ & $\rho_{10} e^{-\frac{1}{4\pi c_{1} \xi_{c}}} e^{-\frac{1}{c_{2} \xi_{c}^{2}}}$ & 
    $\sqrt{\rho_{10} \rho_{20}} e^{-\frac{1}{4\pi c_{1} \xi_{c}}} e^{-\frac{1}{c_{2} \xi_{c}^{2}}}$ & 
    $\rho_{10} \rho_{20} e^{-\frac{4}{c_{2} \xi_{c}^{2}}}$ & 
    $\rho_{10} \rho_{20} e^{-\frac{1}{\pi c_1 \xi_c}}$ & $U\left(1\right)_{1,C} \otimes U\left(1\right)_{2,C}$ & 
    $\langle \hat{\Phi}_1(\mathbf{x}) \rangle, \langle \hat{\Phi}_2(\mathbf{x}) \rangle$ \\
\hline
\hline
\end{tabular}}
\label{correlation1}
\end{table*}

To check if the ground state   spontaneously breaks charge symmetry, we need to calculate   correlation functions~\footnote{The correlation functions calculated in this paper are regularized by the coherence length $\xi_c$.} in the presence of quantum fluctuations.  In the present case,  \textit{two-operator} correlation functions can be written as 
\begin{align}
\langle\hat{\Phi}_{1}^{\dagger}(\mathbf{x})\hat{\Phi}_{1}(\mathbf{0})\rangle= & \rho_{10}e^{-\frac{1}{2}\langle\left[\hat{\theta}_{1}(\mathbf{x})-\hat{\theta}_{1}(\mathbf{0})\right]^{2}\rangle}\nonumber\\
= & \rho_{10}e^{\langle\hat{\theta}_{1}(\mathbf{x})\hat{\theta}_{1}(\mathbf{0})\rangle-\langle\hat{\theta}_{1}^{2}(\mathbf{0})\rangle}\,,\nonumber\\
\langle\hat{\Phi}_{2}^{\dagger}(\mathbf{x})\hat{\Phi}_{2}(\mathbf{0})\rangle= & \rho_{20}e^{-\frac{1}{2}\langle\left[\hat{\theta}_{2}(\mathbf{x})-\hat{\theta}_{2}(\mathbf{0})\right]^{2}\rangle}\nonumber\\
= & \rho_{20}e^{\langle\hat{\theta}_{2}(\mathbf{x})\hat{\theta}_{2}(\mathbf{0})\rangle-\langle\hat{\theta}_{2}^{2}(\mathbf{0})\rangle}\,,\nonumber\\
\langle\hat{\Phi}_{1}^{\dagger}(\mathbf{x})\hat{\Phi}_{2}(\mathbf{0})\rangle= & \sqrt{\rho_{10}\rho_{20}}e^{-\frac{1}{2}\langle\left[\hat{\theta}_{1}(\mathbf{x})-\hat{\theta}_{2}(\mathbf{0})\right]^{2}\rangle}\nonumber\\
= & \sqrt{\rho_{10}\rho_{20}}e^{\langle\hat{\theta}_{1}(\mathbf{x})\hat{\theta}_{2}(\mathbf{0})\rangle-\frac{1}{2}\langle\hat{\theta}_{1}^{2}(\mathbf{0})\rangle-\frac{1}{2}\langle\hat{\theta}_{2}^{2}(\mathbf{0})\rangle}\,.\nonumber
\end{align}
The \textit{four-operator} correlation functions can be written as 
\begin{align}
& \langle\hat{\Phi}_{1}^{\dagger}(\mathbf{x})\hat{\Phi}_{2}(\mathbf{x})\hat{\Phi}_{1}(\mathbf{0})\hat{\Phi}_{2}^{\dagger}(\mathbf{0})\rangle\nonumber\\
= & \rho_{10}\rho_{20}\exp\bigg[\langle\hat{\theta}_{1}(\mathbf{x})\hat{\theta}_{1}(\mathbf{0})\rangle+\langle\hat{\theta}_{2}(\mathbf{x})\hat{\theta}_{2}(\mathbf{0})\rangle\nonumber\\
 & +2\langle\hat{\theta}_{1}(\mathbf{0})\hat{\theta}_{2}(\mathbf{0})\rangle-\langle\hat{\theta}_{1}^{2}(\mathbf{0})\rangle-\langle\hat{\theta}_{2}^{2}(\mathbf{0})\rangle-2\langle\hat{\theta}_{1}(\mathbf{x})\hat{\theta}_{2}(\mathbf{0})\rangle\bigg],\nonumber\\
    & \langle\hat{\Phi}_{1}^{\dagger}(\mathbf{x})\hat{\Phi}_{2}^\dagger(\mathbf{x})\hat{\Phi}_{1}(\mathbf{0})\hat{\Phi}_{2}(\mathbf{0})\rangle\nonumber\\
    = & \rho_{10}\rho_{20}\exp\bigg[\langle\hat{\theta}_{1}(\mathbf{x})\hat{\theta}_{1}(\mathbf{0})\rangle+\langle\hat{\theta}_{2}(\mathbf{x})\hat{\theta}_{2}(\mathbf{0})\rangle\nonumber\\
    & -2\langle\hat{\theta}_{1}(\mathbf{0})\hat{\theta}_{2}(\mathbf{0})\rangle-\langle\hat{\theta}_{1}^{2}(\mathbf{0})\rangle-\langle\hat{\theta}_{2}^{2}(\mathbf{0})\rangle+2\langle\hat{\theta}_{1}(\mathbf{x})\hat{\theta}_{2}(\mathbf{0})\rangle\bigg].\nonumber
  \end{align}
  As usual, the long-distance behaviors of correlation functions reflect certain order parameters of SSB phases.
  Here, $\langle\hat{\Phi}_1^\dagger(\mathbf{x})\hat{\Phi}_1(\mathbf{0})\rangle$, $\langle\hat{\Phi}_1^\dagger(\mathbf{x})\hat{\Phi}_2(\mathbf{x})\hat{\Phi}_2^\dagger(\mathbf{0})\hat{\Phi}_1(\mathbf{0})\rangle$, and $\langle\hat{\Phi}_1^\dagger(\mathbf{x})\hat{\Phi}_2^\dagger(\mathbf{x})\hat{\Phi}_1(\mathbf{0})\hat{\Phi}_2(\mathbf{0})\rangle$ correspond to the order parameters $\langle\hat{\Phi}_1(\mathbf{x})\rangle$, $\langle\hat{\Phi}_1(\mathbf{x})\hat{\Phi}_2^\dagger(\mathbf{x})\rangle$, and $\langle\hat{\Phi}_1(\mathbf{x})\hat{\Phi}_2(\mathbf{x})\rangle$ respectively.
Using the equations above, we can calculate two-operator and four-operator correlation functions in different spatial dimensions. 
The calculation results of correlation functions are shown in Table \ref{correlation1}. The calculation details are in Appendix \ref{detail1}. In these correlation functions, we set $c_{1}=\frac{2\sqrt{g\kappa(\rho_{10}^{2}+\rho_{20}^{2})}}{g}$, $c_{2}=\frac{8\sqrt{g\Gamma\rho_{10}\rho_{20}}}{g}$. The coherence length $\xi_c=\pi\sqrt{\frac{2\Gamma}{g}}$ is given by Eq.~(\ref{coh1}).
During the computation, we require that $\left|\mathbf{x}\right|$ is much larger than the coherence length $\xi_c$ but     much smaller than the system size $L$.

For the later convenience, we  express the total charge symmetry in two equivalent ways: 
\begin{align}
  \mathcal{G}_C=U(1)_{1,C}\otimes U(1)_{2,C}=U(1)_{+,C}\otimes U(1)_{-,C},
\end{align}
where $U(1)_{+,C}$ corresponds to the conserved total charges $\hat{Q}_+=\int d^d\mathbf{x}(\hat{\rho}_1+\hat{\rho}_2)$ of the whole system of two species of bosons, and \textit{the relative charge symmetry} $U(1)_{-,C}$ corresponds to the conserved relative charges (i.e., charge difference) $\hat{Q}_-=\int d^d\mathbf{x}(\hat{\rho}_1-\hat{\rho}_2)$.

 We can see that in one spatial dimension, the correlation functions $\langle\hat{\Phi}_1^\dagger(\mathbf{x})\hat{\Phi}_1(\mathbf{0})\rangle$ and $\langle\hat{\Phi}_1^\dagger(\mathbf{x})\hat{\Phi}_2^\dagger(\mathbf{x})\hat{\Phi}_1(\mathbf{0})\hat{\Phi}_2(\mathbf{0})\rangle$ decay exponentially. 
The correlation function $\langle\hat{\Phi}_1^\dagger(\mathbf{x})\hat{\Phi}_2(\mathbf{x})\hat{\Phi}_2^\dagger(\mathbf{0})\hat{\Phi}_1(\mathbf{0})\rangle$ decays in a power law. This shows that there is no spontaneous breaking of charge symmetry in one spatial dimension.

In two spatial dimensions, the correlation functions $\langle\hat{\Phi}_1^\dagger(\mathbf{x})\hat{\Phi}_1(\mathbf{0})\rangle$ and $\langle\hat{\Phi}_1^\dagger(\mathbf{x})\hat{\Phi}_2^\dagger(\mathbf{x})\hat{\Phi}_1(\mathbf{0})\hat{\Phi}_2(\mathbf{0})\rangle$ decay in a power law. 
The correlation function $\langle\hat{\Phi}_1^\dagger(\mathbf{x})\hat{\Phi}_2(\mathbf{x})\hat{\Phi}_2^\dagger(\mathbf{0})\hat{\Phi}_1(\mathbf{0})\rangle$ saturates to a constant at long distances. 
According to the commutation relations 
\begin{align}
     [\hat{\Phi}_1(\mathbf{x})\hat{\Phi}_2^\dagger(\mathbf{x}),\hat{Q}_+ ]=&0\,,\nonumber\\
     [\hat{\Phi}_1(\mathbf{x})\hat{\Phi}_2^\dagger(\mathbf{x}),\hat{Q}_- ]=&2\hat{\Phi}_1(\mathbf{x})\hat{\Phi}_2^\dagger(\mathbf{x})\,,
\end{align}
the system breaks the relative charge symmetry $U(1)_{-,C}$, and has a true ODLRO. 
In three spatial dimensions and above, all the correlation functions mentioned above are constants at long distances. This means that the charge symmetry is completely broken in three spatial dimensions.

In summary, the charge symmetry is not broken in one spatial dimension, and is completely broken in three spatial dimension and above. In two spatial dimensions,  a partial breaking of the charge symmetry occurs.

\subsection{Arbitary \([d,N,m,m^\prime]\) models of Model Series A}
\label{SecArbitaryM1}

In this subsection, we generalize the above analysis to \([d,N,m,m^\prime]\) models of Model Series A with the conserved quantities [Eqs.~(\ref{type1con1}, \ref{type1con2}, \ref{type1con3}, \ref{type1con4})]. 
The corresponding Hamiltonian density is Eq.~(\ref{Htype1}). 
In this case, we focus on an isotropic coupling constant $K_{a}^{i_{1}\cdots i_{N+1}}=\frac{\kappa}{2}$, $\Gamma_{a,b}^{i_{1}\cdots i_{N}}=\frac{\Gamma}{2}$. 
For simplicity, the chemical potential $\mu_{a}=\mu>0$ is assumed, resulting in classical degenerate ground states with finite uniform density distribution $\rho_{a}=\rho_{0}$. 
We can also derive an effective theory for the quantum fluctuation field $\hat{\theta}_{a}$ after condensation,
\begin{align}
\!\!\!\!\!\mathcal{H}\!=\! & \sum_{a}^{m}\bigg[\frac{g}{2}\hat{\pi}_a^2+\frac{\kappa\rho_{0}^{N+1}}{2}(\nabla^{N+1}\hat{\theta}_{a})^{2}\bigg]\nonumber\\
 & \!\!+\frac{\Gamma\rho_{0}^{2N}}{2}\!\sum_{a\neq b}^{m'}\!\sum_{i_{1},\cdots,i_{N}}^{d}(\partial_{i_{1}}\!\cdots\!\partial_{i_{N}}\hat{\theta}_{a}-\partial_{i_{1}}\!\cdots\!\partial_{i_{N}}\hat{\theta}_{b})^{2}.\!
\end{align}
Through the same processes, we can obtain the Hamiltonian in the momentum space
\begin{align}
H=\frac{1}{2(2\pi)^{d}}\int d^{d}\mathbf{k}\psi^{\dagger} (\mathbf{k}) M (\mathbf{k}) \psi (\mathbf{k}) ,
\end{align}
where 
\begin{align}
\psi (\mathbf{k}) =&\begin{bmatrix}\hat{\pi}_{1} (\mathbf{k})  & \cdots & \hat{\pi}_{m} (\mathbf{k})  & \hat{\theta}_{1} (\mathbf{k})  & \cdots & \hat{\theta}_{m} (\mathbf{k}) \end{bmatrix}^{T},\\
  M (\mathbf{k}) =&\text{diag}(M_{1} (\mathbf{k}) ,M_{2} (\mathbf{k}) ,M_3 (\mathbf{k}) ).
\end{align}
The matrix $M_{1} (\mathbf{k}) $ is a $m\times m$ diagonal matrix whose diagonal elements are $g$. The diagonal elements of the $m'\times m'$ matrix $M_{2} (\mathbf{k}) $ are $[\kappa\rho_{0}^{N+1}\left|\mathbf{k}\right|^{2N+2}+2(m'-1)\Gamma\rho_{0}^{2N}\left|\mathbf{k}\right|^{2N}]$,
and all the off-diagonal elements are $[-2\Gamma\rho_{0}^{2N}\left|\mathbf{k}\right|^{2N}]$. $M_3 (\mathbf{k}) $ is a diagonal $(m-m')\times (m-m')$ matrix whose diagonal elements are $[\kappa\rho_{0}^{N+1}\left|\mathbf{k}\right|^{2N+2}]$.
By diagonalizing this Hamiltonian, we end up with the dispersion relations:
\begin{align}
    \omega_a= & \sqrt{\kappa g\rho_{0}^{N+1}}\left|\mathbf{k}\right|^{N+1}\quad(1\leq a\leq m-m^\prime+1)\,,\\
    \omega_a= & \sqrt{\kappa g\rho_{0}^{N+1}\left|\mathbf{k}\right|^{2N+2}+2m^{\prime}g\Gamma\rho_{0}^{2n}\left|\mathbf{k}\right|^{2N}}\nonumber\\
    \approx & \sqrt{2m^{\prime}g\Gamma\rho_{0}^{2N}}\left|\mathbf{k}\right|^{N}\quad(m-m^\prime+2\leq a\leq m)\,.
\end{align}
The two-operator correlation function can be written as 
\begin{align}
  \langle\hat{\Phi}_{a}^{\dagger}(\mathbf{x})\hat{\Phi}_{a}(\mathbf{0})\rangle= & \rho_{0}e^{-\frac{1}{2}\langle\left[\hat{\theta}_{a}(\mathbf{x})-\hat{\theta}_{a}(\mathbf{0})\right]^{2}\rangle}\nonumber\\
  = & \rho_{0}e^{\langle\hat{\theta}_{a}(\mathbf{x})\hat{\theta}_{a}(\mathbf{0})\rangle-\langle\hat{\theta}_{a}^{2}(\mathbf{0})\rangle}\,.
  \label{dNmcorrelator1}
  \end{align}
The four-operator correlation functions can be written as 
\begin{align}
& \langle\hat{\Phi}_{a}^{\dagger}(\mathbf{x})\hat{\Phi}_{b}(\mathbf{x})\hat{\Phi}_{a}(\mathbf{0})\hat{\Phi}_{b}^{\dagger}(\mathbf{0})\rangle\nonumber\\
= & \rho_{0}^2\exp\bigg[\langle\hat{\theta}_{a}(\mathbf{x})\hat{\theta}_{a}(\mathbf{0})\rangle+\langle\hat{\theta}_{b}(\mathbf{x})\hat{\theta}_{b}(\mathbf{0})\rangle+2\langle\hat{\theta}_{a}(\mathbf{0})\hat{\theta}_{b}(\mathbf{0})\rangle\nonumber\\
 & -\langle\hat{\theta}_{a}^{2}(\mathbf{0})\rangle-\langle\hat{\theta}_{b}^{2}(\mathbf{0})\rangle-2\langle\hat{\theta}_{a}(\mathbf{x})\hat{\theta}_{b}(\mathbf{0})\rangle\bigg]\,,
 \label{dNmcorrelator2}\\
  & \langle\hat{\Phi}_{a}^{\dagger}(\mathbf{x})\hat{\Phi}_{b}^\dagger(\mathbf{x})\hat{\Phi}_{a}(\mathbf{0})\hat{\Phi}_{b}(\mathbf{0})\rangle\nonumber\\
  = & \rho_{0}^2\exp\bigg[\langle\hat{\theta}_{a}(\mathbf{x})\hat{\theta}_{a}(\mathbf{0})\rangle+\langle\hat{\theta}_{b}(\mathbf{x})\hat{\theta}_{b}(\mathbf{0})\rangle-2\langle\hat{\theta}_{a}(\mathbf{0})\hat{\theta}_{b}(\mathbf{0})\rangle\nonumber\\
  & -\langle\hat{\theta}_{a}^{2}(\mathbf{0})\rangle-\langle\hat{\theta}_{b}^{2}(\mathbf{0})\rangle+2\langle\hat{\theta}_{a}(\mathbf{x})\hat{\theta}_{b}(\mathbf{0})\rangle\bigg]\,.
  \label{dNmcorrelator3}
\end{align}
If the species of bosons satisfy the conditions $1\leq a\,,b\leq m'$ and $a\neq b$, the exponent of $e$ in~Eq. (\ref{dNmcorrelator2}) is an integral of a function which is proportional to $(1/\left|\mathbf{k}\right|^{N})$.
The correlation function $\langle\hat{\Phi}_{a}^{\dagger}(\mathbf{x})\hat{\Phi}_{b}(\mathbf{x})\hat{\Phi}_{a}(\mathbf{0})\hat{\Phi}_{b}^{\dagger}(\mathbf{0})\rangle$ saturates to a constant in spatial dimensions $d\geq N+1$.
For the correlation functions $\langle\hat{\Phi}_{a}^{\dagger}(\mathbf{x})\hat{\Phi}_{a}(\mathbf{0})\rangle$ and $\langle\hat{\Phi}_{a}^{\dagger}(\mathbf{x})\hat{\Phi}_{b}^{\dagger}(\mathbf{x})\hat{\Phi}_{a}(\mathbf{0})\hat{\Phi}_{b}(\mathbf{0})\rangle$, in the exponent of $e$, the integrand is the sum of a function proportional to $(1/\left|\mathbf{k}\right|^N)$ and another function proportional to $(1/\left|\mathbf{k}\right|^{N+1})$.
The two correlation functions saturate to a constant only in dimensions $d\geq N+2$.
This implies a partial spontaneous breaking of charge symmetry in spatial dimensions $d=N+1$.
The system breaks the relative charge symmetry of the bosons involved in the hybridization.

\section{Hybrid fractonic superfluids: Model Series B}

\label{EFTofM2} In this section, we focus on the details in Model Series B. 
Based on the Hamiltonian [Eq.~(\ref{Hmdq})] of a hybrid dipole-quadrupole conserving model, we present the ground state along with its Goldstone modes and calculate correlation functions to study the minimal spatial dimension in which this model   break charge symmetry and has a true ODLRO.

\subsection{Classical ground state wavefunctions}

Classically, the energy density $\mathcal{E}$ of the Hamiltonian [Eq.~(\ref{Hmdq})] has the form as: 
\begin{align}
  \mathcal{E}= & \sum_{i,j}^{d}K_{1}^{ij}\left|\phi_{1}\partial_{i}\partial_{j}\phi_{1}-\partial_{i}\phi_{1}\partial_{j}\phi_{2}\right|^{2}\nonumber\\
   & +\sum_{i,j,k}^{d}K_{2}^{ijk}\bigg|\phi_{2}^{2}\partial_{i}\partial_{j}\partial_{k}\phi_{2}-\phi_{2}\partial_{i}\phi_{2}\partial_{j}\partial_{k}\phi_{2}\nonumber\\
   & -\phi_{2}\partial_{j}\phi_{2}\partial_{i}\partial_{k}\phi_{2}-\phi_{2}\partial_{k}\phi_{2}\partial_{i}\partial_{j}\phi_{2}+2\partial_{i}\phi_{2}\partial_{j}\phi_{2}\partial_{k}\phi_{2}\bigg|^{2}\nonumber\\
   & +\sum_{i}^{d}\Gamma_{1,2}^{i}\left|\ell\phi_{1}\phi_{2}\partial_{i}^{2}\phi_{2}-\ell\phi_{1}\partial_{i}\phi_{2}\partial_{i}\phi_{2}-2\partial_{i}\phi_{1}\phi_{2}^{2}\right|^{2}\nonumber\\
   & -\mu_{1}\left|\phi_{1}\right|^{2}-\mu_{2}\left|\phi_{2}\right|^{2}+\frac{g}{2}\left|\phi_{1}\right|^{4}+\frac{g}{2}\left|\phi_{2}\right|^{4}.
\end{align}
If $\mu_{1},\mu_{2}>0$, at $\left|\phi_{1}\right|=\sqrt{\rho_{10}}=\sqrt{\frac{\mu_{1}}{g}}$,
$\left|\phi_{2}\right|=\sqrt{\rho_{20}}=\sqrt{\frac{\mu_{2}}{g}}$,
the potential reaches the minimal value.
Then, the energy density reduces to :
$\mathcal{E}= \sum_{i,j}^{d}K_{1}^{ij}\rho_{10}^{2}(\partial_{i}\partial_{j}\theta_{1})^{2}+\sum_{i,j,k}^{d}K_{2}^{ijk}\rho_{20}^{3}(\partial_{i}\partial_{j}\partial_{k}\theta_{2})^{2} +\sum_{i}^{d}\Gamma_{1,2}^{i}\rho_{10}\rho_{20}^{2}(2\partial_{i}\theta_{1}-\ell\partial_{i}^{2}\theta_{2})^{2}-\frac{\mu_{1}^{2}}{2g}-\frac{\mu_{2}^{2}}{2g}.$
To minimize the kinetic terms, we need to enforce three conditions: $\partial_{i}\partial_{j}\theta_{1}=0$,$\partial_{i}\partial_{j}\partial_{k}\theta_{2}=0$, and $2\partial_{i}\theta_{1}-\ell\partial_{i}^{2}\theta_{2}=0$
for any $i$, $j$, and $k$. The first two conditions are satisfied by  $\theta_{1}=\theta_{10}+\sum_{i}^{d}\beta_{i}^{(1)}x_{i}$ and $\theta_{2}=\theta_{20}+\sum_{i,j}^{d}\beta_{ij}^{(2)}x_{i}x_{j}$.
And the last condition gives another limit of $\beta_{i}^{(1)}$ and $\beta_{ij}^{(2)}$: $2\beta_{i}^{(1)}-\ell\beta_{ii}^{(2)}=0$.
Thus, we can give the ground state,
\begin{align}
& |\text{GS}_{\bm{\beta}^{(1)},\bm{\beta}^{(2)}}^{\{\theta_{a0}\}}\rangle\nonumber\\
= & \bigotimes_{\mathbf{x}}\frac{1}{C}\exp\left[\sqrt{\rho_{10}}e^{i(\theta_{10}+\sum_{i}^{d}\beta_{i}^{(1)}x_{i})}\hat{\Phi}_{1}^{\dagger}(\mathbf{x})\right]\nonumber\\
 & \exp\left[\sqrt{\rho_{20}}e^{i(\theta_{20}+\sum_{i,j}^{d}\beta_{ij}^{(2)}x_{i}x_{j})}\hat{\Phi}_{2}^{\dagger}(\mathbf{x})\right]|0\rangle.
\end{align}
This ground state also has the formation of ODLRO. We can calculate correlation functions in the classical level,
\begin{align}
\langle\hat{\Phi}_{1}^{\dagger}(\mathbf{x})\hat{\Phi}_{1}(\mathbf{0})\rangle= & \rho_{10}e^{-i\sum_{i}^{d}\beta_{i}^{(1)}x_{i}},\\
\langle\hat{\Phi}_{2}^{\dagger}(\mathbf{x})\hat{\Phi}_{2}(\mathbf{0})\rangle= & \rho_{20}e^{-i\sum_{i,j}^{d}\beta_{ij}^{(2)}x_{i}x_{j}}.
\end{align}
The order parameter can be determined with finite expectation value on the ground state,
\begin{align}
\langle\hat{\Phi}_{1}(\mathbf{x})\rangle= & \sqrt{\rho_{10}}e^{i(\theta_{10}+\sum_{i}^{d}\beta_{i}^{(1)}x_{i})},\\
\langle\hat{\Phi}_{2}(\mathbf{x})\rangle= & \sqrt{\rho_{20}}e^{i(\theta_{20}+\sum_{i,j}^{d}\beta_{ij}^{(2)}x_{i}x_{j})}.
\end{align}

\subsection{Goldstone modes and quantum fluctuations}
\label{mainM2}
Considering quantum fluctuations, we have $\hat{\Phi}_{a}(\mathbf{x})=\sqrt{\rho_{a0}+\hat{f}_{a}(\mathbf{x})}e^{i\hat{\theta}_{a}(\mathbf{x})}$, where $\hat{f}_{a}(\mathbf{x})\ll\rho_{a0}$. Through the same processes in Model Series A, we can obtain the effective Hamiltonian under the long-wave approximation,
\begin{align}
  \mathcal{H}= & \frac{g}{2}\hat{\pi}_{1}^{2}+\frac{g}{2}\hat{\pi}_{2}^{2}+\frac{\kappa_{1}\rho_{10}^{2}}{2}\sum_{i,j}^{d}\hat{\theta}_{1}\partial_{i}^{2}\partial_{j}^{2}\hat{\theta}_{1}\nonumber\\
   & -\frac{\kappa_{2}\rho_{20}^{3}}{2}\sum_{i,j,k}^{d}\theta_{2}\partial_{i}^{2}\partial_{j}^{2}\partial_{k}^{2}\hat{\theta}_{2}+\frac{\Gamma\rho_{10}\rho_{20}^{2}}{2}\sum_{i}^{d}\Big(\ell^{2}\hat{\theta}_{2}\partial_{i}^{4}\hat{\theta}_{2}\nonumber\\
   & -4\hat{\theta}_{1}\partial_{i}^{2}\hat{\theta}_{1}+2\ell\hat{\theta}_{1}\partial_{i}^{3}\hat{\theta}_{2}-2\ell\hat{\theta}_{2}\partial_{i}^{3}\hat{\theta}_{1}\Big)\,,
  \label{Hmdq1}
  \end{align}
where we set $K_{1}^{ij}=\frac{\kappa_{1}}{2}$, $K_{2}^{ijk}=\frac{\kappa_{2}}{2}$, and $\Gamma_{1,2}^{i}=\frac{\Gamma}{2}$, $\forall i,j,k$, that is, all the elements of the coefficient tensors $K_{1}^{ij}$, $K_{2}^{ijk}$, and $\Gamma_{1,2}^{i}$ are $\frac{\kappa_{1}}{2}$, $\frac{\kappa_{2}}{2}$, and $\frac{\Gamma}{2}$ respectively.
$\hat{\pi}_{a}=-\hat{f}_a$ are the conjugate momentum densities of $\hat{\theta}_a$.
Continuing the derivation process, we can finally obtain the regularized quantized and diagonalized Hamiltonian,
\begin{align}
\!\!\!H\!=& \frac{1}{(2\pi)^{d}}\!\int \!d^{d}\mathbf{k}D_{1} (\mathbf{k}) \hat{\alpha}^{\dagger} (\mathbf{k}) \hat{\alpha} (\mathbf{k}) \!+\!D_{2} (\mathbf{k}) \hat{\beta}^{\dagger} (\mathbf{k}) \hat{\beta} (\mathbf{k}) ,
\label{Hmdqfinmain}
\end{align}
where the normal ordering is applied. 
The operators in Model Series B above still satisfies the same commutation relations as in Model Series A.
The dispersion relations are
\begin{align}
\omega_{1}= & D_{1} (\mathbf{k}) ,\quad\omega_{2}=D_{2} (\mathbf{k}) .
\end{align}
The dispersion relations of Model Series B are complex and anisotropic and are given in Appendix \ref{M2dispersion}.
We can only approximate the dispersion relations for given spatial dimensions.
One of the dispersion relations is linear. 
To leading order of $\left|\mathbf{k}\right|$, the dispersion relation is isotropic.
\begin{align}
    \omega_1\approx&\sqrt{4g\Gamma\rho_{10}\rho_{20}^2}\left|\mathbf{k}\right|.
\end{align}
In one spatial dimension, the other one of the dispersion relations is given by,
\begin{align}
  \omega_{2}\approx\sqrt{\kappa_{2}g\rho_{20}^{3}+\frac{\kappa_{1}g\rho_{10}^{2}\ell^{2}}{4}}\left|\mathbf{k}\right|^{3}.
\end{align}
In two or more spatial dimensions, $\omega_2$ is anisotropic and possesses a dispersion of second order.
Since the derivation process is similar to that of Model Series A, we put the detailed derivation of Model Series B in Appendix \ref{detailM2}. The derivation of the dispersion relations in different spatial dimensions is given in Appendix \ref{detail2}.
\subsection{Correlation functions}
\label{SecCorrelationM2}
To check the effect of hybridization in Model Series B on ODLRO, we calculate correlation functions. Unlike Model Series A, we only need two-operator correlation functions to characterize the spontaneous breaking of charge symmetry in Model Series B instead of four-operator correlation functions.
The two-operator correlation functions can be written as 
\begin{align}
\langle\hat{\Phi}_{1}^{\dagger}(\mathbf{x})\hat{\Phi}_{1}(\mathbf{0})\rangle= & \rho_{10}e^{-\frac{1}{2}\langle\left[\hat{\theta}_{1}(\mathbf{x})-\hat{\theta}_{1}(\mathbf{0})\right]^{2}\rangle}\nonumber\\
= & \rho_{10}e^{\langle\hat{\theta}_{1}(\mathbf{x})\hat{\theta}_{1}(\mathbf{0})\rangle-\langle\hat{\theta}_{1}^{2}(\mathbf{0})\rangle},\\
\langle\hat{\Phi}_{2}^{\dagger}(\mathbf{x})\hat{\Phi}_{2}(\mathbf{0})\rangle= & \rho_{20}e^{-\frac{1}{2}\langle\left[\hat{\theta}_{2}(\mathbf{x})-\hat{\theta}_{2}(\mathbf{0})\right]^{2}\rangle}\nonumber\\
= & \rho_{20}e^{\langle\hat{\theta}_{2}(\mathbf{x})\hat{\theta}_{2}(\mathbf{0})\rangle-\langle\hat{\theta}_{2}^{2}(\mathbf{0})\rangle}.
\end{align}
Then, we can calculate the correlation function of the two species
of bosons. Due to the anisotropy, we can just give a general trend
of the correlation functions in two or more spatial dimensions. The
calculation results of correlation functions are shown in Table \ref{correlation2}. 
The details of the analytical calculations and numerical calculations are in Appendix \ref{detail2}. In these correlation functions, we set $c_{11}=\frac{4\Gamma\rho_{10}\rho_{20}}{\sqrt{g\Gamma\rho_{10}}}$,
$c_{21}=\frac{4\sqrt{\kappa_{1}\rho_{10}^{2}\ell^{2}+4\kappa_{2}\rho_{20}^{3}}}{\sqrt{g}\ell^{2}}$ and the coherence length $\xi_c=2\pi\sqrt{\frac{\Gamma \rho_{20}^2}{g\rho_{10}}}$.
\begin{table}[htbp]
\centering
\caption{Correlation functions of the hybrid dipole-quadrupole moment conserving model [Eq.~(\ref{Hmdq})] at zero temperature with quantum fluctuations. This table illustrates the broken charge symmetry of this model and the corresponding order parameters at $d=1\,,2\,,3$. The analysis is detailed in Sec.~\ref{SecCorrelationM2}. The parameters \(c_{11}\), \(c_{21}\), \(\rho_{10}\), and \(\xi_c\) are positive real constants, with \(\xi_c\) specifically denoting the coherence length.}
\renewcommand{\arraystretch}{1.6}  
\resizebox{\columnwidth}{!}{  
\begin{tabular}{l c c c c}
\hline
\hline
$d$ & $\langle \hat{\Phi}_{1}^{\dagger}(\mathbf{x}) \hat{\Phi}_{1}(\mathbf{0}) \rangle$ &
     $\langle \hat{\Phi}_{2}^{\dagger}(\mathbf{x}) \hat{\Phi}_{2}(\mathbf{0}) \rangle$ &
     \makecell{Broken Charge\\Symmetry} &
     \makecell{Order\\Parameter} \\
 \hline
$1$ & $\rho_{10} \left( \frac{\left|\mathbf{x}\right|}{\xi_{c}} \right)^{-\frac{c_{11} + c_{21}}{\pi c_{11} c_{21}}}$ & 0 & N/A & N/A \\
 
$2$ & $\rho_{10} e^{-\frac{c_{11} + c_{21}}{2\pi c_{11} c_{21} \xi_{c}}}$ & Power-law decay & $U(1)_{1,C}$ & $\langle \hat{\Phi}_1(\mathbf{x}) \rangle$ \\
 
$3$ & Constant & Constant & $U(1)_{1,C} \otimes U(1)_{2,C}$ & $\langle \hat{\Phi}_1(\mathbf{x}) \rangle, \langle \hat{\Phi}_2(\mathbf{x}) \rangle$ \\
\hline
\hline
\end{tabular}}
\label{correlation2}
\end{table}

Through the calculations, we confirm that the correlation function $\langle\hat{\Phi}_1^\dagger(\mathbf{x})\hat{\Phi}_1(\mathbf{0})\rangle$ decays in a power law in one spatial dimension and saturates a constant in two or more spatial dimensions. The correlation function $\langle\hat{\Phi}_2^\dagger(\mathbf{x})\hat{\Phi}_2(\mathbf{0})\rangle$ decays exponentially in one spatial dimension, decays in a power law in two spatial dimensions, and saturates a constant in three or more spatial dimensions.
The constant behaviors of these two correlation functions represent the spontaneous breaking of $U(1)_{1,C}$ symmetry and $U(1)_{2,C}$ symmetry respectively.
Therefore, this system has a true ODLRO in two spatial dimensions and above.
Compared with dipole-conserving and quadrupole-conserving fractonic superfluid models with no hybridization, the spontaneous breaking of the charge symmetry $U(1)_{1,C}$ and $U(1)_{2,C}$ occurs in lower spatial dimensions.
This decrease in dimension comes from the hybridization of conserved quantities as well as the coupling of the two species of bosons.

\section{Bose-Hubbard lattice models}
\label{Lattice model} 

Just as conventional superfluids can be realized in the small $U$ phase of conventional Bose-Hubbard lattice models, a similar Bose-Hubbard-type lattice model can be constructed, whose low-energy physics in the small $U$ limit reduces to the low-energy effective field theory of fractonic superfluids~\cite{Fractonicsuperfluids1,DBHM}. In the preceding sections, we have studied the HFS phases of matter in continuum space.

In this section, we will construct lattice models that (i) satisfy the aforementioned conservation laws and (ii) reduce to continuum models in Model Series A and B in the small $U$ limit. Therefore,  unless otherwise specified, we will use \textit{Lattice Model Series A} (\textit{Lattice Model Series B}) to denote the Bose-Hubbard lattice models that, in the small $U$ limit, reduce to   continuum models in Model Series A (Model Series B). Upon completing lattice construction, in this section we will also present a third-order perturbation theory to demonstrate how to realize lattice models in strongly tilted optical lattices.

\subsection{Construction of Lattice Model Series}

\begin{figure}
  \centering
  \includegraphics[width=0.48\textwidth]{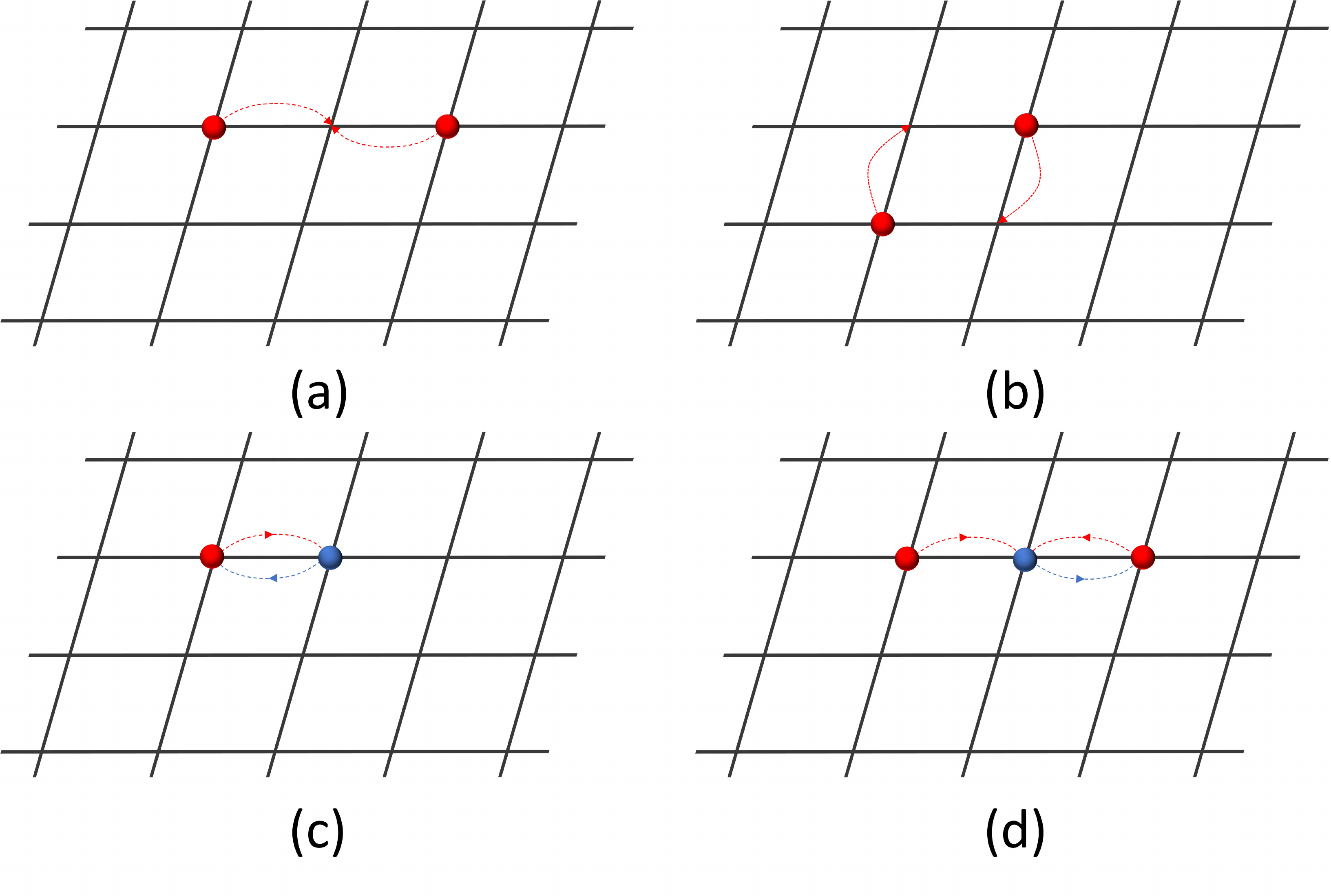}
  \caption{Illustration of the hopping terms [Eqs.~(\ref{M1hopping}, \ref{M2hopping})] on a square lattice. The balls of red and blue colors represent bosons of two distinct species. The hopping shown in (a) corresponds to diagonal $t_a^{(ij)}$ terms in Eq.~(\ref{M1hopping}). The hopping shown in (b) corresponds to off-diagonal $t_a^{(ij)}$ terms in Eq.~(\ref{M1hopping}). The hopping shown in (c) corresponds to $t_{ab}^{(i)}$ terms in Eq.~(\ref{M1hopping}). In (d), the blue ball represents one boson of species $1$, and the red balls represent bosons of species $2$. The hopping shown in (d) corresponds to the $t_{12}^{(i)}$ terms in Eq.~(\ref{M2hopping}).}
  \label{fighoppings}
\end{figure}

We specifically study the lattice model corresponding to the \([d,1,2,2]\) model shown in Eq.~(\ref{Hmdd}).  We put all fields on a square lattice and use difference operators to replace all spatial derivatives:  
\begin{align}
\partial_{i}\hat{\Phi}_{a}(\mathbf{x})\!\longrightarrow & \frac{1}{\eta}(\hat{b}_{a,\mathbf{i}+\hat{x}_i}\!-\!\hat{b}_{a,\mathbf{i}}),\label{differencemethod1111}\\
\!\!\!\partial_{i}\partial_{j}\hat{\Phi}_{a}(\mathbf{x})\!\longrightarrow & \frac{1}{\eta^{2}}(\hat{b}_{a,\mathbf{i}+\hat{x}_i+\hat{x}_j}\!-\!\hat{b}_{a,\mathbf{i}+\hat{x}_i}\!-\!\hat{b}_{a,\mathbf{i}+\hat{x}_j}\!+\!\hat{b}_{a,\mathbf{i}}),
\label{differencemethod2222}
\end{align}
where $\eta$ is the lattice constant. $\hat{x}_i$ and $\hat{x}_j$ ($i,j=1,2,\cdots, d$) denote   spatial unit vectors. 
$\mathbf{i}={\eta}^{-1}(x_1,\cdots,x_d)$ represent lattice coordinates. By applying the above substitution rules,  we can straightforwardly obtain the following lattice model that conserves quantities in Eqs.~(\ref{ddcon00}, \ref{ddcon}):
\begin{align}
H_{\text{lat A}}=H_{\text{hop}}+H_{\text{onsite}},\label{LM1}
\end{align}
where 
\begin{align}
\!\!H_{\text{hop}}= & \sum_{a}^{2}\bigg[\sum_{\mathbf{i},i,j}-t_{a}^{(ij)}\hat{b}_{a,\mathbf{i}+\hat{x}_i+\hat{x}_j}^{\dagger}\hat{b}_{a,\mathbf{i}}^{\dagger}\hat{b}_{a,\mathbf{i}+\hat{x}_i}\hat{b}_{a,\mathbf{i}+\hat{x}_j}\nonumber\\
 & +\sum_{b\neq a}^{2}\sum_{\mathbf{i},i}-t_{ab}^{(i)}\hat{b}_{b,\mathbf{i}+\hat{x}_i}^{\dagger}\hat{b}_{a,\mathbf{i}+\hat{x}_i}\hat{b}_{a,\mathbf{i}}^{\dagger}\hat{b}_{b,\mathbf{i}}\bigg]+\text{h.c.}\,,
 \label{M1hopping}\\
\!\!H_{\text{onsite}}=&\sum_{a}^{2}\bigg[-\mu\sum_{\mathbf{i}}\hat{n}_{a,\mathbf{i}}+\frac{U}{2}\sum_{\mathbf{i}}\hat{n}_{a,\mathbf{i}}(\hat{n}_{a,\mathbf{i}}-1)\bigg].
\end{align}
The $t_{a}^{(ij)}$ terms and the $t_{ab}^{(i)}$ terms ($t_{a}^{(ij)}=t_{a}^{(ji)}>0$, $t_{ab}^{(i)}=t_{ba}^{(i)}>0$) correspond to the $K_{a}^{ij}$ terms and the $\Gamma_{a,b}^{i}$ terms in the \([d,1,2,2]\) model shown in Eq.~(\ref{Hmdd}) respectively.  
These hopping processes are illustrated in Fig.~\ref{fighoppings}(a,b,c) in the case of two dimensional square lattice.
 
Next, we discuss the construction of the lattice model corresponding to the continuum model described by Eq.~(\ref{Hmdq}). Evidently, Model Series B is significantly more complex than Model Series A, and this complexity extends to the construction of its lattice model. For simplicity, we neglect the $K_{2}^{ijk}$ terms.  We further set $\ell=2\eta$ for the sake of convenience and finally  end up with the following lattice model that conserves   charges of each species, the hybrid moments, and other lower moments [Eqs.~(\ref{dqcon1}, \ref{dqcon2}, \ref{dqcon3}, \ref{dqcon4})]: 
\begin{align}
  H_{\text{lat B}}=H_{\text{hop}}+H_{\text{onsite}},
\end{align}
where 
\begin{align}
\!\!\!\!\!\!\!H_{\text{hop}}\!= \!& \bigg[\sum_{\mathbf{i},i,j}-t_{1}^{(ij)}\hat{b}_{1,\mathbf{i}+\hat{x}_i+\hat{x}_j}^{\dagger}\hat{b}_{1,\mathbf{i}}^{\dagger}\hat{b}_{1,\mathbf{i}+\hat{x}_i}\hat{b}_{1,\mathbf{i}+\hat{x}_j}\nonumber\\
 +&\sum_{\mathbf{i},i}-t_{12}^{(i)}\hat{b}_{1,\mathbf{i}+\hat{x}_i}^{\dagger}\hat{b}_{1,\mathbf{i}}\hat{b}_{2,\mathbf{i}}^{\dagger2}\hat{b}_{2,\mathbf{i}+\hat{x}_i}\hat{b}_{2,\mathbf{i}-\hat{x}_i}\bigg]\!+\!\text{h.c.},\label{M2hopping}\\
\!\!\!H_{\text{onsite}}=&\sum_{a}^{2}\bigg[-\mu\sum_{\mathbf{i}}\hat{n}_{a,\mathbf{i}}+\frac{U}{2}\sum_{\mathbf{i}}\hat{n}_{a,\mathbf{i}}(\hat{n}_{a,\mathbf{i}}-1)\bigg].
\end{align}
The $t_1^{(ij)}$ terms and the $t_{12}^{(i)}$ terms correspond to the $K_1^{ij}$ terms and the $\Gamma_{1,2}^i$ terms in  Eq.~(\ref{Hmdq}) respectively.  The hopping processes represented by  the $t_{12}^{(i)}$ terms are illustrated in Fig.~\ref{fighoppings}(d) in the case of two dimensional square lattice.

\subsection{Mean-field phase diagram of Lattice Model Series A}
\label{meanfield}
\begin{figure}[htbp]
  \centering
  \includegraphics[width=0.44\textwidth]{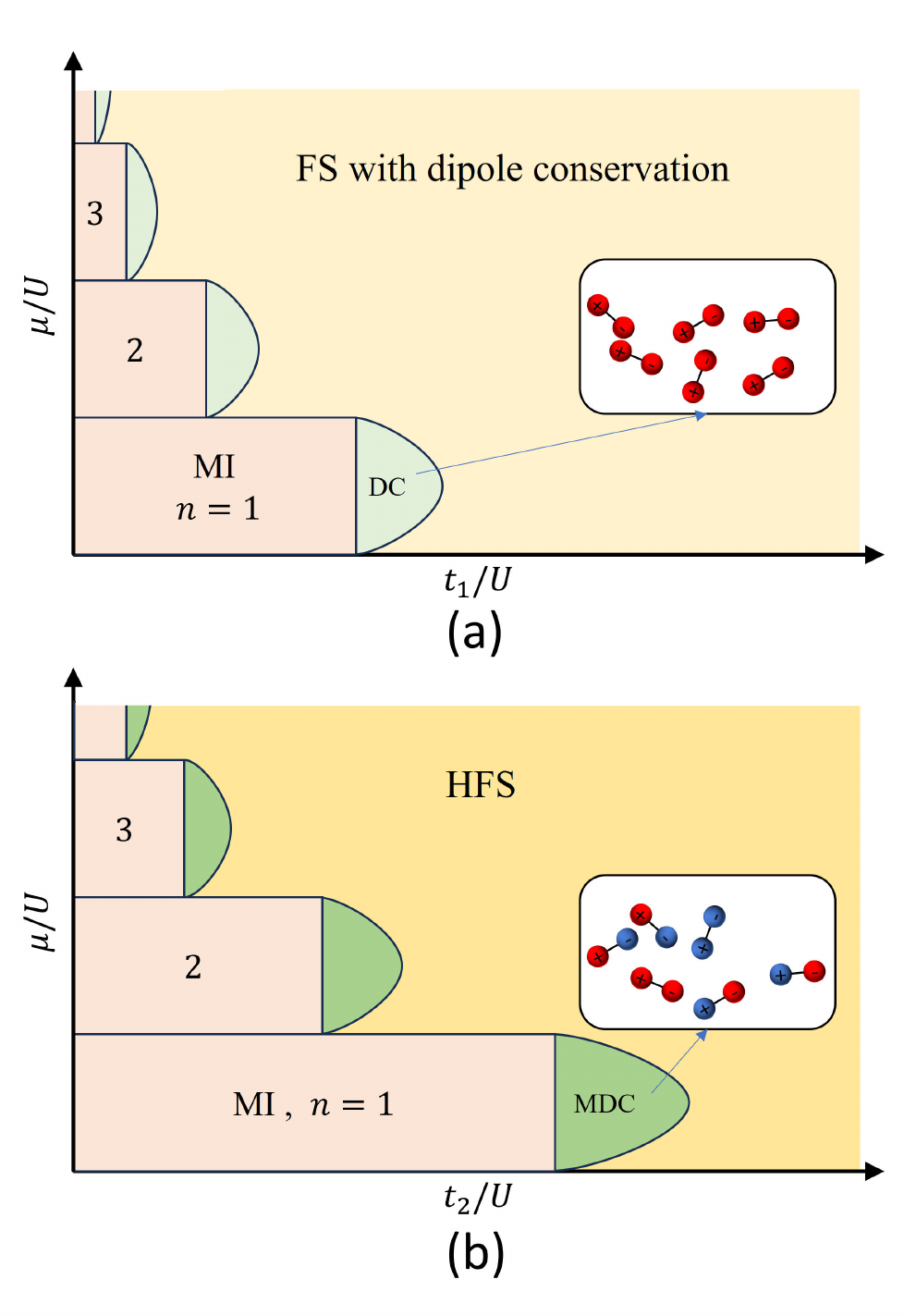}
  \caption{Schematic phase diagrams from mean-field theory by setting (a) $t_2=0$, and (b) $t_1=0$. The pink regions are  Mott insulators, and $n$ represents the average boson filling per site. The light and dark green regions in the diagrams are dipole condensates (DC) and multi-dipole condensates (MDC) respectively. The light and dark yellow regions are fractonic superfluids with dipole conservation \cite{Fractonicsuperfluids1,DBHM} and hybrid fractonic superfluids described by Eq. (\ref{Hmdd}) respectively. In the dipole condensate phase, $\langle \hat{b}_{a,\mathbf{i}+\hat{x}_i}^\dagger \hat{b}_{a,\mathbf{i}}\rangle\neq 0$, $\langle \hat{b}_{1,\mathbf{i}}^\dagger \hat{b}_{2,\mathbf{i}}\rangle= 0$. In the multi-dipole condensate phase, $\langle \hat{b}_{a,\mathbf{i}+\hat{x}_i}^\dagger \hat{b}_{a,\mathbf{i}}\rangle\neq 0$, $\langle \hat{b}_{1,\mathbf{i}}^\dagger \hat{b}_{2,\mathbf{i}}\rangle\neq 0$.}
  \label{phase_t}
\end{figure}

\begin{figure}[htbp]
  \centering
  \includegraphics[width=0.44\textwidth]{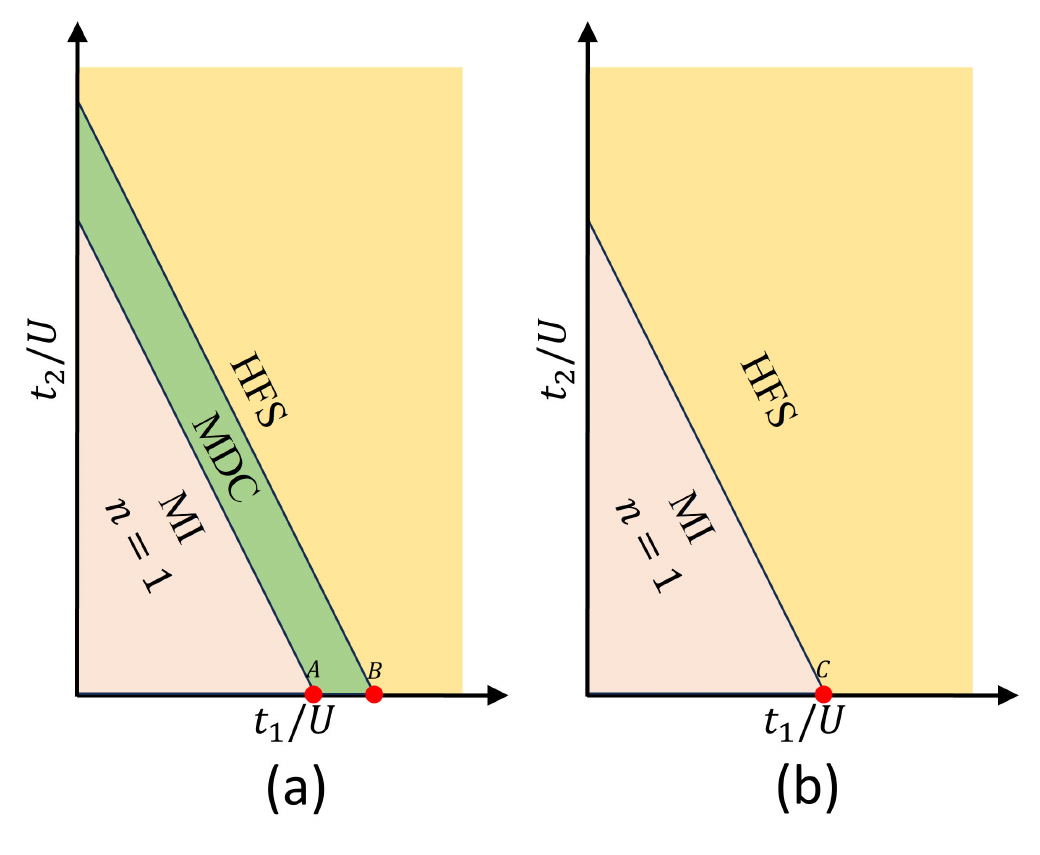}
  \caption{Schematic phase diagrams from mean-field theory by setting (a) $\mu/U=\frac{1}{2}$, and (b) $\mu/U=1$. In (a), along the horizontal axis ($t_2=0$), the region between the point A and B is the DC phase, and the region on the right of point B is fractonic superfluids with dipole conservation. In (b), along the horizontal axis ($t_2=0$), the region on  the right of point C is fractonic superfluids with dipole conservation.
  }
  \label{phase_mu}
\end{figure}

We employ the mean-field approximation to sketch out the schematic phase diagram of Lattice Model Series A [Eq.~(\ref{LM1})], as shown in Fig.~\ref{phase_t} and Fig.~\ref{phase_mu}.
We set $t_{a}^{(ij)}=t_{1}$, $t_{ab}^{(i)}=t_{2}$,
which are the coefficients of the hopping terms of the same species and different species respectively.

At the point $t_2=0$, the Hamiltonian [Eq.~(\ref{LM1})] reduces to two independent dipolar Bose-Hubbard models~\cite{DBHM}, and has a similar phase diagram [Fig.~\ref{phase_t} (a)].
At the limit $t_1= 0$, all particles lose their mobility, and the system is at the Mott insulator (MI) phase with fixed average boson densities of species 1 and 2.
At the small $U$ limit (i.e., large $t_1/U$, $\mu/U$), the system is at the fractonic superfluid (FS) phase which is studied in Ref.~\cite{Fractonicsuperfluids1}.
Between the MI phase and the FS phase, the light green region in Fig.~\ref{phase_t} (a) shows the dipole condensate (DC) phase with non-zero $\langle\hat{b}_{a,\mathbf{i}+\hat{x}_i}^\dagger \hat{b}_{a,\mathbf{i}}\rangle$. 
At the mean-field level, the phase transition between MI and DC occurs at the point $\frac{1}{2dt_1}-\frac{2n(n+1)}{U}= 0$~\cite{DBHM}.

When we take $t_2>0$, the hybridization of bosons of different species drives the emergence of another order parameter $\langle \hat{b}_{1,\mathbf{i}}^\dagger \hat{b}_{2,\mathbf{i}}\rangle$.
Non-zero $\langle \hat{b}_{1,\mathbf{i}}^\dagger \hat{b}_{2,\mathbf{i}}\rangle$ implies the existence of exotic intermediate phases with condensates of dipolar bound states $\hat{b}_{1,\mathbf{i}}^\dagger\hat{b}_{2,\mathbf{i}}$. 
To distinguish it from the DC phase with condensates of dipoles formed by bosons of the same species, the phases are referred to as the \textit{multi-dipole condensates} (MDC).
The physics in the small $U$ limit of this lattice model is described by the HFS [Eq.~(\ref{Hmdd})] in this paper.
A summary of comparison between Bose-Hubbard-type models with different conservation laws is given in Table \ref{BHMtable} in the Introduction.

By applying the dipolar and single-particle mean-field theory \cite{DBHM}, we can determine the locations of the phase boundaries between different phases and complete the schematic phase diagram (Fig.~\ref{phase_t} and Fig.~\ref{phase_mu}).
Detailed steps of the mean-field theory to determine the phase boundary between the MI phase and the MDC phase are given in Appendix \ref{meanfieldapp}.
To derive the phase diagram beyond mean-field theory, it
will be meaningful to employ more advanced treatments, e.g., large-scale Monte-Carlo and tensor network.


\subsection{Realizations of Lattice Model Series A in  strongly tilted optical lattice}

In the following, we discuss how to realize the Hamiltonian~[Eq.~(\ref{LM1})] of Lattice Model Series A in  strongly tilted optical lattices.
We start with a microscopic Hamiltonian with two species of bosons in a tilted optical lattice chain.
Since there is only one direction, we can simplify the notation $\hat{x}_1=1$ and the lattice coordinate $\mathbf{i}$ becomes an integer $\mathbf{i}=1,2,\cdots$. The Hamiltonian can be written as:
\begin{align}
\!\!\!\!  &H_{\text{tilted}}=-t\sum_{a,\mathbf{i}}(\hat{b}_{a,\mathbf{i}}^\dagger\hat{b}_{a,\mathbf{i}+1}+\text{h.c.})-\mu\sum_{a,\mathbf{i}}\hat{n}_{a,\mathbf{i}}\nonumber\\
\!\!\!\! \!&\!+\!\frac{U}{2}\!\sum_{a,\mathbf{i}}\hat{n}_{a,\mathbf{i}}(\hat{n}_{a,\mathbf{i}}-1)\!+\!U^\prime\sum_{\mathbf{i}}\hat{n}_{1,\mathbf{i}}\hat{n}_{2,\mathbf{i}}\!+\!V\sum_{a,\mathbf{i}}\mathbf{i}\hat{n}_{a,\mathbf{i}},\!\!\!
\end{align}
where $t$, $\mu$, and $U$ terms are just terms in Bose-Hubbard model, and the $U^\prime$ term denotes the interaction between bosons of different species. $V$ is the strength of the tilt potential. 
In the limit $V\gg t\,, U$, we perform a third-order perturbation theory.
The effective Hamiltonian becomes
\begin{align}
  &H_{\text{eff}}=\sum_{a,\mathbf{i}}\bigg(-\frac{t^2U}{V^2}\hat{b}_{a,\mathbf{i}}(\hat{b}_{a,\mathbf{i}+1}^{\dagger})^2\hat{b}_{a,\mathbf{i}+2}+\text{h.c.}\bigg)\nonumber\\
  &+\!\sum_{a\neq b\,,\mathbf{i}}\!\bigg(\frac{t^2U^\prime}{V^2}\hat{b}_{a,\mathbf{i}}^\dagger \hat{b}_{a,\mathbf{i}+1}\hat{b}_{b,\mathbf{i}+1}\hat{b}_{b,\mathbf{i}+2}^\dagger-\frac{t^2U^\prime}{V^2} \hat{b}_{a,\mathbf{i}}^\dagger \hat{b}_{a,\mathbf{i}+1}\hat{b}_{b,\mathbf{i}}\hat{b}_{b,\mathbf{i}+1}^\dagger\nonumber\\
  &+\text{h.c.}\bigg)+\!\sum_{a,\mathbf{i}}\!\Big(\!-\mu-\frac{U}{2}+V\mathbf{i}\Big)\hat{n}_{a,\mathbf{i}}+\!\sum_{a,\mathbf{i}}\!\bigg(\frac{U}{2}-\frac{2t_0^2U}{V^2}\bigg)\hat{n}_{a,\mathbf{i}}^2\nonumber\\
  &+\frac{4t_0^2U}{V^2}\hat{n}_{a,\mathbf{i}}\hat{n}_{a,\mathbf{i}+1}+\sum_{a\neq b\,,\mathbf{i}}\bigg[\Big(\frac{U^\prime}{2}-\frac{2t_0^2U^\prime}{V^2}\Big)\hat{n}_{a,\mathbf{i}}\hat{n}_{b,\mathbf{i}}\nonumber\\
  &+\frac{2t_0^2U^\prime}{V^2}\hat{n}_{a,\mathbf{i}}\hat{n}_{b,\mathbf{i}+1}\bigg]\,.
  \label{tiltedeff}
\end{align}
The detailed derivation in Eq.~(\ref{tiltedeff}) is given in Appendix~\ref{detailtilted}.

\section{Summary and Outlook}
\label{Summary and Outlook} 
In conclusion, in this paper we have completed  the third part in the trilogy on fractonic superfluids~\cite{Fractonicsuperfluids1,Fractonicsuperfluids2}, exploring the novel symmetries arising from higher-moment conservation and the associated symmetry-breaking phenomena. Just like the prior two papers about fractonic superfluids, the main goal of this paper is to lay the foundation for hybrid fractonic superfluids, leaving various promising discussions for future studies. Therefore, we anticipate further investigations, such as those akin to Refs.~\cite{NSofFractonicsuperfluids,fracton23,Fractonicsuperfluidsdefect,reviewofFractonicsuperfluids}, to deepen our understanding on new symmetries and symmetry-breaking phases in the presence of higher moment conservation. Let us elaborate in more details:

By endowing algebraic structures \cite{fracton20} with conserved hybrid moments, a rich variety of effective field theories can be constructed.
It is compelling to derive and numerically solve the hydrodynamic equations \cite{NSofFractonicsuperfluids} of systems with conservation of hybrid moments.
We can also employ renormalization group analysis to determine whether systems with conserved hybrid moments can emerge at low energies \cite{fracton23}.
The finite-temperature phase transitions \cite{Fractonicsuperfluidsdefect} in these hybrid-moment-conserving models are also worth investigating, where the configuration of symmetry defects (i.e., various generalized topological vortices) and hierarchical Kosterlitz-Thouless transitions potentially get enriched by moment hybridization.
The results of the partial spontaneous breaking of charge symmetry will generalize the Hohenberg-Mermin-Wagner theorems~\cite{fracton30,HMWT}.  
As mobility restrictions lead to unconventional types of toric codes~\cite{PhysRevB.105.045128}, it is interesting to explore further generalizations of discrete gauge theory and toric codes through additional possibilities of mobility constraints arising from conservations, such as the hybridization discussed in the present work. 
Detecting such states in the context of stabilizer codes and cellular automata, e.g., in Refs.~\cite{PRXQuantum.5.030328,SSPT2,2024arXiv241011942L}, will be an intriguing direction. 
 In principle, these Bose-Hubbard models constructed in this paper  can be  extended to fermionic Hubbard models for realizing non-Fermi liquids~\cite{fracton24}. 
Numerical simulations, such as quantum Monte Carlo, DMRG, and tensor network methods, can be applied to study these lattice models. These lattice models have the potential to be  academically attractive in cold atom communities, such as strongly tilted optical lattices discussed in Refs.~\cite{PhysRevB.101.174204,PhysRevX.10.011042,scherg_observing_2021,PhysRevLett.130.010201} and general Josephson effects in Ref.~\cite{fracton45}. 
We can construct some other lattice systems with hybrid moment conservation which are experimentally realizable in tilted optical lattices.
By incorporating the concept of conserved hybridized moments, diffusion in spin models \cite{diffusionofhigher-moment} can also be explored.
The conservation of hybrid moments may induce different fragmentation behaviors in Hilbert space~\cite{PhysRevB.101.174204}.
Finally, it will be interesting to investigate how the moment hybridization affect  the dilaton spectrum in an unusual Ginzburg-Landau theory~\cite{fracton14}.

 \acknowledgements
We thank Long Zhang for helpful discussions. This work was partially supported by the National Natural Science Foundation of China (NSFC) under Grant Nos.~12474149 and 12074438. The calculations reported were performed on resources provided by the Guangdong Provincial Key Laboratory of Magnetoelectric Physics and Devices (No. 2022B1212010008).


%

\onecolumngrid

\appendix

\section{Derivations of Goldstone modes and dispersion relations for Model Series A}\label{appendix_s}
 
\subsection{Derivation of the Hamiltonian in the momentum space}\label{Hmddmom}

Based on the Hamiltonian [Eq.~(\ref{Hmdd1})], we can derive the Hamiltonian of the momentum space [Eq.~(\ref{Hmddmommain})] in the following steps. In the momentum space, we can perform a Fourier transform,
\begin{align}
H= & \frac{1}{(2\pi)^{2d}}\int d^{d}\mathbf{x}d^{d}\mathbf{k}d^{d}\mathbf{k^{\prime}}e^{i\mathbf{k}\cdot\mathbf{x}+i\mathbf{k^{\prime}}\cdot\mathbf{x}}\sum_{a}^{2}\Bigg[\frac{g}{2}\hat{\pi}_{a} (\mathbf{k}) \hat{\pi}_{a}(\mathbf{k^{\prime}})+\frac{\kappa\rho_{a0}^{2}\left|\mathbf{k}\right|^{4}}{2}\hat{\theta}_{a} (\mathbf{k}) \hat{\theta}_{a}(\mathbf{k^{\prime}})\nonumber\\
+ & \frac{\Gamma\rho_{a0}\rho_{b0}\left|\mathbf{k}\right|^{2}}{2}\sum_{b\neq a}^{2}(\hat{\theta}_{a} (\mathbf{k}) \hat{\theta}_{a}(\mathbf{k^{\prime}})+\hat{\theta}_{b} (\mathbf{k}) \hat{\theta}_{b}(\mathbf{k^{\prime}})-\hat{\theta}_{a} (\mathbf{k}) \hat{\theta}_{b}(\mathbf{k^{\prime}})-\hat{\theta}_{b} (\mathbf{k}) \hat{\theta}_{a}(\mathbf{k^{\prime}}))\Bigg].
\end{align}

Integrating over the real space, we can obtain a $\delta$ function
so that only the integral of the momentum space is left in the Hamiltonian,
\begin{align}
H= & \frac{1}{(2\pi)^{d}}\int d^{d}\mathbf{k}d^{d}\mathbf{k^{\prime}}\delta(\mathbf{k}+\mathbf{k}^{\prime})\sum_{a}^{2}\Bigg[\frac{g}{2}\hat{\pi}_{a} (\mathbf{k}) \hat{\pi}_{a}(\mathbf{k^{\prime}})+\frac{\kappa\rho_{a0}^{2}\left|\mathbf{k}\right|^{4}}{2}\hat{\theta}_{a} (\mathbf{k}) \hat{\theta}_{a}(\mathbf{k^{\prime}})\nonumber\\
+ & \frac{\Gamma\rho_{a0}\rho_{b0}\left|\mathbf{k}\right|^{2}}{2}\sum_{b\neq a}^{2}(\hat{\theta}_{a} (\mathbf{k}) \hat{\theta}_{a}(\mathbf{k^{\prime}})+\hat{\theta}_{b} (\mathbf{k}) \hat{\theta}_{b}(\mathbf{k^{\prime}})-\hat{\theta}_{a} (\mathbf{k}) \hat{\theta}_{b}(\mathbf{k^{\prime}})-\hat{\theta}_{b} (\mathbf{k}) \hat{\theta}_{a}(\mathbf{k^{\prime}}))\Bigg]\nonumber\\
= & \frac{1}{(2\pi)^{d}}\int d^{d}\mathbf{k}\sum_{a}^{2}\Bigg[\frac{g}{2}\hat{\pi}_{a} (\mathbf{k}) \hat{\pi}_{a} (-\mathbf{k}) +\frac{\kappa\rho_{a0}^{2}\left|\mathbf{k}\right|^{4}}{2}\hat{\theta}_{a} (\mathbf{k}) \hat{\theta}_{a} (-\mathbf{k}) \nonumber\\
+ & \frac{\Gamma\rho_{a0}\rho_{b0}\left|\mathbf{k}\right|^{2}}{2}\sum_{b\neq a}^{2}(\hat{\theta}_{a} (\mathbf{k}) \hat{\theta}_{a} (-\mathbf{k}) +\hat{\theta}_{b} (\mathbf{k}) \hat{\theta}_{b} (-\mathbf{k}) -\hat{\theta}_{a} (\mathbf{k}) \hat{\theta}_{b} (-\mathbf{k}) -\hat{\theta}_{b} (\mathbf{k}) \hat{\theta}_{a} (-\mathbf{k}) )\Bigg].
\end{align}
The Hamiltonian [Eq.~(\ref{Hmddmommain})] can be obtained by writing
it in the matrix form and using the notation in the main text,
\begin{align}
H= & \frac{1}{2(2\pi)^{d}}\int d^{d}\mathbf{k}\sum_{a}^{2}\Bigg[g\hat{\pi}_{a} (\mathbf{k}) \hat{\pi}_{a} (-\mathbf{k}) +\kappa\rho_{a0}^{2}\left|\mathbf{k}\right|^{4}\hat{\theta}_{a} (\mathbf{k}) \hat{\theta}_{a} (-\mathbf{k}) \nonumber\\
+ & \Gamma\rho_{a0}\rho_{b0}\left|\mathbf{k}\right|^{2}\sum_{b\neq a}^{2}(\hat{\theta}_{a} (\mathbf{k}) \hat{\theta}_{a} (-\mathbf{k}) +\hat{\theta}_{b} (\mathbf{k}) \hat{\theta}_{b} (-\mathbf{k}) -\hat{\theta}_{a} (\mathbf{k}) \hat{\theta}_{b} (-\mathbf{k}) -\hat{\theta}_{b} (\mathbf{k}) \hat{\theta}_{a} (-\mathbf{k}) )\Bigg]\nonumber\\
= & \frac{1}{2(2\pi)^{d}}\int d^{d}\mathbf{k}\psi^{\dagger} (\mathbf{k}) \begin{bmatrix}g & 0 & 0 & 0\\
0 & g & 0 & 0\\
0 & 0 & \kappa\rho_{10}^{2}\left|\mathbf{k}\right|^{4}+2\Gamma\rho_{10}\rho_{20}\left|\mathbf{k}\right|^{2} & -2\Gamma\rho_{10}\rho_{20}\left|\mathbf{k}\right|^{2}\\
0 & 0 & -2\Gamma\rho_{10}\rho_{20}\left|\mathbf{k}\right|^{2} & \kappa\rho_{20}^{2}\left|\mathbf{k}\right|^{4}+2\Gamma\rho_{10}\rho_{20}\left|\mathbf{k}\right|^{2}
\end{bmatrix}\psi (\mathbf{k}) \nonumber\\
= & \frac{1}{2(2\pi)^{d}}\int d^{d}\mathbf{k}\psi^{\dagger} (\mathbf{k}) M (\mathbf{k}) \psi (\mathbf{k}) .
\end{align}
\subsection{Matrix elements of the transformation matrix $T (\mathbf{k}) $}\label{M1Tmatrix}
In the main text, we give the commutation relations that the operators need to satisfy. To keep the commutation relations and diagonalize the matrix $M (\mathbf{k}) $, the transformation matrix $T (\mathbf{k}) $ needs to obey the equation below,
\begin{align}
  T (\mathbf{k}) \Lambda_1T^\dagger (\mathbf{k}) =\Lambda_2
\end{align}
where 
\begin{align}
  \Lambda_1=\begin{bmatrix}
    0 & 0 & -i & 0\\
    0 & 0 & 0 & -i\\
    i & 0 & 0 & 0\\
    0 & i & 0 & 0
  \end{bmatrix},\quad
  \Lambda_2=\text{diag}(1,1,-1,-1).
\end{align}
The transformation matrix $T (\mathbf{k}) $ is given by 
\begin{align}
    T (\mathbf{k}) =\begin{bmatrix}
      T_{11} (\mathbf{k})  & T_{12} (\mathbf{k})  & T_{13} (\mathbf{k})  & T_{14} (\mathbf{k}) \\
      T_{21} (\mathbf{k})  & T_{22} (\mathbf{k})  & T_{23} (\mathbf{k})  & T_{24} (\mathbf{k}) \\
      T_{31} (\mathbf{k})  & T_{32} (\mathbf{k})  & T_{33} (\mathbf{k})  & T_{34} (\mathbf{k}) \\
      T_{41} (\mathbf{k})  & T_{42} (\mathbf{k})  & T_{43} (\mathbf{k})  & T_{44} (\mathbf{k}) \\
    \end{bmatrix},
\end{align}
where the matrix elements are
\begin{align}
    T_{11} (\mathbf{k}) =&-T_{31} (\mathbf{k}) =(i g (\kappa  \left|\mathbf{k}\right|^4 (\rho _{20}^2-\rho _{10}^2)+\sqrt{\kappa ^2 \left|\mathbf{k}\right|^8 (\rho _{10}^2-\rho _{20}^2){}^2+16 \Gamma ^2 k^4 \rho _{10}^2 \rho _{20}^2}))/\nonumber\\
    &\Bigg\{\sqrt[4]{2} \bigg[g (16 \Gamma ^2 \left|\mathbf{k}\right|^4 \rho _{10}^2 \rho _{20}^2+(\kappa  \left|\mathbf{k}\right|^4 (\rho _{20}^2-\rho _{10}^2)+\sqrt{\kappa ^2 k^8 (\rho _{10}^2-\rho _{20}^2){}^2+16 \Gamma ^2 \left|\mathbf{k}\right|^4 \rho _{10}^2 \rho _{20}^2}){}^2) \nonumber\\
    &\sqrt{g (\kappa  \left|\mathbf{k}\right|^4 (\rho _{10}^2+\rho _{20}^2)+4 \Gamma  \left|\mathbf{k}\right|^2 \rho _{10} \rho _{20}-\sqrt{\kappa ^2 \left|\mathbf{k}\right|^8 (\rho _{10}^2-\rho _{20}^2){}^2+16 \Gamma ^2 \left|\mathbf{k}\right|^4 \rho _{10}^2 \rho _{20}^2})}\bigg]^{\frac{1}{2}}\Bigg\},
\end{align}
\begin{align}
    T_{12} (\mathbf{k}) =&-T_{32} (\mathbf{k}) =(4ig\Gamma\rho_{10}\rho_{20}\left|\mathbf{k}\right|^2)/\nonumber\\
    &\Bigg\{\sqrt[4]{2} \bigg[g (16 \Gamma ^2 \left|\mathbf{k}\right|^4 \rho _{10}^2 \rho _{20}^2+(\kappa  \left|\mathbf{k}\right|^4 (\rho _{20}^2-\rho _{10}^2)+\sqrt{\kappa ^2 k^8 (\rho _{10}^2-\rho _{20}^2){}^2+16 \Gamma ^2 \left|\mathbf{k}\right|^4 \rho _{10}^2 \rho _{20}^2}){}^2) \nonumber\\
    &\sqrt{g (\kappa  \left|\mathbf{k}\right|^4 (\rho _{10}^2+\rho _{20}^2)+4 \Gamma  \left|\mathbf{k}\right|^2 \rho _{10} \rho _{20}-\sqrt{\kappa ^2 \left|\mathbf{k}\right|^8 (\rho _{10}^2-\rho _{20}^2){}^2+16 \Gamma ^2 \left|\mathbf{k}\right|^4 \rho _{10}^2 \rho _{20}^2})}\bigg]^{\frac{1}{2}}\Bigg\},
\end{align}
\begin{align}
    T_{13} (\mathbf{k}) =&T_{33} (\mathbf{k}) =\Bigg[(\kappa  \left|\mathbf{k}\right|^4 (\rho _{20}^2-\rho _{10}^2)+\sqrt{\kappa ^2 \left|\mathbf{k}\right|^8 (\rho _{10}^2-\rho _{20}^2){}^2+16 \Gamma ^2 \left|\mathbf{k}\right|^4 \rho _{10}^2 \rho _{20}^2}) \nonumber\\
    &\sqrt[4]{\frac{\kappa  \left|\mathbf{k}\right|^4 (\rho _{10}^2+\rho _{20}^2)+4 \Gamma  \left|\mathbf{k}\right|^2 \rho _{10} \rho _{20}-\sqrt{\kappa ^2 \left|\mathbf{k}\right|^8 (\rho _{10}^2-\rho _{20}^2){}^2+16 \Gamma ^2 \left|\mathbf{k}\right|^4 \rho _{10}^2 \rho _{20}^2}}{g}}\Bigg]/\nonumber\\
    &(2^{3/4} \sqrt{16 \Gamma ^2 \left|\mathbf{k}\right|^4 \rho _{10}^2 \rho _{20}^2+(\kappa  \left|\mathbf{k}\right|^4 (\rho _{20}^2-\rho _{10}^2)+\sqrt{\kappa ^2 \left|\mathbf{k}\right|^8 (\rho _{10}^2-\rho _{20}^2){}^2+16 \Gamma ^2 \left|\mathbf{k}\right|^4 \rho _{10}^2 \rho _{20}^2})^2}),
\end{align}
\begin{align}
    &T_{14} (\mathbf{k}) =T_{34} (\mathbf{k}) \nonumber\\
    =&\Bigg(2 \Gamma  \left|\mathbf{k}\right|^2 \rho _{10} \rho _{20} \sqrt[4]{\frac{2 \kappa  \left|\mathbf{k}\right|^4 (\rho _{10}^2+\rho _{20}^2)+8 \Gamma  \left|\mathbf{k}\right|^2 \rho _{10} \rho _{20}-2 \sqrt{\kappa ^2 \left|\mathbf{k}\right|^8 (\rho _{10}^2-\rho _{20}^2){}^2+16 \Gamma ^2 \left|\mathbf{k}\right|^4 \rho _{10}^2 \rho _{20}^2}}{g}}\Bigg)/\nonumber\\
    &(\sqrt{16 \Gamma ^2 \left|\mathbf{k}\right|^4 \rho _{10}^2 \rho _{20}^2+(\kappa  \left|\mathbf{k}\right|^4 (\rho _{20}^2-\rho _{10}^2)+\sqrt{\kappa ^2 \left|\mathbf{k}\right|^8 (\rho _{10}^2-\rho _{20}^2){}^2+16 \Gamma ^2 \left|\mathbf{k}\right|^4 \rho _{10}^2 \rho _{20}^2}){}^2})
\end{align}
\begin{align}
    T_{21} (\mathbf{k}) =&-T_{41} (\mathbf{k}) =-\Bigg[i g (\kappa  \left|\mathbf{k}\right|^4 (\rho _{10}^2-\rho _{20}^2)+\sqrt{\kappa ^2 \left|\mathbf{k}\right|^8 (\rho _{10}^2-\rho _{20}^2){}^2+16 \Gamma ^2 \left|\mathbf{k}\right|^4 \rho _{10}^2 \rho _{20}^2})\Bigg]/\nonumber\\
    &\Bigg\{2^{3/4} \left|\mathbf{k}\right|^2 \bigg[g \Big(16 \Gamma ^2 \rho _{10}^2 \rho _{20}^2+\kappa ^2 \left|\mathbf{k}\right|^4 (\rho _{10}^2-\rho _{20}^2){}^2+\kappa  (\rho _{10}^2-\rho_{20}^2) \sqrt{\kappa ^2 \left|\mathbf{k}\right|^8 (\rho _{10}^2-\rho _{20}^2){}^2+16 \Gamma ^2 \left|\mathbf{k}\right|^4 \rho _{10}^2 \rho _{20}^2}\Big)\nonumber\\
    &\sqrt{g (\kappa  \left|\mathbf{k}\right|^4 (\rho _{10}^2+\rho _{20}^2)+4 \Gamma  \left|\mathbf{k}\right|^2 \rho _{10} \rho _{20}+\sqrt{\kappa ^2 \left|\mathbf{k}\right|^8 (\rho _{10}^2-\rho _{20}^2){}^2+16 \Gamma ^2 \left|\mathbf{k}\right|^4 \rho _{10}^2 \rho _{20}^2})}\bigg]^{\frac{1}{2}}\Bigg\},
\end{align}
\begin{align}
    T_{22} (\mathbf{k}) =&-T_{42} (\mathbf{k}) =(2 i \sqrt[4]{2} \Gamma  g \rho _{10} \rho _{20})/\nonumber\\
    &\Bigg[\sqrt[4]{g (\kappa  \left|\mathbf{k}\right|^4 (\rho _{10}^2+\rho _{20}^2)+4 \Gamma  \left|\mathbf{k}\right|^2 \rho _{10} \rho _{20}+\sqrt{\kappa ^2 k^8 (\rho _{10}^2-\rho _{20}^2){}^2+16 \Gamma ^2 \left|\mathbf{k}\right|^4 \rho _{10}^2 \rho _{20}^2})} \nonumber\\
    &\sqrt{g (16 \Gamma ^2 \rho _{10}^2 \rho _{20}^2+\kappa ^2 \left|\mathbf{k}\right|^4 (\rho _{10}^2-\rho _{20}^2){}^2+\kappa(\rho _{10}^2-\rho _{20}^2) \sqrt{\kappa ^2 \left|\mathbf{k}\right|^8 (\rho _{10}^2-\rho _{20}^2){}^2+16 \Gamma ^2 \left|\mathbf{k}\right|^4 \rho _{10}^2 \rho _{20}^2} )}\Bigg],
\end{align}
\begin{align}
    &T_{23} (\mathbf{k}) =T_{43} (\mathbf{k}) \nonumber\\
    =&\frac{\kappa  \left|\mathbf{k}\right|^4 (\rho _{20}^2-\rho _{10}^2)-\sqrt{\kappa ^2 \left|\mathbf{k}\right|^8 (\rho _{10}^2-\rho _{20}^2){}^2+16 \Gamma ^2 \left|\mathbf{k}\right|^4 \rho _{10}^2 \rho _{20}^2}}{2 \sqrt[4]{2} \left|\mathbf{k}\right|^2 \sqrt{\frac{g (16 \Gamma ^2 \rho _{10}^2 \rho _{20}^2+\kappa ^2 \left|\mathbf{k}\right|^4 (\rho _{10}^2-\rho _{20}^2){}^2+\kappa  \rho _{10}^2 \sqrt{\kappa ^2 \left|\mathbf{k}\right|^8 (\rho _{10}^2-\rho _{20}^2){}^2+16 \Gamma ^2 \left|\mathbf{k}\right|^4 \rho _{10}^2 \rho _{20}^2}-\kappa  \rho _{20}^2 \sqrt{\kappa ^2 \left|\mathbf{k}\right|^8 (\rho _{10}^2-\rho _{20}^2){}^2+16 \Gamma ^2 \left|\mathbf{k}\right|^4 \rho _{10}^2 \rho _{20}^2})}{\sqrt{g (\kappa  \left|\mathbf{k}\right|^4 (\rho _{10}^2+\rho _{20}^2)+4 \Gamma  \left|\mathbf{k}\right|^2 \rho _{10} \rho _{20}+\sqrt{\kappa ^2 \left|\mathbf{k}\right|^8 (\rho _{10}^2-\rho _{20}^2){}^2+16 \Gamma ^2 \left|\mathbf{k}\right|^4 \rho _{10}^2 \rho _{20}^2})}}}},
\end{align}
\begin{align}
    &T_{24} (\mathbf{k}) =T_{44} (\mathbf{k}) \nonumber\\
    =&\frac{2^{3/4} \Gamma  \rho _{10} \rho _{20}}{\sqrt{\frac{g (16 \Gamma ^2 \rho _{10}^2 \rho _{20}^2+\kappa ^2 \left|\mathbf{k}\right|^4 (\rho _{10}^2-\rho _{20}^2){}^2+\kappa  \rho _{10}^2 \sqrt{\kappa ^2 \left|\mathbf{k}\right|^8 (\rho _{10}^2-\rho _{20}^2){}^2+16 \Gamma ^2 \left|\mathbf{k}\right|^4 \rho _{10}^2 \rho _{20}^2}-\kappa  \rho _{20}^2 \sqrt{\kappa ^2 \left|\mathbf{k}\right|^8 (\rho _{10}^2-\rho _{20}^2){}^2+16 \Gamma ^2 \left|\mathbf{k}\right|^4 \rho _{10}^2 \rho _{20}^2})}{\sqrt{g (\kappa  \left|\mathbf{k}\right|^4 (\rho _{10}^2+\rho _{20}^2)+4 \Gamma  \left|\mathbf{k}\right|^2 \rho _{10} \rho _{20}+\sqrt{\kappa ^2 \left|\mathbf{k}\right|^8 (\rho _{10}^2-\rho _{20}^2){}^2+16 \Gamma ^2 \left|\mathbf{k}\right|^4 \rho _{10}^2 \rho _{20}^2})}}}}.
\end{align}

\subsection{Derivation of the Hamiltonian represented by the creation and annihilation operators}\label{Hmddfin}
After diagonalization, the Hamiltonian [Eq.~(\ref{M1diagH})] becomes 
\begin{align}
    H= & \frac{1}{2(2\pi)^{d}}\phi^{\dagger} (\mathbf{k}) (T^{-1} (\mathbf{k}) )^{\dagger}M (\mathbf{k}) T^{-1} (\mathbf{k}) \phi (\mathbf{k}) \nonumber\\
    = & \frac{1}{2(2\pi)^{d}}\int d^{d}\mathbf{k}D_{1} (\mathbf{k}) (\hat{\alpha}^{\dagger} (\mathbf{k}) \hat{\alpha} (\mathbf{k}) +\hat{\alpha} (-\mathbf{k}) \hat{\alpha}^{\dagger} (-\mathbf{k}) )+D_{2} (\mathbf{k}) (\hat{\beta}^{\dagger} (\mathbf{k}) \hat{\beta} (\mathbf{k}) +\hat{\beta} (-\mathbf{k}) \hat{\beta}^{\dagger} (-\mathbf{k}) ).
\end{align}
Because of $D_{1} (\mathbf{k}) =D_{1} (-\mathbf{k}) $, $D_{2} (\mathbf{k}) =D_{2} (-\mathbf{k}) $, we rewrite the Hamiltonian as 
\begin{align}
H= & \frac{1}{2(2\pi)^{d}}\int d^{d}\mathbf{k}D_{1} (\mathbf{k}) (\hat{\alpha}^{\dagger} (\mathbf{k}) \hat{\alpha} (\mathbf{k}) +\hat{\alpha} (\mathbf{k}) \hat{\alpha}^{\dagger} (\mathbf{k}) )+D_{2} (\mathbf{k}) (\hat{\beta}^{\dagger} (\mathbf{k}) \hat{\beta} (\mathbf{k}) +\hat{\beta} (\mathbf{k}) \hat{\beta}^{\dagger} (\mathbf{k}) )\nonumber\\
= & \frac{1}{(2\pi)^{d}}\int d^{d}\mathbf{k}D_{1} (\mathbf{k}) \hat{\alpha}^{\dagger} (\mathbf{k}) \hat{\alpha} (\mathbf{k}) +D_{2} (\mathbf{k}) \hat{\beta}^{\dagger} (\mathbf{k}) \hat{\beta} (\mathbf{k}) +\int d^{d}\mathbf{k}\frac{D_{1} (\mathbf{k}) +D_{2} (\mathbf{k}) }{2}\delta^{(d)}(\mathbf{0})\nonumber\\
= & \frac{1}{(2\pi)^{d}}\int d^{d}\mathbf{k}D_{1} (\mathbf{k}) \hat{\alpha}^{\dagger} (\mathbf{k}) \hat{\alpha} (\mathbf{k}) +D_{2} (\mathbf{k}) \hat{\beta}^{\dagger} (\mathbf{k}) \hat{\beta} (\mathbf{k}) +\frac{\tilde{V}}{(2\pi)^{d}}\int d^{d}\mathbf{k}\frac{D_{1}(\mathbf{x})+D_{2} (\mathbf{k}) }{2},
\end{align}
where $\tilde{V}$ is the volume of the $d$-dimensional space, the last term is the vacuum zero point energy. Then, we obtain the dispersion relations of Goldstone mode [Eq.~(\ref{M1dispersion})].

\section{The calculation details of correlation function of Model Series A}

\label{detail1} In this appendix, we will give the calculation details of correlation functions of Model Series A.
From the main text and Appendix \ref{M1Tmatrix}, we have the Bogoliubov transformation from $\hat{\theta}$ and $\hat{\pi}$ to the independent modes $\hat{\alpha}$ and $\hat{\beta}$,
\begin{align}
\psi (\mathbf{k}) = & \begin{bmatrix}\hat{\pi}_{1} (\mathbf{k})  & \hat{\pi}_{2} (\mathbf{k})  & \hat{\theta}_{1} (\mathbf{k})  & \hat{\theta}_{2} (\mathbf{k}) \end{bmatrix}^{T}=T^{-1} (\mathbf{k}) \phi (\mathbf{k}) =\Lambda_{1}T^{\dagger} (\mathbf{k}) \Lambda_{2}\phi (\mathbf{k}) \nonumber\\
= & \begin{bmatrix}-iT_{13}^{\ast} (\mathbf{k})  & -iT_{23}^{\ast} (\mathbf{k})  & iT_{33}^{\ast} (\mathbf{k})  & iT_{43}^{\ast} (\mathbf{k}) \\
-iT_{14}^{\ast} (\mathbf{k})  & -iT_{24}^{\ast} (\mathbf{k})  & iT_{34}^{\ast} (\mathbf{k})  & iT_{44}^{\ast} (\mathbf{k}) \\
iT_{11}^{\ast} (\mathbf{k})  & iT_{21}^{\ast} (\mathbf{k})  & -iT_{31}^{\ast} (\mathbf{k})  & -iT_{41}^{\ast} (\mathbf{k}) \\
iT_{12}^{\ast} (\mathbf{k})  & iT_{22}^{\ast} (\mathbf{k})  & -iT_{32}^{\ast} (\mathbf{k})  & -iT_{42}^{\ast} (\mathbf{k}) 
\end{bmatrix}\begin{bmatrix}\hat{\alpha} (\mathbf{k}) \\
\hat{\beta} (\mathbf{k}) \\
\hat{\alpha}^{\dagger} (-\mathbf{k}) \\
\hat{\beta}^{\dagger} (-\mathbf{k}) 
\end{bmatrix}.
\end{align}
In the main text, we express two-operator and four-operator correlation functions of the boson operator $\hat{\Phi}_a$ in terms of two-operator correlation functions of the phase field operator $\hat{\theta}_a$.
Two-operator correlation functions of the phase field operator $\hat{\theta}_a$ can be given in terms of the matrix elements of the matrix $T (\mathbf{k}) $ in the following procedure,
\begin{align}
\langle\hat{\theta}_{1}(\mathbf{x})\hat{\theta}_{1}(\mathbf{0})\rangle= & \langle\int\frac{d^{d}\mathbf{k}d^{d}\mathbf{k}^{\prime}}{(2\pi)^{2d}}e^{i\mathbf{k}\cdot\mathbf{x}}\hat{\theta}_{1} (\mathbf{k}) \hat{\theta}_{1}\ (\mathbf{k}') \rangle=\langle\int\frac{d^{d}\mathbf{k}d^{d}\mathbf{k}^{\prime}}{(2\pi)^{2d}}e^{i\mathbf{k}\cdot\mathbf{x}}\hat{\theta}_{1} (\mathbf{k}) \hat{\theta}_{1}^{\dagger}(-\mathbf{k}^{\prime})\rangle\nonumber\\
= & \int\frac{d^{d}\mathbf{k}d^{d}\mathbf{k}^{\prime}}{(2\pi)^{2d}}e^{i\mathbf{k}\cdot\mathbf{x}}\langle\left[iT_{11}^{\ast} (\mathbf{k}) \hat{\alpha} (\mathbf{k}) +iT_{21}^{\ast} (\mathbf{k}) \hat{\beta} (\mathbf{k}) -iT_{31}^{\ast} (\mathbf{k}) \hat{\alpha}^{\dagger} (-\mathbf{k}) -iT_{41}^{\ast} (\mathbf{k}) \hat{\beta}^{\dagger} (-\mathbf{k}) \right]\nonumber\\
 & \left[-iT_{11}(-\mathbf{k}^{\prime})\hat{\alpha}^{\dagger}(-\mathbf{k}^{\prime})-iT_{21}(-\mathbf{k}^{\prime})\hat{\beta}^{\dagger}(-\mathbf{k}^{\prime})+iT_{31}(-\mathbf{k}^{\prime})\hat{\alpha}\ (\mathbf{k}') +iT_{41}(-\mathbf{k}^{\prime})\hat{\beta}\ (\mathbf{k}') \right]\rangle.
\end{align}
For the expectation value of the creation and annihilation operators
in the ground state, we have 
\begin{align}
& \langle\hat{\alpha}\ (\mathbf{k}') \hat{\alpha} (\mathbf{k}) \rangle=\langle\hat{\beta}\ (\mathbf{k}') \hat{\beta} (\mathbf{k}) \rangle=\langle\hat{\alpha}^{\dagger}(-\mathbf{k}^{\prime})\hat{\alpha} (\mathbf{k}) \rangle=\langle\hat{\beta}^{\dagger}(-\mathbf{k}^{\prime})\hat{\beta} (\mathbf{k}) \rangle=0,
\end{align}
\begin{align}
\langle\hat{\alpha} (\mathbf{k}) \hat{\alpha}^{\dagger}(-\mathbf{k}^{\prime})\rangle=\langle\hat{\beta} (\mathbf{k}) \hat{\beta}^{\dagger}(-\mathbf{k}^{\prime})\rangle=(2\pi)^{d}\delta^{(d)}(\mathbf{k}+\mathbf{k}^{\prime}).
\end{align}
Then we can obtain the correlation function 
\begin{align}
\langle\hat{\theta}_{1}(\mathbf{x})\hat{\theta}_{1}(\mathbf{0})\rangle=\int\frac{d^{d}\mathbf{k}}{(2\pi)^{d}}e^{i\mathbf{k}\cdot\mathbf{x}}(\left|T_{11} (\mathbf{k}) \right|^{2}+\left|T_{21} (\mathbf{k}) \right|^{2}).
\end{align}
Similarly, we can get 
\begin{align}
\langle\hat{\theta}_{2}(\mathbf{x})\hat{\theta}_{2}(\mathbf{0})\rangle=\int\frac{d^{d}\mathbf{k}}{(2\pi)^{d}}e^{i\mathbf{k}\cdot\mathbf{x}}(\left|T_{12} (\mathbf{k}) \right|^{2}+\left|T_{22} (\mathbf{k}) \right|^{2}).
\end{align}
The correlation function of different species of phase field operators is given by
\begin{align}
\langle\hat{\theta}_{1}(\mathbf{x})\hat{\theta}_{2}(\mathbf{0})\rangle= & \langle\int\frac{d^{d}\mathbf{k}d^{d}\mathbf{k}^{\prime}}{(2\pi)^{2d}}e^{i\mathbf{k}\cdot\mathbf{x}}\hat{\theta}_{1} (\mathbf{k}) \hat{\theta}_{2}^{\dagger} (-\mathbf{k}) \rangle\nonumber\\
= & \int\frac{d^{d}\mathbf{k}d^{d}\mathbf{k}^{\prime}}{(2\pi)^{2d}}e^{i\mathbf{k}\cdot\mathbf{x}}\langle\left[iT_{11}^{\ast} (\mathbf{k}) \hat{\alpha} (\mathbf{k}) +iT_{21}^{\ast} (\mathbf{k}) \hat{\beta} (\mathbf{k}) -iT_{31}^{\ast} (\mathbf{k}) \hat{\alpha}^{\dagger} (-\mathbf{k}) -iT_{41}^{\ast} (\mathbf{k}) \hat{\beta}^{\dagger} (-\mathbf{k}) \right]\nonumber\\
 & \left[-iT_{12}(-\mathbf{k}^{\prime})\hat{\alpha}^{\dagger}(-\mathbf{k}^{\prime})-iT_{22}(-\mathbf{k}^{\prime})\hat{\beta}^{\dagger}(-\mathbf{k}^{\prime})+iT_{32}(-\mathbf{k}^{\prime})\hat{\alpha}\ (\mathbf{k}') +iT_{42}(-\mathbf{k}^{\prime})\hat{\beta}\ (\mathbf{k}') \right]\rangle\nonumber\\
= & \int\frac{d^{d}\mathbf{k}}{(2\pi)^{d}}e^{i\mathbf{k}\cdot\mathbf{x}}\left[T_{11}^{\ast} (\mathbf{k}) T_{12} (\mathbf{k}) +T_{21}^{\ast} (\mathbf{k}) T_{22} (\mathbf{k}) \right].
\end{align}

For the $T (\mathbf{k}) $ matrix of Model Series A in Appendix \ref{M1Tmatrix}, we can give some approximations,
\begin{align}
\left|T_{11} (\mathbf{k}) \right|^{2}\approx & \frac{g}{2\sqrt{2g\kappa(\rho_{10}^{2}+\rho_{20}^{2})}\left|\mathbf{k}\right|^{2}}=\frac{1}{c_{1}\left|\mathbf{k}\right|^{2}},\nonumber\\
\left|T_{12} (\mathbf{k}) \right|^{2}\approx & \frac{g}{2\sqrt{2g\kappa(\rho_{10}^{2}+\rho_{20}^{2})}\left|\mathbf{k}\right|^{2}}=\frac{1}{c_{1}\left|\mathbf{k}\right|^{2}},\nonumber\\
\left|T_{21} (\mathbf{k}) \right|^{2}\approx & \frac{g}{8\sqrt{g\Gamma\rho_{10}\rho_{20}}\left|\mathbf{k}\right|}=\frac{1}{c_{2}\left|\mathbf{k}\right|},\nonumber\\
\left|T_{22} (\mathbf{k}) \right|^{2}\approx & \frac{g}{8\sqrt{g\Gamma\rho_{10}\rho_{20}}\left|\mathbf{k}\right|}=\frac{1}{c_{2}\left|\mathbf{k}\right|},\nonumber\\
T_{11}^{\ast} (\mathbf{k}) T_{12} (\mathbf{k}) \approx & \frac{g}{2\sqrt{2g\kappa(\rho_{10}^{2}+\rho_{20}^{2})}\left|\mathbf{k}\right|^{2}}=\frac{1}{c_{1}\left|\mathbf{k}\right|^{2}},\nonumber\\
T_{21}^{\ast} (\mathbf{k}) T_{22} (\mathbf{k}) \approx & -\frac{g}{8\sqrt{g\Gamma\rho_{10}\rho_{20}}\left|\mathbf{k}\right|}=-\frac{1}{c_{2}\left|\mathbf{k}\right|}.
\end{align}
Based on the above derivation, we can calculate correlation functions in different spatial dimensions. 

\subsection{One spatial dimension}

In one spatial dimension, we can plug in the results of the above
approximation. The correlation functions of the operator $\hat{\theta}$
are 
\begin{align}
\langle\hat{\theta}_{1}(x)\hat{\theta}_{1}(0)\rangle=\langle\hat{\theta}_{2}(x)\hat{\theta}_{2}(0)\rangle= & \int\frac{dk}{2\pi}e^{ikx}\left(\frac{1}{c_{1}\left|k\right|^{2}}+\frac{1}{c_{2}\left|k\right|}\right)\nonumber\\
= & \frac{1}{2\pi c_{1}}\int dk\frac{e^{ikx}}{\left|k\right|^{2}}+\frac{1}{2\pi c_{2}}\int dk\frac{e^{ikx}}{\left|k\right|}\nonumber\\
= & \frac{1}{2\pi c_{1}}\left(\int_{-\infty}^{-\frac{2\pi}{L}}dk\frac{e^{ikx}}{k^{2}}+\int_{\frac{2\pi}{L}}^{\infty}dk\frac{e^{ikx}}{k^{2}}\right)+\frac{1}{2\pi c_{2}}\left(\int_{-\infty}^{-\frac{2\pi}{L}}dk\frac{e^{ikx}}{-k}+\int_{\frac{2\pi}{L}}^{\infty}\frac{e^{ikx}}{k}\right).
\end{align}
We need to compute these integrals separately. Considering $L>0$, $x>0$, we have the first two integrals 
\begin{align}
\int_{-\infty}^{-\frac{2\pi}{L}}dk\frac{e^{ikx}}{k^{2}}+\int_{\frac{2\pi}{L}}^{\infty}dk\frac{e^{ikx}}{k^{2}}=\frac{1}{2\pi}\left[e^{-2i\pi x}L+e^{2i\pi x}L+2i\pi x\Gamma(0,-\frac{2i\pi x}{L})-2i\pi x\Gamma(0,-\frac{2i\pi x}{L})\right]\overset{L\rightarrow\infty}{=}\frac{L}{\pi}-\pi x,
\end{align}
where $\Gamma(a,z)=\int_{z}^{\infty}t^{a-1}e^{-t}dt$ is
the incomplete gamma function. The last two integrals are 
\begin{align}
\int_{-\infty}^{-\frac{2\pi}{L}}dk\frac{e^{ikx}}{-k}+\int_{\frac{2\pi}{L}}^{\infty}\frac{e^{ikx}}{k}=\Gamma(0,-\frac{2i\pi x}{L})+\Gamma(0,\frac{2i\pi x}{L})\overset{L\rightarrow\infty}{=}-2\left[\gamma+\ln(\frac{2\pi x}{L})\right],
\end{align}
where $\gamma$ is the Euler-Mascheroni constant. Then, we can get the correlation functions. 
\begin{align}
\langle\hat{\theta}_{1}(x)\hat{\theta}_{1}(0)\rangle=\langle\hat{\theta}_{2}(x)\hat{\theta}_{2}(0)\rangle\overset{L\rightarrow\infty}{=} & \frac{1}{2\pi c_{1}}\left(\frac{L}{\pi}-\pi\left|x\right|\right)+\frac{1}{\pi c_{2}}\left[-\gamma+\ln(\frac{L}{2\pi\left|x\right|})\right]
\end{align}
Similarly, we have 
\begin{align}
\langle\hat{\theta}_{1}(x)\hat{\theta}_{2}(0)\rangle\overset{L\rightarrow\infty}{=} & \frac{1}{2\pi c_{1}}\left(\frac{L}{\pi}-\pi\left|x\right|\right)-\frac{1}{\pi c_{2}}\left[-\gamma+\ln(\frac{L}{2\pi\left|x\right|})\right].
\end{align}
To get $\langle\hat{\theta}_{a}^{2}(0)\rangle=\lim_{x\rightarrow0}\langle\theta_{a}(x)\theta_{a}(0)\rangle$,
we just need to replace $\left|x\right|$ by $\xi_{c}$. $\xi_{c}$
is the coherence length,
\begin{align}
\langle\hat{\theta}_{1}^{2}(0)\rangle=\langle\hat{\theta}_{2}^{2}(0)\rangle= & \frac{1}{2\pi c_{1}}\left(\frac{L}{\pi}-\pi\xi_{c}\right)+\frac{1}{\pi c_{2}}\left[-\gamma+\ln(\frac{L}{2\pi\xi_{c}})\right],
\end{align}
\begin{align}
\langle\hat{\theta}_{1}(0)\hat{\theta}_{2}(0)\rangle= & \frac{1}{2\pi c_{1}}\left(\frac{L}{\pi}-\pi\xi_{c}\right)-\frac{1}{\pi c_{2}}\left[-\gamma+\ln(\frac{L}{2\pi\xi_{c}})\right].
\end{align}
Then, we can calculate correlation functions of $\hat{\Phi}$. The
two-operator correlation functions are 
\begin{align}
\langle\hat{\Phi}_{1}^{\dagger}(x)\hat{\Phi}_{1}(0)\rangle= & \rho_{10}e^{\langle\hat{\theta}_{1}(x)\hat{\theta}_{1}(0)\rangle-\langle\hat{\theta}_{1}^{2}(0)\rangle}=\rho_{10}e^{-\frac{\left|x\right|}{2c_{1}}+\frac{\xi_{c}}{2c_{1}}}\left(\frac{\left|x\right|}{\xi_{c}}\right)^{-\frac{1}{\pi c_{2}}}\approx\rho_{10}e^{-\frac{\left|x\right|}{2c_{1}}}\left(\frac{\left|x\right|}{\xi_{c}}\right)^{-\frac{1}{\pi c_{2}}},
\end{align}
\begin{align}
\langle\hat{\Phi}_{2}^{\dagger}(x)\hat{\Phi}_{2}(0)\rangle= & \rho_{20}e^{\langle\hat{\theta}_{2}(x)\hat{\theta}_{2}(0)\rangle-\langle\hat{\theta}_{2}^{2}(0)\rangle}=\rho_{20}e^{-\frac{\left|x\right|}{2c_{1}}+\frac{\xi_{c}}{2c_{1}}}\left(\frac{\left|x\right|}{\xi_{c}}\right)^{-\frac{1}{\pi c_{2}}}\approx\rho_{20}e^{-\frac{\left|x\right|}{2c_{1}}}\left(\frac{\left|x\right|}{\xi_{c}}\right)^{-\frac{1}{\pi c_{2}}},
\end{align}
\begin{align}
\langle\hat{\Phi}_{1}^{\dagger}(x)\hat{\Phi}_{2}(0)\rangle= & \sqrt{\rho_{10}\rho_{20}}e^{\langle\hat{\theta}_{1}(\mathbf{x})\hat{\theta}_{2}(\mathbf{0})\rangle-\frac{1}{2}\langle\hat{\theta}_{1}^{2}(\mathbf{0})\rangle-\frac{1}{2}\langle\hat{\theta}_{2}^{2}(\mathbf{0})\rangle}=\sqrt{\rho_{10}\rho_{20}}e^{-\frac{\left|x\right|}{2c_{1}}+\frac{\xi_{c}}{2c_{1}}+\frac{2\gamma}{\pi c_{2}}}\left(\frac{L^{2}}{4\pi^{2}\xi_{c}\left|x\right|}\right)^{-\frac{1}{\pi c_{2}}}\overset{L\rightarrow\infty}{=}0.
\end{align}
The four-operator correlation functions are
\begin{align}
& \langle\hat{\Phi}_{1}^{\dagger}(\mathbf{x})\hat{\Phi}_{2}(\mathbf{x})\hat{\Phi}_{1}(\mathbf{0})\hat{\Phi}_{2}^{\dagger}(\mathbf{0})\rangle\nonumber\\
= & \rho_{10}\rho_{20}\exp\bigg(\langle\hat{\theta}_{1}(\mathbf{x})\hat{\theta}_{1}(\mathbf{0})\rangle+\langle\hat{\theta}_{2}(\mathbf{x})\hat{\theta}_{2}(\mathbf{0})\rangle+2\langle\hat{\theta}_{1}(\mathbf{0})\hat{\theta}_{2}(\mathbf{0})\rangle-\langle\hat{\theta}_{1}^{2}(\mathbf{0})\rangle-\langle\hat{\theta}_{2}^{2}(\mathbf{0})\rangle-2\langle\hat{\theta}_{1}(\mathbf{x})\hat{\theta}_{2}(\mathbf{0})\rangle\bigg)\nonumber\\
= & \rho_{10}\rho_{20}(\frac{\left|x\right|}{\xi_{c}})^{-\frac{4}{\pi c_{2}}},
\end{align}
\begin{align}
    & \langle\hat{\Phi}_{1}^{\dagger}(\mathbf{x})\hat{\Phi}_{2}^\dagger(\mathbf{x})\hat{\Phi}_{1}(\mathbf{0})\hat{\Phi}_{2}(\mathbf{0})\rangle\nonumber\\
    = & \rho_{10}\rho_{20}\exp\bigg(\langle\hat{\theta}_{1}(\mathbf{x})\hat{\theta}_{1}(\mathbf{0})\rangle+\langle\hat{\theta}_{2}(\mathbf{x})\hat{\theta}_{2}(\mathbf{0})\rangle-2\langle\hat{\theta}_{1}(\mathbf{0})\hat{\theta}_{2}(\mathbf{0})\rangle-\langle\hat{\theta}_{1}^{2}(\mathbf{0})\rangle-\langle\hat{\theta}_{2}^{2}(\mathbf{0})\rangle+2\langle\hat{\theta}_{1}(\mathbf{x})\hat{\theta}_{2}(\mathbf{0})\rangle\bigg)\nonumber\\
    =&\rho_{10}\rho_{20}e^{\frac{2\xi_c}{c_1}}e^{-\frac{2}{c_1}\left|x\right|}\approx\rho_{10}\rho_{20}e^{-\frac{2}{c_1}\left|x\right|}.
\end{align}
Since all these correlation functions do not saturate to constants, Model Series A does not break the charge symmetry in one spatial dimension. Since the correlation function $\langle\hat{\Phi}_{1}^{\dagger}(\mathbf{x})\hat{\Phi}_{2}(\mathbf{x})\hat{\Phi}_{1}(\mathbf{0})\hat{\Phi}_{2}^{\dagger}(\mathbf{0})\rangle$ is power-law decay, the system is algebraically ordered in one spatial dimension.

\subsection{Two spatial dimensions}

In two spatial dimensions, we can plug in the results of the above approximation. The correlation functions of the operator $\hat{\theta}$ are 
\begin{align}
\langle\hat{\theta}_{1}(\mathbf{x})\hat{\theta}_{1}(\mathbf{0})\rangle=\langle\hat{\theta}_{2}(\mathbf{x})\hat{\theta}_{2}(\mathbf{0})\rangle= & \int\frac{d^{2}\mathbf{k}}{(2\pi)^{2}}e^{i\mathbf{k}\cdot\mathbf{x}}\left(\frac{1}{c_{1}\left|\mathbf{k}\right|^{2}}+\frac{1}{c_{2}\left|\mathbf{k}\right|}\right)\nonumber\\
= & \frac{1}{4\pi^{2}c_{1}}\int dkd\theta\frac{e^{ik\left|\mathbf{x}\right|\cos\theta}}{k}+\frac{1}{4\pi^{2}c_{2}}\int dkd\theta e^{ik\left|\mathbf{x}\right|\cos\theta}\nonumber\\
= & \frac{1}{4\pi^{2}c_{1}}\int_{0}^{2\pi}d\theta\int_{\frac{2\pi}{L}}^{\infty}dk\frac{e^{ik\left|\mathbf{x}\right|\cos\theta}}{k}+\frac{1}{4\pi^{2}c_{2}}\int_{0}^{2\pi}d\theta\int_{0}^{\infty}dke^{ik\left|\mathbf{x}\right|\cos\theta}.
\end{align}
We need to compute these integrals separately,
\begin{align}
\int_{0}^{2\pi}d\theta\int_{\frac{2\pi}{L}}^{\infty}dk\frac{e^{ik\left|\mathbf{x}\right|\cos\theta}}{k}\overset{L\rightarrow\infty}{=}-2\pi\left[\gamma+\ln(\frac{\pi\left|\mathbf{x}\right|}{L})\right],
\end{align}
\begin{align}
\int_{0}^{2\pi}d\theta\int_{0}^{\infty}dke^{ik\left|\mathbf{x}\right|\cos\theta}=2\pi\int_{0}^{\infty}dkJ_{0}(k\left|\mathbf{x}\right|)=\frac{2\pi}{\left|\mathbf{x}\right|},
\end{align}
where $J_{n}(z)$ is the Bessel function of the first kind.
Then, we can get the correlation functions. 
\begin{align}\langle\hat{\theta}_{1}(\mathbf{x})\hat{\theta}_{1}(\mathbf{0})\rangle=\langle\hat{\theta}_{2}(\mathbf{x})\hat{\theta}_{2}(\mathbf{0})\rangle\overset{L\rightarrow\infty}{=} & \frac{1}{2\pi c_{1}}\left[-\gamma+\ln(\frac{L}{\pi\left|\mathbf{x}\right|})\right]+\frac{1}{2\pi c_{2}\left|\mathbf{x}\right|}
\end{align}
Similarly, we have 
\begin{align}
\langle\hat{\theta}_{1}(\mathbf{x})\hat{\theta}_{2}(\mathbf{0})\rangle\overset{L\rightarrow\infty}{=}\frac{1}{2\pi c_{1}}\left[-\gamma+\ln(\frac{L}{\pi\left|\mathbf{x}\right|})\right]-\frac{1}{2\pi c_{2}\left|\mathbf{x}\right|}.
\end{align}
To get $\langle\hat{\theta}_{a}^{2}(0)\rangle=\lim_{x\rightarrow0}\langle\theta_{a}(x)\theta_{a}(0)\rangle$,
we just need to replace $\left|x\right|$ by $\xi_{c}$,
\begin{align}
\langle\hat{\theta}_{1}^{2}(\mathbf{0})\rangle=\langle\hat{\theta}_{2}^{2}(\mathbf{0})\rangle= & \frac{1}{2\pi c_{1}}\left[-\gamma+\ln(\frac{L}{\pi\xi_{c}})\right]+\frac{1}{2\pi c_{2}\xi_{c}},
\end{align}
\begin{align}
\langle\hat{\theta}_{1}(\mathbf{0})\hat{\theta}_{2}(\mathbf{0})\rangle=\frac{1}{2\pi c_{1}}\left[-\gamma+\ln(\frac{L}{\pi\xi_{c}})\right]-\frac{1}{2\pi c_{2}\xi_{c}}.
\end{align}
Then, we can calculate correlation functions of $\hat{\Phi}$. The
two-operator correlation functions are 
\begin{align}
\langle\hat{\Phi}_{1}^{\dagger}(\mathbf{x})\hat{\Phi}_{1}(\mathbf{0})\rangle= & \rho_{10}e^{\langle\hat{\theta}_{1}(\mathbf{x})\hat{\theta}_{1}(\mathbf{0})\rangle-\langle\hat{\theta}_{1}^{2}(\mathbf{0})\rangle}=\rho_{10}\left(\frac{\left|\mathbf{x}\right|}{\xi_{c}}\right)^{-\frac{1}{2\pi c_{1}}}e^{\frac{1}{2\pi c_{2}\left|\mathbf{x}\right|}-\frac{1}{2\pi c_{2}\xi_{c}}},
\end{align}
\begin{align}
\langle\hat{\Phi}_{2}^{\dagger}(\mathbf{x})\hat{\Phi}_{2}(\mathbf{0})\rangle= & \rho_{20}e^{\langle\hat{\theta}_{2}(\mathbf{x})\hat{\theta}_{2}(\mathbf{0})\rangle-\langle\hat{\theta}_{2}^{2}(\mathbf{0})\rangle}=\rho_{20}\left(\frac{\left|\mathbf{x}\right|}{\xi_{c}}\right)^{-\frac{1}{2\pi c_{1}}}e^{\frac{1}{2\pi c_{2}\left|\mathbf{x}\right|}-\frac{1}{2\pi c_{2}\xi_{c}}},
\end{align}
\begin{align}
\langle\hat{\Phi}_{1}^{\dagger}(\mathbf{x})\hat{\Phi}_{2}(\mathbf{0})\rangle= & \sqrt{\rho_{10}\rho_{20}}e^{\langle\hat{\theta}_{1}(\mathbf{x})\hat{\theta}_{2}(\mathbf{0})\rangle-\frac{1}{2}\langle\hat{\theta}_{1}^{2}(\mathbf{0})\rangle-\frac{1}{2}\langle\hat{\theta}_{2}^{2}(\mathbf{0})\rangle}=\sqrt{\rho_{10}\rho_{20}}\left(\frac{\left|\mathbf{x}\right|}{\xi_{c}}\right)^{-\frac{1}{2\pi c_{1}}}e^{-\frac{1}{2\pi c_{2}\left|\mathbf{x}\right|}-\frac{1}{2\pi c_{2}\xi_{c}}}.
\end{align}
The four-operator correlation functions are
\begin{align}
& \langle\hat{\Phi}_{1}^{\dagger}(\mathbf{x})\hat{\Phi}_{2}(\mathbf{x})\hat{\Phi}_{1}(\mathbf{0})\hat{\Phi}_{2}^{\dagger}(\mathbf{0})\rangle\nonumber\\
= & \rho_{10}\rho_{20}\exp\bigg(\langle\hat{\theta}_{1}(\mathbf{x})\hat{\theta}_{1}(\mathbf{0})\rangle+\langle\hat{\theta}_{2}(\mathbf{x})\hat{\theta}_{2}(\mathbf{0})\rangle+2\langle\hat{\theta}_{1}(\mathbf{0})\hat{\theta}_{2}(\mathbf{0})\rangle-\langle\hat{\theta}_{1}^{2}(\mathbf{0})\rangle-\langle\hat{\theta}_{2}^{2}(\mathbf{0})\rangle-2\langle\hat{\theta}_{1}(\mathbf{x})\hat{\theta}_{2}(\mathbf{0})\rangle\bigg)\nonumber\\
= & \rho_{10}\rho_{20}e^{\frac{2}{\pi c_{1}\left|\mathbf{x}\right|}-\frac{2}{\pi c_{1}\xi_{c}}},
\end{align}
\begin{align}
    & \langle\hat{\Phi}_{1}^{\dagger}(\mathbf{x})\hat{\Phi}_{2}^\dagger(\mathbf{x})\hat{\Phi}_{1}(\mathbf{0})\hat{\Phi}_{2}(\mathbf{0})\rangle\nonumber\\
    = & \rho_{10}\rho_{20}\exp\bigg(\langle\hat{\theta}_{1}(\mathbf{x})\hat{\theta}_{1}(\mathbf{0})\rangle+\langle\hat{\theta}_{2}(\mathbf{x})\hat{\theta}_{2}(\mathbf{0})\rangle-2\langle\hat{\theta}_{1}(\mathbf{0})\hat{\theta}_{2}(\mathbf{0})\rangle-\langle\hat{\theta}_{1}^{2}(\mathbf{0})\rangle-\langle\hat{\theta}_{2}^{2}(\mathbf{0})\rangle+2\langle\hat{\theta}_{1}(\mathbf{x})\hat{\theta}_{2}(\mathbf{0})\rangle\bigg)\nonumber\\
    =&\rho_{10}\rho_{20}\left(\frac{\left|\mathbf{x}\right|}{\xi_c}\right)^{-\frac{2}{\pi c_1}}.
\end{align}
Since the correlation function $\langle\hat{\Phi}_{1}^{\dagger}(\mathbf{x})\hat{\Phi}_{2}(\mathbf{x})\hat{\Phi}_{1}(\mathbf{0})\hat{\Phi}_{2}^{\dagger}(\mathbf{0})\rangle$ saturates to a constant when $\left|\mathbf{x}\right|\rightarrow\infty$, Model Series A spontaneously breaks the relative charge symmetry $U(1)_{-,C}$ and has a true ODLRO in two spatial dimensions. 

\subsection{Three spatial dimensions}

In three spatial dimensions, we can plug in the results of the above
approximation. The correlation functions of the operator $\hat{\theta}$
are 
\begin{align}
\langle\hat{\theta}_{1}(\mathbf{x})\hat{\theta}_{1}(\mathbf{0})\rangle=\langle\hat{\theta}_{2}(\mathbf{x})\hat{\theta}_{2}(\mathbf{0})\rangle= & \int\frac{d^{3}\mathbf{k}}{(2\pi)^{3}}e^{i\mathbf{k}\cdot\mathbf{x}}\left(\frac{1}{c_{1}\left|\mathbf{k}\right|^{2}}+\frac{1}{c_{2}\left|\mathbf{k}\right|}\right)\nonumber\\
= & \frac{1}{8\pi^{3}c_{1}}\int_{0}^{\pi}d\psi\int_{0}^{2\pi}d\theta\int_{0}^{\infty}dk(\sin\psi e^{ik\left|\mathbf{x}\right|\cos\theta\sin\psi})\nonumber\\
 & +\frac{1}{8\pi^{3}c_{2}}\int_{0}^{\pi}d\psi\int_{0}^{2\pi}d\theta\int_{0}^{\frac{2\pi}{\xi_{c}}}dk(k\sin\psi e^{ik\left|\mathbf{x}\right|\cos\theta\sin\psi}).
\end{align}
We need to compute these integrals separately,
\begin{align}
\int_{0}^{\pi}d\psi\int_{0}^{2\pi}d\theta\int_{0}^{\infty}dk(\sin\psi e^{ik\left|\mathbf{x}\right|\cos\theta\sin\psi})=\int_{0}^{\infty}dk\frac{4\pi\sin(k\left|\mathbf{x}\right|)}{k\left|\mathbf{x}\right|}=\frac{2\pi^{2}}{\left|\mathbf{x}\right|},
\end{align}
\begin{align}
\int_{0}^{\pi}d\psi\int_{0}^{2\pi}d\theta\int_{0}^{\frac{2\pi}{\xi_{c}}}dk(k\sin\psi e^{ik\left|\mathbf{x}\right|\cos\theta\sin\psi})=\int_{0}^{\frac{2\pi}{\xi_{c}}}dk\frac{4\pi k\sin(k\left|\mathbf{x}\right|)}{k\left|\mathbf{x}\right|}=\frac{8\pi\sin^{2}(\frac{\pi\left|\mathbf{x}\right|}{\xi_{c}})}{\left|\mathbf{x}\right|^{2}}.
\end{align}
Then, we can get the correlation functions,
\begin{align}
\langle\hat{\theta}_{1}(\mathbf{x})\hat{\theta}_{1}(\mathbf{0})\rangle=\langle\hat{\theta}_{2}(\mathbf{x})\hat{\theta}_{2}(\mathbf{0})\rangle= & \frac{1}{4\pi c_{1}\left|\mathbf{x}\right|}+\frac{\sin^{2}(\frac{\pi\left|\mathbf{x}\right|}{\xi_{c}})}{\pi^{2}c_{2}\left|\mathbf{x}\right|^{2}}.
\end{align}
Similarly, we have 
\begin{align}
\langle\hat{\theta}_{1}(\mathbf{x})\hat{\theta}_{2}(\mathbf{0})\rangle=\frac{1}{4\pi c_{1}\left|\mathbf{x}\right|}-\frac{\sin^{2}(\frac{\pi\left|\mathbf{x}\right|}{\xi_{c}})}{\pi^{2}c_{2}\left|\mathbf{x}\right|^{2}}.
\end{align}
Since the coherence length is introduced by the previous integration,
we can't just set $\left|\mathbf{x}\right|=\xi_{c}$ to get $\langle\hat{\theta}_{a}^{2}(\mathbf{0})\rangle$.
We need to calculate the integral,
\begin{align}
\int\frac{d^{3}\mathbf{k}}{(2\pi)^{3}}\frac{1}{c_{2}\left|\mathbf{k}\right|}=\frac{1}{8\pi^{3}c_{2}}\int_{0}^{\frac{2\pi}{\xi_{c}}}dk(4\pi k^{2}\frac{1}{k})=\frac{1}{2\pi^{2}c_{2}}\int_{0}^{\frac{2\pi}{\xi_{c}}}kdk=\frac{1}{c_{2}\xi_{c}^{2}}.
\end{align}
Then, we can get 
\begin{align}
\langle\hat{\theta}_{1}^{2}(\mathbf{0})\rangle=\langle\hat{\theta}_{2}^{2}(\mathbf{0})\rangle= & \frac{1}{4\pi c_{1}\xi_{c}}+\frac{1}{c_{2}\xi_{c}^{2}},
\end{align}
\begin{align}
\langle\hat{\theta}_{1}(\mathbf{0})\hat{\theta}_{2}(\mathbf{0})\rangle=\frac{1}{4\pi c_{1}\xi_{c}}-\frac{1}{c_{2}\xi_{c}^{2}}.
\end{align}
Then, we can calculate correlation functions of $\hat{\Phi}$. The two-operator correlation functions are 
\begin{align}
\langle\hat{\Phi}_{1}^{\dagger}(\mathbf{x})\hat{\Phi}_{1}(\mathbf{0})\rangle= & \rho_{10}e^{\langle\hat{\theta}_{1}(\mathbf{x})\hat{\theta}_{1}(\mathbf{0})\rangle-\langle\hat{\theta}_{1}^{2}(\mathbf{0})\rangle}\overset{\left|\mathbf{x}\right|\rightarrow\infty}{=}\rho_{10}e^{-\frac{1}{4\pi c_{1}\xi_{c}}-\frac{1}{c_{2}\xi_{c}^{2}}},
\end{align}
\begin{align}
\langle\hat{\Phi}_{2}^{\dagger}(\mathbf{x})\hat{\Phi}_{2}(\mathbf{0})\rangle= & \rho_{20}e^{\langle\hat{\theta}_{2}(\mathbf{x})\hat{\theta}_{2}(\mathbf{0})\rangle-\langle\hat{\theta}_{2}^{2}(\mathbf{0})\rangle}\overset{\left|\mathbf{x}\right|\rightarrow\infty}{=}\rho_{20}e^{-\frac{1}{4\pi c_{1}\xi_{c}}-\frac{1}{c_{2}\xi_{c}^{2}}},
\end{align}
\begin{align}
\langle\hat{\Phi}_{1}^{\dagger}(\mathbf{x})\hat{\Phi}_{2}(\mathbf{0})\rangle= & \sqrt{\rho_{10}\rho_{20}}e^{\langle\hat{\theta}_{1}(\mathbf{x})\hat{\theta}_{2}(\mathbf{0})\rangle-\frac{1}{2}\langle\hat{\theta}_{1}^{2}(\mathbf{0})\rangle-\frac{1}{2}\langle\hat{\theta}_{2}^{2}(\mathbf{0})\rangle}\overset{\left|\mathbf{x}\right|\rightarrow\infty}{=}\sqrt{\rho_{10}\rho_{20}}e^{-\frac{1}{4\pi c_{1}\xi_{c}}-\frac{1}{c_{2}\xi_{c}^{2}}}.
\end{align}
The four-operator correlation functions are
\begin{align}
& \langle\hat{\Phi}_{1}^{\dagger}(\mathbf{x})\hat{\Phi}_{2}(\mathbf{x})\hat{\Phi}_{1}(\mathbf{0})\hat{\Phi}_{2}^{\dagger}(\mathbf{0})\rangle\nonumber\\
= & \rho_{10}\rho_{20}\exp\bigg(\langle\hat{\theta}_{1}(\mathbf{x})\hat{\theta}_{1}(\mathbf{0})\rangle+\langle\hat{\theta}_{2}(\mathbf{x})\hat{\theta}_{2}(\mathbf{0})\rangle+2\langle\hat{\theta}_{1}(\mathbf{0})\hat{\theta}_{2}(\mathbf{0})\rangle-\langle\hat{\theta}_{1}^{2}(\mathbf{0})\rangle-\langle\hat{\theta}_{2}^{2}(\mathbf{0})\rangle-2\langle\hat{\theta}_{1}(\mathbf{x})\hat{\theta}_{2}(\mathbf{0})\rangle\bigg)\nonumber\\
\overset{\left|\mathbf{x}\right|\rightarrow\infty}{=} & \rho_{10}\rho_{20}e^{-\frac{4}{c_{2}\xi_{c}^{2}}},
\end{align}
\begin{align}
    & \langle\hat{\Phi}_{1}^{\dagger}(\mathbf{x})\hat{\Phi}_{2}^\dagger(\mathbf{x})\hat{\Phi}_{1}(\mathbf{0})\hat{\Phi}_{2}(\mathbf{0})\rangle\nonumber\\
    = & \rho_{10}\rho_{20}\exp\bigg(\langle\hat{\theta}_{1}(\mathbf{x})\hat{\theta}_{1}(\mathbf{0})\rangle+\langle\hat{\theta}_{2}(\mathbf{x})\hat{\theta}_{2}(\mathbf{0})\rangle-2\langle\hat{\theta}_{1}(\mathbf{0})\hat{\theta}_{2}(\mathbf{0})\rangle-\langle\hat{\theta}_{1}^{2}(\mathbf{0})\rangle-\langle\hat{\theta}_{2}^{2}(\mathbf{0})\rangle+2\langle\hat{\theta}_{1}(\mathbf{x})\hat{\theta}_{2}(\mathbf{0})\rangle\bigg)\nonumber\\
    \overset{\left|\mathbf{x}\right|\rightarrow\infty}{=}&\rho_{10}\rho_{20}e^{-\frac{1}{\pi c_1\xi_c}}.
\end{align}
Since all the correlation functions above saturate to constants when $\left|\mathbf{x}\right|\rightarrow\infty$, Model Series A breaks the charge symmetry and has a true ODLRO in three spatial dimensions. 
\section{Derivations of Goldstone modes and dispersion relations for Model Series B}
\label{detailM2}
In this appendix, we give the detailed derivation of Goldstone mode and dispersion relation for Model Series B in Sec. \ref{mainM2}. 
\subsection{Derivation of the effective Hamiltonian density}
Considering quantum fluctuations, we have $\hat{\Phi}_{a}(\mathbf{x})=\sqrt{\rho_{a0}+\hat{f}_{a}(\mathbf{x})}e^{i\hat{\theta}_{a}(\mathbf{x})}$, where $\hat{f}_{a}(\mathbf{x})\ll\rho_{a0}$.
The commutation relation between $\hat{\theta}_a$ and $\hat{f}_a$ is given by $[\hat{\theta}_a(\mathbf{x}),\hat{f}_b(\mathbf{y})]=-i\delta^{(d)}(\mathbf{x}-\mathbf{y})\delta_{ab}$. Therefore, the conjugate momentum operator of $\hat{\theta}_a$ is $\hat{\pi}_a=-\hat{f}_a$, and the operators $\hat{\theta}_a$ and $\hat{\pi}_a$ satisfy the commutation relations, $[\hat{\theta}_{a}(\mathbf{x}),\hat{\pi}_{b}(\mathbf{y})]= i\delta_{ab}\delta^{d}(\mathbf{x}-\mathbf{y})$.
We can derive the effective Hamiltonian,
\begin{align}
  \mathcal{H}= 
    &\frac{g}{2}\hat{\pi}_{1}^{2}+\frac{g}{2}\hat{\pi}_{2}^{2}+\sum_{i,j}^{d}K_{1}^{ij}\left[\rho_{10}^{2}(\partial_{i}\partial_{j}\hat{\theta}_{1})^{2}+\frac{(\partial_{i}\partial_{j}\hat{\pi}_{1})^{2}}{4}\right]\nonumber\\
    &+\sum_{i,j,k}^{d}K_{2}^{ijk}\bigg[\rho_{20}^{3}(\partial_{i}\partial_{j}\partial_{k}\hat{\theta}_{2})^{2}+\frac{1}{4}\rho_{20}(\partial_{i}\partial_{j}\partial_{k}\hat{\pi}_{2})^{2}+\frac{1}{2}(\partial_{i}\hat{\pi}_{2}\partial_{j}\partial_{k}\hat{\pi}_{2}+\partial_{j}\hat{\pi}_{2}\partial_{i}\partial_{k}\hat{\pi}_{2}+\partial_{k}\hat{\pi}_{2}\partial_{i}\partial_{j}\hat{\pi}_{2})\partial_{i}\partial_{j}\partial_{k}\hat{\pi}_{2}\bigg]\nonumber\\
    &+\sum_{i}^{d}\Gamma_{1,2}^{i}\bigg[4\rho_{10}\rho_{20}^{2}(\partial_{i}\hat{\theta}_{1})^{2}-4\ell\rho_{10}\rho_{20}^{2}\partial_{i}\hat{\theta}_{1}\partial_{i}^{2}\hat{\theta}_{2}+\ell^{2}\rho_{10}\rho_{20}^{2}(\partial_{i}^{2}\hat{\theta}_{2})^{2}+\frac{\rho_{20}^{2}}{\rho_{10}}(\partial_{i}\hat{\rho}_{1})^{2}-\ell\partial_{i}\hat{\pi}_{1}(\partial_{i}\hat{\rho}_{2})^{2}\nonumber\\
    &-\ell\rho_{20}\partial_{i}\hat{\pi}_{1}\partial_{i}^{2}\hat{\pi}_{2}+\frac{\ell^{2}\rho_{10}}{4\rho_{20}^{2}}(\partial_{i}\hat{\pi}_{2})^{4}+\frac{\ell^{2}\rho_{10}}{2\rho_{20}}(\partial_{i}\hat{\pi}_{2})^{2}\partial_{i}^{2}\hat{\pi}_{2}+\frac{1}{4}\ell^{2}\rho_{10}(\partial_{i}^{2}\hat{\pi}_{2})^{2}\bigg].\label{effHM2}
  \end{align}
We can obtain the time evolution of $\hat{\theta}_a$,
\begin{align}
  \dot{\hat{\theta}}_{1}= & g\hat{\pi}_{1}+\sum_{i,j}^{d}\frac{1}{2}K_{1}^{ij}\partial_{i}^{2}\partial_{j}^{2}\hat{\pi}_{1}+\sum_{i}^{d}-2\Gamma_{1,2}^{i}\frac{\rho_{20}^{2}}{\rho_{10}}\partial_{i}^{2}\hat{\pi}_{1}+2\Gamma_{1,2}^{i}\ell\partial_{i}\hat{\pi}_{2}\partial_{i}^{2}\hat{\pi}_{2}+\Gamma_{1,2}^{i}\ell\rho_{20}\partial_{i}^{3}\hat{\pi}_{2}\\
  \dot{\hat{\theta}}_{2}= & g\hat{\pi}_{2}+\sum_{i,j,k}^{d}\frac{1}{2}K_{2}^{ijk}\bigg[-\rho_{20}\partial_{i}^{2}\partial_{j}^{2}\partial_{k}^{2}\hat{\pi}_{2}-3(\partial_{i}\partial_{j}\partial_{k}\hat{\pi}_{2})^{2}-2\partial_{i}\partial_{j}\hat{\pi}_{2}\partial_{i}\partial_{j}\partial_{k}^{2}\hat{\pi}_{2}-2\partial_{i}\partial_{k}^{2}\hat{\pi}_{2}\partial_{i}\partial_{j}^{2}\hat{\pi}_{2}-2\partial_{i}\partial_{k}\hat{\pi}_{2}\partial_{i}\partial_{j}^{2}\partial_{k}\hat{\pi}_{2}\nonumber\\
  & -2\partial_{j}^{2}\partial_{k}\hat{\pi}_{2}\partial_{i}^{2}\partial_{k}\hat{\pi}_{2}-2\partial_{j}\partial_{k}^{2}\hat{\pi}_{2}\partial_{i}^{2}\partial_{j}\hat{\pi}_{2}-2\partial_{j}\partial_{k}\hat{\pi}_{2}\partial_{i}^{2}\partial_{j}\partial_{k}\hat{\pi}_{2}-\partial_{i}^{2}\hat{\pi}_{2}\partial_{j}^{2}\partial_{k}^{2}\hat{\pi}_{2}-\partial_{j}^{2}\hat{\pi}_{2}\partial_{i}^{2}\partial_{k}^{2}\hat{\pi}_{2}-\partial_{k}^{2}\hat{\pi}_{2}\partial_{i}^{2}\partial_{j}^{2}\hat{\pi}_{2}\bigg]\nonumber\\
  & +\sum_{i}^{d}\Gamma_{1,2}^{i}\bigg[2\ell\partial_{i}^{2}\hat{\pi}_{1}\partial_{i}\hat{\pi}_{2}+2\ell\partial_{i}\hat{\pi}_{1}\partial_{i}^{2}\hat{\pi}_{2}-\ell\rho_{20}\partial_{i}^{3}\hat{\pi}_{1}-\frac{3\ell^{2}\rho_{10}}{\rho_{20}^{2}}(\partial_{i}\hat{\pi}_{2})^{2}\partial_{i}^{2}\hat{\pi}_{2}-\frac{1}{2}\ell^{2}\rho_{10}\partial_{i}^{4}\hat{\pi}_{2}\bigg].
\end{align}
In the long-wave length limit, we have
\begin{align}
  g\hat{\pi}_{1}\gg & \sum_{i,j}^{d}\frac{1}{2}K_{1}^{ij}\partial_{i}^{2}\partial_{j}^{2}\hat{\pi}_{1}+\sum_{i}^{d}-2\Gamma_{1,2}^{i}\frac{\rho_{20}^{2}}{\rho_{10}}\partial_{i}^{2}\hat{\pi}_{1}+2\Gamma_{1,2}^{i}\ell\partial_{i}\hat{\pi}_{2}\partial_{i}^{2}\hat{\pi}_{2}+\Gamma_{1,2}^{i}\ell\rho_{20}\partial_{i}^{3}\hat{\pi}_{2}\\
  g\hat{\pi}_{2}\gg & \sum_{i,j,k}^{d}\frac{1}{2}K_{2}^{ijk}\bigg[-\rho_{20}\partial_{i}^{2}\partial_{j}^{2}\partial_{k}^{2}\hat{\pi}_{2}-3(\partial_{i}\partial_{j}\partial_{k}\hat{\pi}_{2})^{2}-2\partial_{i}\partial_{j}\hat{\pi}_{2}\partial_{i}\partial_{j}\partial_{k}^{2}\hat{\pi}_{2}-2\partial_{i}\partial_{k}^{2}\hat{\pi}_{2}\partial_{i}\partial_{j}^{2}\hat{\pi}_{2}-2\partial_{i}\partial_{k}\hat{\pi}_{2}\partial_{i}\partial_{j}^{2}\partial_{k}\hat{\pi}_{2}\nonumber\\
  & -2\partial_{j}^{2}\partial_{k}\hat{\pi}_{2}\partial_{i}^{2}\partial_{k}\hat{\pi}_{2}-2\partial_{j}\partial_{k}^{2}\hat{\pi}_{2}\partial_{i}^{2}\partial_{j}\hat{\pi}_{2}-2\partial_{j}\partial_{k}\hat{\pi}_{2}\partial_{i}^{2}\partial_{j}\partial_{k}\hat{\pi}_{2}-\partial_{i}^{2}\hat{\pi}_{2}\partial_{j}^{2}\partial_{k}^{2}\hat{\pi}_{2}-\partial_{j}^{2}\hat{\pi}_{2}\partial_{i}^{2}\partial_{k}^{2}\hat{\pi}_{2}-\partial_{k}^{2}\hat{\pi}_{2}\partial_{i}^{2}\partial_{j}^{2}\hat{\pi}_{2}\bigg]\nonumber\\
  & +\sum_{i}^{d}\Gamma_{1,2}^{i}\bigg[2\ell\partial_{i}^{2}\hat{\pi}_{1}\partial_{i}\hat{\pi}_{2}+2\ell\partial_{i}\hat{\pi}_{1}\partial_{i}^{2}\hat{\pi}_{2}-\ell\rho_{20}\partial_{i}^{3}\hat{\pi}_{1}-\frac{3l^{2}\rho_{10}}{\rho_{20}^{2}}(\partial_{i}\hat{\pi}_{2})^{2}\partial_{i}^{2}\hat{\pi}_{2}-\frac{1}{2}\ell^{2}\rho_{10}\partial_{i}^{4}\hat{\pi}_{2}\bigg].
\label{lwll}
\end{align}
Here we just need to consider the lowest order and get 
\begin{align}
g\gg\sum_{i}^{d}2\Gamma_{1,2}^{i}\frac{\rho_{20}^{2}}{\rho_{10}}k_{i}^{2}.\label{app1}
\end{align}
If we set an isotropic $K_{1}^{ij}=\frac{\kappa_{1}}{2}$, $K_{2}^{ijk}=\frac{\kappa_{2}}{2}$, and $\Gamma_{1,2}^{i}=\frac{\Gamma}{2}$, we can get the coherence length $\xi_{c}=2\pi\sqrt{\frac{\Gamma\rho_{20}^{2}}{g\rho_{10}}}$.
Then, we can drop the higher-order terms in Eq.~(\ref{effHM2}), and get the effective Hamiltonian,
\begin{align}
  \mathcal{H}= & \frac{g}{2}\hat{\pi}_1^2+\frac{g}{2}\hat{\pi}_2^2+\frac{\kappa_1\rho_{10}^{2}}{2}(\partial_{i}\partial_{j}\hat{\theta}_{1})^{2}+\frac{\kappa_2\rho_{20}^{3}}{2}(\partial_{i}\partial_{j}\partial_{k}\hat{\theta}_{2})^{2}+\frac{\Gamma\rho_{10}\rho_{20}^{2}}{2}(\ell\partial_{i}^{2}\hat{\theta}_{2}-2\partial_{i}\hat{\theta}_{1})^{2}.
\end{align}
\subsection{Derivation of the Hamiltonian in the momentum space}
In the momentum space, we can perform a Fourier transform,
\begin{align}
  H= & \frac{1}{(2\pi)^{2d}}\int d^{d}\mathbf{x}d^{d}\mathbf{k}d^{d}\mathbf{k}^{\prime}e^{i\mathbf{k}\cdot\mathbf{x}+i\mathbf{k}^{\prime}\cdot\mathbf{x}}\Bigg[\frac{g}{2}\hat{\pi}_{1} (\mathbf{k}) \hat{\pi}_{1}\ (\mathbf{k}') +\frac{g}{2}\hat{\pi}_{2} (\mathbf{k}) \hat{\pi}_{2}\ (\mathbf{k}') +\frac{\kappa_{1}\rho_{10}^{2}}{2}\sum_{i,j}^{d}k_{i}^{2}k_{j}^{2}\hat{\theta}_{1} (\mathbf{k}) \hat{\theta}_{1}\ (\mathbf{k}') \nonumber\\
  & +\frac{\kappa_{2}\rho_{20}^{3}}{2}\sum_{i,j,k}^{d}k_{i}^{2}k_{j}^{2}k_{k}^{2}\hat{\theta}_{2} (\mathbf{k}) \hat{\theta}_{2}\ (\mathbf{k}') +\frac{\Gamma\rho_{10}\rho_{20}^{2}}{2}\sum_{i}^{d}\bigg[\ell^{2}k_{i}^{4}\hat{\theta}_{2} (\mathbf{k}) \hat{\theta}_{2}\ (\mathbf{k}') +4k_{i}^{2}\hat{\theta}_{1} (\mathbf{k}) \hat{\theta}_{1}\ (\mathbf{k}') +2i\ell k_{i}^{3}\hat{\theta}_{1} (\mathbf{k}) \hat{\theta}_{2}\ (\mathbf{k}') \nonumber\\
  & -2i\ell k_{i}^{3}\hat{\theta}_{2} (\mathbf{k}) \hat{\theta}_{1}\ (\mathbf{k}') \bigg]\Bigg].
\end{align}

Integrating over the real space, we can obtain a $\delta$ function so that only the integral of the momentum space is left in the Hamiltonian,
\begin{align}
  H= & \frac{1}{(2\pi)^{d}}\int d^{d}\mathbf{k}d^{d}\mathbf{k}^{\prime}\delta(\mathbf{k}+\mathbf{k}^{\prime})\Bigg[\frac{g}{2}\hat{\pi}_{1} (\mathbf{k}) \hat{\pi}_{1}\ (\mathbf{k}') +\frac{g}{2}\hat{\pi}_{2} (\mathbf{k}) \hat{\pi}_{2}\ (\mathbf{k}') +\frac{\kappa_{1}\rho_{10}^{2}\left|\mathbf{k}\right|^{4}}{2}\hat{\theta}_{1} (\mathbf{k}) \hat{\theta}_{1}\ (\mathbf{k}') +\frac{\kappa_{2}\rho_{20}^{3}\left|\mathbf{k}\right|^{6}}{2}\hat{\theta}_{2} (\mathbf{k}) \hat{\theta}_{2}\ (\mathbf{k}') \nonumber\\
  & +\frac{\Gamma\rho_{10}\rho_{20}^{2}}{2}\sum_{i}^{d}\bigg[\ell^{2}k_{i}^{4}\hat{\theta}_{2} (\mathbf{k}) \hat{\theta}_{2}\ (\mathbf{k}') +4k_{i}^{2}\hat{\theta}_{1} (\mathbf{k}) \hat{\theta}_{1}\ (\mathbf{k}') +2i\ell k_{i}^{3}\hat{\theta}_{1} (\mathbf{k}) \hat{\theta}_{2}\ (\mathbf{k}') -2i\ell k_{i}^{3}\hat{\theta}_{2} (\mathbf{k}) \hat{\theta}_{1}\ (\mathbf{k}') \bigg]\Bigg]\nonumber\\
 = & \frac{1}{(2\pi)^{d}}\int d^{d}\mathbf{k}\Bigg[\frac{g}{2}\hat{\pi}_{1} (\mathbf{k}) \hat{\pi}_{1} (-\mathbf{k}) +\frac{g}{2}\hat{\pi}_{2} (\mathbf{k}) \hat{\pi}_{2} (-\mathbf{k}) +\frac{\kappa_{1}\rho_{10}^{2}\left|\mathbf{k}\right|^{4}}{2}\hat{\theta}_{1} (\mathbf{k}) \hat{\theta}_{1} (-\mathbf{k}) +\frac{\kappa_{2}\rho_{20}^{3}\left|\mathbf{k}\right|^{6}}{2}\hat{\theta}_{2} (\mathbf{k}) \hat{\theta}_{2} (-\mathbf{k}) \nonumber\\
  & +\frac{\Gamma\rho_{10}\rho_{20}^{2}}{2}\sum_{i}^{d}\bigg[\ell^{2}k_{i}^{4}\hat{\theta}_{2} (\mathbf{k}) \hat{\theta}_{2} (-\mathbf{k}) +4k_{i}^{2}\hat{\theta}_{1} (\mathbf{k}) \hat{\theta}_{1} (-\mathbf{k}) +2i\ell k_{i}^{3}\hat{\theta}_{1} (\mathbf{k}) \hat{\theta}_{2} (-\mathbf{k}) -2i\ell k_{i}^{3}\hat{\theta}_{2} (\mathbf{k}) \hat{\theta}_{1} (-\mathbf{k}) \bigg]\Bigg].
\end{align}
Using the same method we used to compute Model 1, we can write the Hamiltonian in matrix form,
\begin{align}
  H= & \frac{1}{2(2\pi)^{d}}\int d^{d}\mathbf{k}\Bigg[g\hat{\pi}_{1} (\mathbf{k}) \hat{\pi}_{1} (-\mathbf{k}) +g\hat{\pi}_{2} (\mathbf{k}) \hat{\pi}_{2} (-\mathbf{k}) \nonumber\\
   & +\begin{bmatrix}\hat{\theta}_{1} (-\mathbf{k})  & \hat{\theta}_{2} (-\mathbf{k}) \end{bmatrix}\begin{bmatrix}\kappa_{1}\rho_{10}^{2}\left|\mathbf{k}\right|^{4}+4\Gamma\rho_{10}\rho_{20}^{2}\left|\mathbf{k}\right|^{2} & -2i\Gamma\rho_{10}\rho_{20}^{2}\ell\sum_{i}^{d}k_{i}^{3}\\
  2i\Gamma\rho_{10}\rho_{20}^{2}\ell\sum_{i}^{d}k_{i}^{3} & \kappa_{2}\rho_{20}^{3}\left|\mathbf{k}\right|^{6}+\Gamma\rho_{10}\rho_{20}^{2}\ell^{2}\sum_{i}^{d}k_{i}^{4}
  \end{bmatrix}\begin{bmatrix}\hat{\theta}_{1} (\mathbf{k}) \\
  \hat{\theta}_{2} (\mathbf{k}) 
  \end{bmatrix}\Bigg]\nonumber\\
  = & \frac{1}{2(2\pi)^{d}}\int d^{d}\mathbf{k}\psi^{\dagger} (\mathbf{k}) \begin{bmatrix}g & 0 & 0 & 0\\
  0 & g & 0 & 0\\
  0 & 0 & \kappa_{1}\rho_{10}^{2}\left|\mathbf{k}\right|^{4}+4\Gamma\rho_{10}\rho_{20}^{2}\left|\mathbf{k}\right| & -2i\Gamma\rho_{10}\rho_{20}^{2}\ell\sum_{i}^{d}k_{i}^{3}\\
  0 & 0 & 2i\Gamma\rho_{10}\rho_{20}^{2}\ell\sum_{i}^{d}k_{i}^{3} & \kappa_{2}\rho_{20}^{3}\left|\mathbf{k}\right|^{6}+\Gamma\rho_{10}\rho_{20}^{2}\ell^{2}\sum_{i}^{d}k_{i}^{4}
  \end{bmatrix}\psi (\mathbf{k}) \nonumber\\
  = & \frac{1}{2(2\pi)^{d}}\int d^{d}\mathbf{k}\psi^{\dagger} (\mathbf{k}) M (\mathbf{k}) \psi (\mathbf{k}) .
  \label{Hmdqmom}
  \end{align}
In the momentum space, the operators $\hat{\theta}_a (\mathbf{k}) $ and the conjugate momentum densities $\hat{\pi}_b(\mathbf{k}')$ satisfy the commutation relations, $\left[\hat{\theta}_{a} (\mathbf{k}) ,\hat{\pi}_{b}\ (\mathbf{k}') \right]= i(2\pi)^{d}\delta_{ab}\delta^{(d)}(\mathbf{k}+\mathbf{k}^{\prime})$.
\subsection{Bogoliubov transformation and the transformation matrix $T (\mathbf{k}) $}
In this Hamiltonian, two species of bosons couple to each other. By the Bogoliubov transformation, we can convert the coupling bosons into two independent modes. 
We introduce a transformation matrix $T (\mathbf{k}) $ which changes the basis from the basis $\psi (\mathbf{k}) $ into the basis $\phi (\mathbf{k}) $,
\begin{align}
\phi (\mathbf{k}) =\begin{bmatrix}\hat{\alpha} (\mathbf{k})  & \hat{\beta} (\mathbf{k})  & \hat{\alpha}^{\dagger} (-\mathbf{k})  & \hat{\beta}^{\dagger} (-\mathbf{k}) \end{bmatrix}^{T}=T (\mathbf{k}) \psi (\mathbf{k}) ,
\end{align}
where $\hat{\alpha} (\mathbf{k}) $ and $\hat{\beta} (\mathbf{k}) $ are annihilation operators of two independent modes, and they satisfy the commutation relations. 
\begin{align}
& \left[\hat{\alpha} (\mathbf{k}) ,\hat{\alpha}^{\dagger}\ (\mathbf{k}') \right]=\left[\hat{\beta} (\mathbf{k}) ,\hat{\beta}^{\dagger}\ (\mathbf{k}') \right]=(2\pi)^{d}\delta^{(d)}(\mathbf{k}-\mathbf{k}^{\prime})\\
 & \left[\hat{\alpha} (\mathbf{k}) ,\hat{\alpha}\ (\mathbf{k}') \right]=\left[\hat{\beta} (\mathbf{k}) ,\hat{\beta}\ (\mathbf{k}') \right]=\left[\hat{\alpha} (\mathbf{k}) ,\hat{\beta}\ (\mathbf{k}') \right]=\left[\hat{\alpha} (\mathbf{k}) ,\hat{\beta}^{\dagger}\ (\mathbf{k}') \right]=\left[\hat{\beta} (\mathbf{k}) ,\hat{\alpha}^{\dagger}\ (\mathbf{k}') \right]=0
\end{align}
To keep the commutation relations and diagonalize the matrix $M (\mathbf{k}) $, the transformation matrix $T (\mathbf{k}) $ needs to obey the equation below,
\begin{align}
  T (\mathbf{k}) \Lambda_1T^\dagger (\mathbf{k}) =\Lambda_2\\
\end{align}
where 
\begin{align}
  \Lambda_1=\begin{bmatrix}
    0 & 0 & -i & 0\\
    0 & 0 & 0 & -i\\
    i & 0 & 0 & 0\\
    0 & i & 0 & 0
  \end{bmatrix},\quad
  \Lambda_2=\text{diag}(1,1,-1,-1).
\end{align}
The diagonalized Hamiltonian becomes 
\begin{align}
    H= & \frac{1}{2(2\pi)^{d}}\phi^{\dagger} (\mathbf{k}) (T^{-1} (\mathbf{k}) )^{\dagger}M (\mathbf{k}) T^{-1} (\mathbf{k}) \phi (\mathbf{k}) ,
\end{align}
where the matrix $(T^{-1} (\mathbf{k}) )^{\dagger}M (\mathbf{k}) T^{-1} (\mathbf{k}) =\text{diag}(D_1 (\mathbf{k}) ,D_2 (\mathbf{k}) ,D_3 (\mathbf{k}) ,D_4 (\mathbf{k}) )$ is a diagonal matrix with the diagonal elements:
\begin{align}
  &D_{1} (\mathbf{k}) =D_{3} (\mathbf{k}) = \Bigg\{\frac{g}{2} \bigg(\kappa _1 \left|\mathbf{k}\right|^4 \rho _{10}^2+\kappa _2 \left|\mathbf{k}\right|^6 \rho _{20}^3+4 \Gamma  \left|\mathbf{k}\right|^2 \rho _{10} \rho _{20}^2+\Gamma  \ell^2 \rho _{10} \rho _{20}^2 \sum_i^d k_i^4\nonumber\\
  &+\sqrt{(\kappa _1 \left|\mathbf{k}\right|^4 \rho _{10}^2-\rho _{20}^2 (\kappa _2 \left|\mathbf{k}\right|^6 \rho _{20}+\Gamma  \rho _{10} (\ell^2 \sum_i^d k_i^4-4 \left|\mathbf{k}\right|^2))){}^2+16 \Gamma ^2 \ell^2 \rho _{10}^2 \rho _{20}^4 (\sum_i^d k_i^3)^2}\bigg)\Bigg\}^{\frac{1}{2}}\nonumber\\
  &D_{2} (\mathbf{k}) =D_{4} (\mathbf{k}) = \Bigg\{\frac{g}{2} \bigg(\kappa _1 \left|\mathbf{k}\right|^4 \rho _{10}^2+\kappa _2 \left|\mathbf{k}\right|^6 \rho _{20}^3+4 \Gamma  \left|\mathbf{k}\right|^2 \rho _{10} \rho _{20}^2+\Gamma  \ell^2 \rho _{10} \rho _{20}^2 \sum_i^d k_i^4\nonumber\\
  &-\sqrt{(\kappa _1 \left|\mathbf{k}\right|^4 \rho _{10}^2-\rho _{20}^2 (\kappa _2 \left|\mathbf{k}\right|^6 \rho _{20}+\Gamma  \rho _{10} (\ell^2 \sum_i^d k_i^4-4 \left|\mathbf{k}\right|^2))){}^2+16 \Gamma ^2 \ell^2 \rho _{10}^2 \rho _{20}^4 (\sum_i^d k_i^3)^2}\bigg)\Bigg\}^{\frac{1}{2}}\label{M2dispersion}
\end{align}
The transformation matrix $T (\mathbf{k}) $ is given by 
\begin{align}
    T (\mathbf{k}) =\begin{bmatrix}
      T_{11} (\mathbf{k})  & T_{12} (\mathbf{k})  & T_{13} (\mathbf{k})  & T_{14} (\mathbf{k}) \\
      T_{21} (\mathbf{k})  & T_{22} (\mathbf{k})  & T_{23} (\mathbf{k})  & T_{24} (\mathbf{k}) \\
      T_{31} (\mathbf{k})  & T_{32} (\mathbf{k})  & T_{33} (\mathbf{k})  & T_{34} (\mathbf{k}) \\
      T_{41} (\mathbf{k})  & T_{42} (\mathbf{k})  & T_{43} (\mathbf{k})  & T_{44} (\mathbf{k}) \\
    \end{bmatrix},
\end{align}
where the matrix elements are
\begin{align}
  T_{11} (\mathbf{k}) =&-T_{31} (\mathbf{k}) =-\Bigg\{g \bigg[-\kappa _2 \left|\mathbf{k}\right|^6 \rho _{20}^3+\kappa _1 \left|\mathbf{k}\right|^4 \rho _{10}^2+4 \Gamma  \left|\mathbf{k}\right|^2 \rho _{10} \rho _{20}^2-\Gamma  \ell^2 \rho _{10} \rho _{20}^2 \sum_i^dk_i^4\nonumber\\
  &+\sqrt{(\kappa _1 \left|\mathbf{k}\right|^4 \rho _{10}^2-\rho _{20}^2 (\kappa _2 \left|\mathbf{k}\right|^6 \rho _{20}+\Gamma  \rho _{10} (\ell^2 \sum_i^dk_i^4-4 \left|\mathbf{k}\right|^2))){}^2+16 \Gamma ^2 \ell^2 \rho _{10}^2 \rho _{20}^4 (\sum_i^dk_i^3)^2}\bigg]\Bigg\}/\nonumber\\
  &\Bigg\{2 \sqrt[4]{2} \Gamma  \ell \rho _{10} \rho _{20}^2 \sum_i^dk_i^3 \Bigg[g \Bigg(4+\Bigg(\bigg(-\kappa _2 \left|\mathbf{k}\right|^6 \rho _{20}^3+\kappa _1 \left|\mathbf{k}\right|^4 \rho _{10}^2+4 \Gamma  \left|\mathbf{k}\right|^2 \rho _{10} \rho _{20}^2-\Gamma  \ell^2 \rho _{10} \rho _{20}^2 \sum_i^dk_i^4\nonumber\\
  &+\sqrt{(\kappa _1 \left|\mathbf{k}\right|^4 \rho _{10}^2-\rho _{20}^2 (\kappa _2 \left|\mathbf{k}\right|^6 \rho _{20}+\Gamma  \rho _{10} (\ell^2 \sum_i^dk_i^4-4 k^2))){}^2+16 \Gamma ^2 \ell^2 \rho _{10}^2 \rho _{20}^4 \Big(\sum_i^dk_i^3\Big)^2}\bigg){}^2\Bigg)/\nonumber\\
  &\Big(4 \Gamma ^2 \ell^2 \rho _{10}^2 \rho _{20}^4 \Big(\sum_i^dk_i^3\Big)^2\Big)\Bigg)\bigg(g \Big(\kappa _2 \left|\mathbf{k}\right|^6 \rho _{20}^3+\kappa _1 \left|\mathbf{k}\right|^4 \rho _{10}^2+4 \Gamma  \left|\mathbf{k}\right|^2 \rho _{10} \rho _{20}^2+\Gamma  \ell^2 \rho _{10} \rho _{20}^2 \sum_i^dk_i^4\nonumber\\
  &+\sqrt{(\kappa _1 \left|\mathbf{k}\right|^4 \rho _{10}^2-\rho _{20}^2 (\kappa _2 \left|\mathbf{k}\right|^6 \rho _{20}+\Gamma  \rho _{10} (\ell^2 \sum_i^dk_i^4-4 \left|\mathbf{k}\right|^2))){}^2+16 \Gamma ^2 \ell^2 \rho _{10}^2 \rho _{20}^4 \Big(\sum_i^dk_i^3\Big)^2}\Big)\bigg)^{\frac{1}{2}}\Bigg]^{\frac{1}{2}}\Bigg\},
\end{align}
\begin{align}
  T_{12} (\mathbf{k}) =&-T_{32} (\mathbf{k}) =\Bigg(i 2^{3/4} g\Bigg)/\Bigg\{\Bigg[g \Big(4+\Bigg(\bigg(-\kappa _2 \left|\mathbf{k}\right|^6 \rho _{20}^3+\kappa _1 \left|\mathbf{k}\right|^4 \rho _{10}^2+4 \Gamma  \left|\mathbf{k}\right|^2 \rho _{10} \rho _{20}^2-\Gamma  \ell^2 \rho _{10} \rho _{20}^2 \sum_i^dk_i^4\nonumber\\
  &+\sqrt{(\kappa _1 \left|\mathbf{k}\right|^4 \rho _{10}^2-\rho _{20}^2 (\kappa _2 \left|\mathbf{k}\right|^6 \rho _{20}+\rho _{10} (\Gamma  \ell^2 \sum_i^dk_i^4-4 \Gamma  \left|\mathbf{k}\right|^2))){}^2+16 \Gamma ^2 \ell^2 \rho _{10}^2 \rho _{20}^4 \Big(\sum_i^dk_i^3\Big)^2}\bigg)^2\Bigg)\nonumber\\
  &/{4 \Gamma ^2 \ell^2 \rho _{10}^2 \rho _{20}^4 \Big(\sum_i^dk_i^3\Big)^2}\Big) \Bigg[g \big(\kappa _2 \left|\mathbf{k}\right|^6 \rho _{20}^3+\kappa _1 \left|\mathbf{k}\right|^4 \rho _{10}^2+4 \Gamma  \left|\mathbf{k}\right|^2 \rho _{10} \rho _{20}^2+\Gamma  \ell^2 \rho _{10} \rho _{20}^2 \sum_i^dk_i^4\nonumber\\
  &+\sqrt{(\kappa _1 \left|\mathbf{k}\right|^4 \rho _{10}^2-\rho _{20}^2 (\kappa _2 \left|\mathbf{k}\right|^6 \rho _{20}+\rho _{10} (\Gamma  \ell^2 \sum_i^dk_i^4-4 \Gamma  \left|\mathbf{k}\right|^2))){}^2+16 \Gamma ^2 \ell^2 \rho _{10}^2 \rho _{20}^4 \Big(\sum_i^dk_i^3\Big)^2}\big)\Bigg]^{\frac{1}{2}}\Bigg]^{\frac{1}{2}}\Bigg\},
\end{align}
\begin{align}
  T_{13} (\mathbf{k}) =&T_{33} (\mathbf{k}) =\Bigg\{i \Bigg[-\kappa _2 \left|\mathbf{k}\right|^6 \rho _{20}^3+\kappa _1 \left|\mathbf{k}\right|^4 \rho _{10}^2+4 \Gamma  \left|\mathbf{k}\right|^2 \rho _{10} \rho _{20}^2-\Gamma  \ell^2 \rho _{10} \rho _{20}^2 \sum_i^dk_i^4\nonumber\\
  &+\sqrt{(\kappa _1 \left|\mathbf{k}\right|^4 \rho _{10}^2-\rho _{20}^2 (\kappa _2 \left|\mathbf{k}\right|^6 \rho _{20}+\rho _{10} (\Gamma  \ell^2 \sum_i^dk_i^4-4 \Gamma  \left|\mathbf{k}\right|^2))){}^2+16 \Gamma ^2 \ell^2 \rho _{10}^2 \rho _{20}^4 \Big(\sum_i^dk_i^3\Big)^2}\Bigg]\nonumber\\
  &\Bigg(g \bigg(\kappa _2 \left|\mathbf{k}\right|^6 \rho _{20}^3+\kappa _1 \left|\mathbf{k}\right|^4 \rho _{10}^2+4 \Gamma  \left|\mathbf{k}\right|^2 \rho _{10} \rho _{20}^2+\Gamma  \ell^2 \rho _{10} \rho _{20}^2 \sum_i^dk_i^4\nonumber\\
  &+\sqrt{(\kappa _1 \left|\mathbf{k}\right|^4 \rho _{10}^2-\rho _{20}^2 (\kappa _2 \left|\mathbf{k}\right|^6 \rho _{20}+\rho _{10} (\Gamma  \ell^2 \sum_i^dk_i^4-4 \Gamma  \left|\mathbf{k}\right|^2))){}^2+16 \Gamma ^2 \ell^2 \rho _{10}^2 \rho _{20}^4 \Big(\sum_i^dk_i^3\Big)^2}\bigg)\Bigg)^{\frac{1}{4}}\Bigg\}/\nonumber\\
  &\Bigg\{2\ 2^{3/4} \Gamma  \ell \rho _{10} \rho _{20}^2 \sum_i^dk_i^3 \Bigg[g \Bigg(\bigg(\Big(-\kappa _2 \left|\mathbf{k}\right|^6 \rho _{20}^3+\kappa _1 \left|\mathbf{k}\right|^4 \rho _{10}^2+4 \Gamma  \left|\mathbf{k}\right|^2 \rho _{10} \rho _{20}^2-\Gamma  \ell^2 \rho _{10} \rho _{20}^2 \sum_i^dk_i^4\nonumber\\
  &+\sqrt{(\kappa _1 \left|\mathbf{k}\right|^4 \rho _{10}^2-\rho _{20}^2 (\kappa _2 \left|\mathbf{k}\right|^6 \rho _{20}+\rho _{10} (\Gamma  \ell^2 \sum_i^dk_i^4-4 \Gamma  \left|\mathbf{k}\right|^2))){}^2+16 \Gamma ^2 \ell^2 \rho _{10}^2 \rho _{20}^4 \Big(\sum_i^dk_i^3\Big)^2}\Big)^2\bigg)/\nonumber\\
  &\bigg(4 \Gamma ^2 \ell^2 \rho _{10}^2 \rho _{20}^4 \Big(\sum_i^dk_i^3\Big)^2\bigg)+4\Bigg)\Bigg]^{\frac{1}{2}}\Bigg\},
\end{align}
\begin{align}
  T_{14} (\mathbf{k}) =&T_{34} (\mathbf{k}) =\Bigg\{\sqrt[4]{2} \Bigg[g \Bigg(\kappa _2 \left|\mathbf{k}\right|^6 \rho _{20}^3+\kappa _1 \left|\mathbf{k}\right|^4 \rho _{10}^2+4 \Gamma  k^2 \rho _{10} \rho _{20}^2+\Gamma  \ell^2 \rho _{10} \rho _{20}^2 \sum_i^dk_i^4\nonumber\\
  &+\sqrt{(\kappa _1 \left|\mathbf{k}\right|^4 \rho _{10}^2-\rho _{20}^2 (\kappa _2 \left|\mathbf{k}\right|^6 \rho _{20}+\rho _{10} (\Gamma  \ell^2 \sum_i^dk_i^4-4 \Gamma  \left|\mathbf{k}\right|^2))){}^2+16 \Gamma ^2 \ell^2 \rho _{10}^2 \rho _{20}^4 \Big(\sum_i^dk_i^3\Big)^2}\Bigg)\Bigg]^{\frac{1}{4}}\Bigg\}/\nonumber\\
  &\Bigg\{\Bigg[g \Bigg(\bigg(\Big(-\kappa _2 \left|\mathbf{k}\right|^6 \rho _{20}^3+\kappa _1 \left|\mathbf{k}\right|^4 \rho _{10}^2+4 \Gamma  \left|\mathbf{k}\right|^2 \rho _{10} \rho _{20}^2-\Gamma  \ell^2 \rho _{10} \rho _{20}^2 \sum_i^dk_i^4\nonumber\\
  &+\sqrt{(\kappa _1 \left|\mathbf{k}\right|^4 \rho _{10}^2-\rho _{20}^2 (\kappa _2 \left|\mathbf{k}\right|^6 \rho _{20}+\rho _{10} (\Gamma  \ell^2 \sum_i^dk_i^4-4 \Gamma  \left|\mathbf{k}\right|^2))){}^2+16 \Gamma ^2 \ell^2 \rho _{10}^2 \rho _{20}^4 \Big(\sum_i^dk_i^3\Big)^2}\Big){}^2\bigg)/\nonumber\\
  &\bigg(4 \Gamma ^2 \ell^2 \rho _{10}^2 \rho _{20}^4 \Big(\sum_i^dk_i^3\Big)^2\bigg)+4\Bigg)\Bigg]^{\frac{1}{2}}\Bigg\},
\end{align}
\begin{align}
  T_{21} (\mathbf{k}) =&-T_{41} (\mathbf{k}) =2^{-\frac{3}{4}}\text{sgn}(\sum_i^dk_i^3)\Bigg\{\Bigg[g \Bigg(\bigg(\kappa _2 \left|\mathbf{k}\right|^6 \rho _{20}^3-\kappa _1 \left|\mathbf{k}\right|^4 \rho _{10}^2-4 \Gamma  \left|\mathbf{k}\right|^2 \rho _{10} \rho _{20}^2+\Gamma  \ell^2 \rho _{10} \rho _{20}^2 \sum_i^dk_i^4\nonumber\\
  &+\sqrt{(\kappa _1 k^4 \rho _{10}^2-\rho _{20}^2 (\kappa _2 \left|\mathbf{k}\right|^6 \rho _{20}+\rho _{10} (\Gamma  \ell^2 \sum_i^dk_i^4-4 \Gamma  \left|\mathbf{k}\right|^2))){}^2+16 \Gamma ^2 \ell^2 \rho _{10}^2 \rho _{20}^4 \Big(\sum_i^dk_i^3\Big)^2}\bigg)\nonumber\\
  &(\rho _{20}^2 (\kappa _2 \left|\mathbf{k}\right|^6 \rho _{20}+\rho _{10} (\Gamma  \ell^2 \sum_i^dk_i^4-4 \Gamma  \left|\mathbf{k}\right|^2))-\kappa _1 \left|\mathbf{k}\right|^4 \rho _{10}^2)+16 \Gamma ^2 \ell^2 \rho _{10}^2 \rho _{20}^4 \Big(\sum_i^dk_i^3\Big)^2\Bigg)\Bigg]/\nonumber\\
  &\Bigg[((\kappa _1 \left|\mathbf{k}\right|^4 \rho _{10}^2-\rho _{20}^2 (\kappa _2 \left|\mathbf{k}\right|^6 \rho _{20}+\rho _{10} (\Gamma  \ell^2 \sum_i^dk_i^4-4 \Gamma  \left|\mathbf{k}\right|^2))){}^2+16 \Gamma ^2 \ell^2 \rho _{10}^2 \rho _{20}^4 \Big(\sum_i^dk_i^3\Big)^2)\nonumber\\
  &\Bigg(g \bigg(\kappa _2 \left|\mathbf{k}\right|^6 \rho _{20}^3+\kappa _1 \left|\mathbf{k}\right|^4 \rho _{10}^2+4 \Gamma  \left|\mathbf{k}\right|^2 \rho _{10} \rho _{20}^2+\Gamma  \ell^2 \rho _{10} \rho _{20}^2 \sum_i^dk_i^4\nonumber\\
  &-\sqrt{(\kappa _1 \left|\mathbf{k}\right|^4 \rho _{10}^2-\rho _{20}^2 (\kappa _2 \left|\mathbf{k}\right|^6 \rho _{20}+\rho _{10} (\Gamma  \ell^2 \sum_i^dk_i^4-4 \Gamma  \left|\mathbf{k}\right|^2))){}^2+16 \Gamma ^2 \ell^2 \rho _{10}^2 \rho _{20}^4 \Big(\sum_i^dk_i^3\Big)^2}\bigg)\Bigg)^{\frac{1}{2}}\Bigg]\Bigg\}^{\frac{1}{2}},
\end{align}
\begin{align}
  T_{22} (\mathbf{k}) =&-T_{42} (\mathbf{k}) =\Bigg(2 i \sqrt[4]{2} \Gamma  \ell \rho _{10} \rho _{20}^2 \left| \sum_i^dk_i^3\right| \Bigg)/\nonumber\\
  &\Bigg\{\Bigg[\kappa _2 \left|\mathbf{k}\right|^6 \rho _{20}^3+\kappa _1 \left|\mathbf{k}\right|^4 \rho _{10}^2+4 \Gamma  \left|\mathbf{k}\right|^2 \rho _{10} \rho _{20}^2+\Gamma  \ell^2 \rho _{10} \rho _{20}^2 \sum_i^dk_i^4\nonumber\\
  &-\sqrt{(\kappa _1 \left|\mathbf{k}\right|^4 \rho _{10}^2-\rho _{20}^2 (\kappa _2 \left|\mathbf{k}\right|^6 \rho _{20}+\rho _{10} (\Gamma  \ell^2 \sum_i^dk_i^4-4 \Gamma  \left|\mathbf{k}\right|^2))){}^2+16 \Gamma ^2 \ell^2 \rho _{10}^2 \rho _{20}^4 \Big(\sum_i^dk_i^3\Big)^2}\Bigg]^{\frac{1}{2}}\nonumber\\
  &\Bigg[\Bigg(16 \Gamma ^2 \ell^2 \rho _{10}^2 \rho _{20}^4 \Big(\sum_i^dk_i^3\Big)^2-(\kappa _1 \left|\mathbf{k}\right|^4 \rho _{10}^2-\rho _{20}^2 (\kappa _2 k^6 \rho _{20}+\rho _{10} (\Gamma  \ell^2 \sum_i^dk_i^4-4 \Gamma  \left|\mathbf{k}\right|^2)))\nonumber\\
  &\bigg(\kappa _2 \left|\mathbf{k}\right|^6 \rho _{20}^3-\kappa _1 \left|\mathbf{k}\right|^4 \rho _{10}^2-4 \Gamma  \left|\mathbf{k}\right|^2 \rho _{10} \rho _{20}^2+\Gamma  \ell^2 \rho _{10} \rho _{20}^2 \sum_i^dk_i^4\nonumber\\
  &+\sqrt{(\kappa _1 \left|\mathbf{k}\right|^4 \rho _{10}^2-\rho _{20}^2 (\kappa _2 \left|\mathbf{k}\right|^6 \rho _{20}+\rho _{10} (\Gamma  \ell^2 \sum_i^dk_i^4-4 \Gamma  \left|\mathbf{k}\right|^2))){}^2+16 \Gamma ^2 \ell^2 \rho _{10}^2 \rho _{20}^4 \Big(\sum_i^dk_i^3\Big)^2}\bigg)\Bigg)/\nonumber\\
  &\Bigg(\bigg(g \Big(\kappa _2 \left|\mathbf{k}\right|^6 \rho _{20}^3+\kappa _1 \left|\mathbf{k}\right|^4 \rho _{10}^2+4 \Gamma  \left|\mathbf{k}\right|^2 \rho _{10} \rho _{20}^2+\Gamma  \ell^2 \rho _{10} \rho _{20}^2 \sum_i^dk_i^4\nonumber\\
  &-\sqrt{(\kappa _1 \left|\mathbf{k}\right|^4 \rho _{10}^2-\rho _{20}^2 (\kappa _2 \left|\mathbf{k}\right|^6 \rho _{20}+\rho _{10} (\Gamma  \ell^2 \sum_i^dk_i^4-4 \Gamma  \left|\mathbf{k}\right|^2))){}^2+16 \Gamma ^2 \ell^2 \rho _{10}^2 \rho _{20}^4 \Big(\sum_i^dk_i^3\Big)^2}\Big)\bigg)^{\frac{1}{2}}\Bigg)\Bigg]^{\frac{1}{2}}\Bigg\},
\end{align}
\begin{align}
  T_{23} (\mathbf{k}) =&T_{43} (\mathbf{k}) =-\Bigg\{i  \text{sgn}(\sum_i^dk_i^3)  \Bigg[\kappa _2 \left|\mathbf{k}\right|^6 \rho _{20}^3-\kappa _1 \left|\mathbf{k}\right|^4 \rho _{10}^2-4 \Gamma  \left|\mathbf{k}\right|^2 \rho _{10} \rho _{20}^2+\Gamma  \ell^2 \rho _{10} \rho _{20}^2 \sum_i^dk_i^4\nonumber\\
  &+\sqrt{(\kappa _1 \left|\mathbf{k}\right|^4 \rho _{10}^2-\rho _{20}^2 (\kappa _2 \left|\mathbf{k}\right|^6 \rho _{20}+\rho _{10} (\Gamma  \ell^2 \sum_i^dk_i^4-4 \Gamma  \left|\mathbf{k}\right|^2))){}^2+16 \Gamma ^2 \ell^2 \rho _{10}^2 \rho _{20}^4 \Big(\sum_i^dk_i^3\Big)^2}\Bigg]\Bigg\}/\nonumber\\
  &\Bigg\{2 \sqrt[4]{2}  \Bigg[-\Bigg(g \bigg((\kappa _1 \left|\mathbf{k}\right|^4 \rho _{10}^2-\rho _{20}^2 (\kappa _2 \left|\mathbf{k}\right|^6 \rho _{20}+\rho _{10} (\Gamma  \ell^2 \sum_i^dk_i^4-4 \Gamma  \left|\mathbf{k}\right|^2)))\nonumber\\
  &\Big(\kappa _2 \left|\mathbf{k}\right|^6 \rho _{20}^3-\kappa _1 \left|\mathbf{k}\right|^4 \rho _{10}^2-4 \Gamma  \left|\mathbf{k}\right|^2 \rho _{10} \rho _{20}^2+\Gamma  \ell^2 \rho _{10} \rho _{20}^2 \sum_i^dk_i^4\nonumber\\
  &+\sqrt{(\kappa _1 \left|\mathbf{k}\right|^4 \rho _{10}^2-\rho _{20}^2 (\kappa _2 \left|\mathbf{k}\right|^6 \rho _{20}+\rho _{10} (\Gamma  \ell^2 \sum_i^dk_i^4-4 \Gamma  \left|\mathbf{k}\right|^2))){}^2+16 \Gamma ^2 \ell^2 \rho _{10}^2 \rho _{20}^4 \Big(\sum_i^dk_i^3\Big)^2}\Big)\nonumber\\
  &-16 \Gamma ^2 \ell^2 \rho _{10}^2 \rho _{20}^4 \Big(\sum_i^dk_i^3\Big)^2\bigg)\Bigg)/\Bigg(\bigg(g \Big(\kappa _2 \left|\mathbf{k}\right|^6 \rho _{20}^3+\kappa _1 \left|\mathbf{k}\right|^4 \rho _{10}^2+4 \Gamma  \left|\mathbf{k}\right|^2 \rho _{10} \rho _{20}^2+\Gamma  \ell^2 \rho _{10} \rho _{20}^2 \sum_i^dk_i^4\nonumber\\
  &-\sqrt{(\kappa _1 \left|\mathbf{k}\right|^4 \rho _{10}^2-\rho _{20}^2 (\kappa _2 \left|\mathbf{k}\right|^6 \rho _{20}+\rho _{10} (\Gamma  \ell^2 \sum_i^dk_i^4-4 \Gamma  \left|\mathbf{k}\right|^2))){}^2+16 \Gamma ^2 \ell^2 \rho _{10}^2 \rho _{20}^4 \Big(\sum_i^dk_i^3\Big)^2}\Big)\bigg)^{\frac{1}{2}}\Bigg)\Bigg]^{\frac{1}{2}}\Bigg\},
\end{align}
\begin{align}
  T_{24} (\mathbf{k}) =&T_{44} (\mathbf{k}) =\Bigg(2^{3/4} \Gamma  \ell \rho _{10} \rho _{20}^2 \left| \sum_i^dk_i^3\right| \Bigg)/\nonumber\\
  &\Bigg\{\Bigg[-\Bigg(g \bigg((\kappa _1 \left|\mathbf{k}\right|^4 \rho _{10}^2-\rho _{20}^2 (\kappa _2 \left|\mathbf{k}\right|^6 \rho _{20}+\rho _{10} (\Gamma  \ell^2 \sum_i^dk_i^4-4 \Gamma  \left|\mathbf{k}\right|^2)))\nonumber\\
  &\Big(\kappa _2 \left|\mathbf{k}\right|^6 \rho _{20}^3-\kappa _1 \left|\mathbf{k}\right|^4 \rho _{10}^2-4 \Gamma  \left|\mathbf{k}\right|^2 \rho _{10} \rho _{20}^2+\Gamma  \ell^2 \rho _{10} \rho _{20}^2 \sum_i^dk_i^4\nonumber\\
  &+\sqrt{(\kappa _1 \left|\mathbf{k}\right|^4 \rho _{10}^2-\rho _{20}^2 (\kappa _2 \left|\mathbf{k}\right|^6 \rho _{20}+\rho _{10} (\Gamma  \ell^2 \sum_i^dk_i^4-4 \Gamma  \left|\mathbf{k}\right|^2))){}^2+16 \Gamma ^2 \ell^2 \rho _{10}^2 \rho _{20}^4 \Big(\sum_i^dk_i^3\Big)^2}\Big)\nonumber\\
  &-16 \Gamma ^2 \ell^2 \rho _{10}^2 \rho _{20}^4 \Big(\sum_i^dk_i^3\Big)^2\bigg)\Bigg)/\Bigg(\bigg(g \Big(\kappa _2 \left|\mathbf{k}\right|^6 \rho _{20}^3+\kappa _1 \left|\mathbf{k}\right|^4 \rho _{10}^2+4 \Gamma  \left|\mathbf{k}\right|^2 \rho _{10} \rho _{20}^2+\Gamma  \ell^2 \rho _{10} \rho _{20}^2 \sum_i^dk_i^4\nonumber\\
  &-\sqrt{(\kappa _1 \left|\mathbf{k}\right|^4 \rho _{10}^2-\rho _{20}^2 (\kappa _2 \left|\mathbf{k}\right|^6 \rho _{20}+\rho _{10} (\Gamma  \ell^2 \sum_i^dk_i^4-4 \Gamma  \left|\mathbf{k}\right|^2))){}^2+16 \Gamma ^2 \ell^2 \rho _{10}^2 \rho _{20}^4 \Big(\sum_i^dk_i^3\Big)^2}\Big)\bigg)^{\frac{1}{2}}\Bigg)\Bigg]^{\frac{1}{2}}\Bigg\}.
\end{align}
\subsection{Derivation of the Hamiltonian represented by the creation and annihilation operators}
After diagonalized, the Hamiltonian becomes 
\begin{align}
    H= & \frac{1}{2(2\pi)^{d}}\int d^{d}\mathbf{k}D_{1} (\mathbf{k}) (\hat{\alpha}^{\dagger} (\mathbf{k}) \hat{\alpha} (\mathbf{k}) +\hat{\alpha} (-\mathbf{k}) \hat{\alpha}^{\dagger} (-\mathbf{k}) )+D_{2} (\mathbf{k}) (\hat{\beta}^{\dagger} (\mathbf{k}) \hat{\beta} (\mathbf{k}) +\hat{\beta} (-\mathbf{k}) \hat{\beta}^{\dagger} (-\mathbf{k}) ).
\end{align}
Because of $D_{1} (\mathbf{k}) =D_{1} (-\mathbf{k}) $,
$D_{2} (\mathbf{k}) =D_{2} (-\mathbf{k}) $, we rewrite the Hamiltonian as 
\begin{align}
H= & \frac{1}{2(2\pi)^{d}}\int d^{d}\mathbf{k}D_{1} (\mathbf{k}) (\hat{\alpha}^{\dagger} (\mathbf{k}) \hat{\alpha} (\mathbf{k}) +\hat{\alpha} (\mathbf{k}) \hat{\alpha}^{\dagger} (\mathbf{k}) )+D_{2} (\mathbf{k}) (\hat{\beta}^{\dagger} (\mathbf{k}) \hat{\beta} (\mathbf{k}) +\hat{\beta} (\mathbf{k}) \hat{\beta}^{\dagger} (\mathbf{k}) )\nonumber\\
= & \frac{1}{(2\pi)^{d}}\int d^{d}\mathbf{k}D_{1} (\mathbf{k}) \hat{\alpha}^{\dagger} (\mathbf{k}) \hat{\alpha} (\mathbf{k}) +D_{2} (\mathbf{k}) \hat{\beta}^{\dagger} (\mathbf{k}) \hat{\beta} (\mathbf{k}) +\int d^{d}\mathbf{k}\frac{D_{1} (\mathbf{k}) +D_{2} (\mathbf{k}) }{2}\delta^{(d)}(\mathbf{0})\nonumber\\
= & \frac{1}{(2\pi)^{d}}\int d^{d}\mathbf{k}D_{1} (\mathbf{k}) \hat{\alpha}^{\dagger} (\mathbf{k}) \hat{\alpha} (\mathbf{k}) +D_{2} (\mathbf{k}) \hat{\beta}^{\dagger} (\mathbf{k}) \hat{\beta} (\mathbf{k}) +\frac{\tilde{V}}{(2\pi)^{d}}\int d^{d}\mathbf{k}\frac{D_{1}(\mathbf{x})+D_{2} (\mathbf{k}) }{2},
\end{align}
where $\tilde{V}$ is the volume of the $d$-dimensional space, the last term is the vacuum zero point energy. 
The dispersion relations are
\begin{align}
\omega_{1}= & D_{1} (\mathbf{k}) ,\quad\omega_{2}=D_{2} (\mathbf{k}) .
\end{align}

\section{The calculation details of correlation functions of Model Series B}

\label{detail2} From the main text, we have the Bogoliubov transformation from $\hat{\theta}$ and $\hat{\pi}$ to the independent modes $\hat{\alpha}$ and $\hat{\beta}$,
\begin{align}
\psi (\mathbf{k}) = & \begin{bmatrix}\hat{\pi}_{1} (\mathbf{k})  & \hat{\pi}_{2} (\mathbf{k})  & \hat{\theta}_{1} (\mathbf{k})  & \hat{\theta}_{2} (\mathbf{k}) \end{bmatrix}^{T}=T^{-1} (\mathbf{k}) \phi (\mathbf{k}) =\Lambda_{1}T^{\dagger} (\mathbf{k}) \Lambda_{2}\phi (\mathbf{k}) \nonumber\\
= & \begin{bmatrix}-iT_{13}^{\ast} (\mathbf{k})  & -iT_{23}^{\ast} (\mathbf{k})  & iT_{33}^{\ast} (\mathbf{k})  & iT_{43}^{\ast} (\mathbf{k}) \\
-iT_{14}^{\ast} (\mathbf{k})  & -iT_{24}^{\ast} (\mathbf{k})  & iT_{34}^{\ast} (\mathbf{k})  & iT_{44}^{\ast} (\mathbf{k}) \\
iT_{11}^{\ast} (\mathbf{k})  & iT_{21}^{\ast} (\mathbf{k})  & -iT_{31}^{\ast} (\mathbf{k})  & -iT_{41}^{\ast} (\mathbf{k}) \\
iT_{12}^{\ast} (\mathbf{k})  & iT_{22}^{\ast} (\mathbf{k})  & -iT_{32}^{\ast} (\mathbf{k})  & -iT_{42}^{\ast} (\mathbf{k}) 
\end{bmatrix}\begin{bmatrix}\hat{\alpha} (\mathbf{k}) \\
\hat{\beta} (\mathbf{k}) \\
\hat{\alpha}^{\dagger} (-\mathbf{k}) \\
\hat{\beta}^{\dagger} (-\mathbf{k}) 
\end{bmatrix}.
\end{align}
The two-operator correlation function can be written as 
\begin{align}
\langle\hat{\Phi}_{1}(\mathbf{x})\hat{\Phi}_{1}(\mathbf{0})\rangle= & \rho_{10}e^{-\frac{1}{2}\langle\left[\hat{\theta}_{1}(\mathbf{x})-\hat{\theta}_{1}(\mathbf{0})\right]^{2}\rangle}=\rho_{10}e^{\langle\hat{\theta}_{1}(\mathbf{x})\hat{\theta}_{1}(\mathbf{0})\rangle-\langle\hat{\theta}_{1}^{2}(\mathbf{0})\rangle},
\end{align}
where 
\begin{align}
\langle\hat{\theta}_{1}(\mathbf{x})\hat{\theta}_{1}(\mathbf{0})\rangle= & \langle\int\frac{d^{d}\mathbf{k}d^{d}\mathbf{k}^{\prime}}{(2\pi)^{2d}}e^{i\mathbf{k}\cdot\mathbf{x}}\hat{\theta}_{1} (\mathbf{k}) \hat{\theta}_{1}\ (\mathbf{k}') \rangle=\langle\int\frac{d^{d}\mathbf{k}d^{d}\mathbf{k}^{\prime}}{(2\pi)^{2d}}e^{i\mathbf{k}\cdot\mathbf{x}}\hat{\theta}_{1} (\mathbf{k}) \hat{\theta}_{1}^{\dagger}(-\mathbf{k}^{\prime})\rangle\nonumber\\
= & \int\frac{d^{d}\mathbf{k}d^{d}\mathbf{k}^{\prime}}{(2\pi)^{2d}}e^{i\mathbf{k}\cdot\mathbf{x}}\langle\left[iT_{11}^{\ast} (\mathbf{k}) \hat{\alpha} (\mathbf{k}) +iT_{21}^{\ast} (\mathbf{k}) \hat{\beta} (\mathbf{k}) -iT_{31}^{\ast} (\mathbf{k}) \hat{\alpha}^{\dagger} (-\mathbf{k}) -iT_{41}^{\ast} (\mathbf{k}) \hat{\beta}^{\dagger} (-\mathbf{k}) \right]\nonumber\\
 & \left[-iT_{11}(-\mathbf{k}^{\prime})\hat{\alpha}^{\dagger}(-\mathbf{k}^{\prime})-iT_{21}(-\mathbf{k}^{\prime})\hat{\beta}^{\dagger}(-\mathbf{k}^{\prime})+iT_{31}(-\mathbf{k}^{\prime})\hat{\alpha}\ (\mathbf{k}') +iT_{41}(-\mathbf{k}^{\prime})\hat{\beta}\ (\mathbf{k}') \right]\rangle.
\end{align}
For the expectation value of the creation and annihilation operators
in the ground state, we have 
\begin{align}
\langle\hat{\alpha}\ (\mathbf{k}') \hat{\alpha} (\mathbf{k}) \rangle=\langle\hat{\beta}\ (\mathbf{k}') \hat{\beta} (\mathbf{k}) \rangle=\langle\hat{\alpha}^{\dagger}(-\mathbf{k}^{\prime})\hat{\alpha} (\mathbf{k}) \rangle=\langle\hat{\beta}^{\dagger}(-\mathbf{k}^{\prime})\hat{\beta} (\mathbf{k}) \rangle=0,
\end{align}
\begin{align}
\langle\hat{\alpha} (\mathbf{k}) \hat{\alpha}^{\dagger}(-\mathbf{k}^{\prime})\rangle=\langle\hat{\beta} (\mathbf{k}) \hat{\beta}^{\dagger}(-\mathbf{k}^{\prime})\rangle=(2\pi)^{d}\delta^{(d)}(\mathbf{k}+\mathbf{k}^{\prime}).
\end{align}
Then we can obtain the correlation function,
\begin{align}
\langle\hat{\theta}_{1}(\mathbf{x})\hat{\theta}_{1}(\mathbf{0})\rangle=\int\frac{d^{d}\mathbf{k}}{(2\pi)^{d}}e^{i\mathbf{k}\cdot\mathbf{x}}(\left|T_{11} (\mathbf{k}) \right|^{2}+\left|T_{21} (\mathbf{k}) \right|^{2}).
\end{align}
Similarly, we can get 
\begin{align}
\langle\hat{\theta}_{2}(\mathbf{x})\hat{\theta}_{2}(\mathbf{0})\rangle=\int\frac{d^{d}\mathbf{k}}{(2\pi)^{d}}e^{i\mathbf{k}\cdot\mathbf{x}}(\left|T_{12} (\mathbf{k}) \right|^{2}+\left|T_{22} (\mathbf{k}) \right|^{2}).
\end{align}
Therefore, to calculate correlation functions, we need to calculate
$\left|T_{11} (\mathbf{k}) \right|^{2}$, $\left|T_{21} (\mathbf{k}) \right|^{2}$,
$\left|T_{12} (\mathbf{k}) \right|^{2}$, and $\left|T_{22} (\mathbf{k}) \right|^{2}$ in different spatial dimensions.

\subsection{One spatial dimension}

In one spatial dimension, the dispersion relations are 
\begin{align}
  \omega_{1}\approx & \sqrt{4g\Gamma\rho_{10}\rho_{20}^{2}}\left|k\right|,\quad\omega_{2}\approx\sqrt{\kappa_{2}g\rho_{20}^{3}+\frac{\kappa_{1}g\rho_{10}^{2}\ell^{2}}{4}}\left|k\right|^{3}
  \end{align}
We can approximate these four terms,
\begin{align}
  \left|T_{11} (\mathbf{k}) \right|^{2}\approx\frac{\sqrt{g\Gamma\rho_{10}}}{4\Gamma\rho_{10}\rho_{20}}\frac{1}{\left|k\right|}=\frac{1}{c_{11}\left|k\right|}, & \quad\left|T_{12} (\mathbf{k}) \right|^{2}\approx\frac{\sqrt{g\Gamma\rho_{10}}\ell^{2}}{16\Gamma\rho_{10}\rho_{20}}\left|k\right|=\frac{\left|k\right|}{c_{12}},\nonumber\\
  \left|T_{21} (\mathbf{k}) \right|^{2}\approx\frac{\sqrt{g}\ell^{2}}{4\sqrt{\kappa_{1}\rho_{10}^{2}\ell^{2}+4\kappa_{2}\rho_{20}^{3}}}\frac{1}{\left|k\right|}=\frac{1}{c_{21}\left|k\right|}, & \quad\left|T_{22} (\mathbf{k}) \right|^{2}\approx\frac{\sqrt{g}}{\sqrt{\kappa_{1}\rho_{10}^{2}\ell^{2}+4\kappa_{2}\rho_{20}^{3}}}\frac{1}{\left|k\right|^{3}}=\frac{1}{c_{22}\left|k\right|^{3}}.
  \end{align}
The correlation function of $\hat{\theta}_{1}$ is 
\begin{align}
\langle\hat{\theta}_{1}(x)\hat{\theta}_{1}(0)\rangle= & \int\frac{dk}{2\pi}e^{ikx}\left(\frac{1}{c_{11}\left|k\right|}+\frac{1}{c_{21}\left|k\right|}\right)=\frac{1}{2\pi}\left(\frac{1}{c_{11}}+\frac{1}{c_{21}}\right)\int dk\frac{e^{ikx}}{\left|k\right|}\nonumber\\
= & \frac{1}{2\pi}\left(\frac{1}{c_{11}}+\frac{1}{c_{21}}\right)\left(\int_{-\infty}^{-\frac{2\pi}{L}}dk\frac{e^{ikx}}{-k}+\int_{\frac{2\pi}{L}}^{\infty}dk\frac{e^{ikx}}{k}\right).
\end{align}
Considering $L>0$, $x>0$, we have the two integrals 
\begin{align}
\int_{-\infty}^{-\frac{2\pi}{L}}dk\frac{e^{ikx}}{-k}+\int_{\frac{2\pi}{L}}^{\infty}\frac{e^{ikx}}{k}=\Gamma(0,-\frac{2i\pi x}{L})+\Gamma(0,\frac{2i\pi x}{L})\overset{L\rightarrow\infty}{=}-2\left[\gamma+\ln(\frac{2\pi x}{L})\right],
\end{align}
where $\gamma$ is the Euler-Mascheroni constant. Then, we can get the correlation functions,
\begin{align}
\langle\hat{\theta}_{1}(x)\hat{\theta}_{1}(0)\rangle\overset{L\rightarrow\infty}{=} & \frac{1}{2\pi}\left(\frac{1}{c_{11}}+\frac{1}{c_{21}}\right)\left[-2(\gamma+\ln(\frac{2\pi\left|x\right|}{L}))\right]=\frac{c_{11}+c_{21}}{\pi c_{11}c_{21}}\left[-\gamma+\ln(\frac{L}{2\pi\left|x\right|})\right].
\end{align}
To get $\langle\hat{\theta}_{1}^{2}(0)\rangle=\lim_{x\rightarrow0}\langle\theta_{1}(x)\theta_{a}(0)\rangle$,
we just need to replace $\left|x\right|$ by $\xi_{c}$. $\xi_{c}$ is the coherence length,
\begin{align}
\langle\hat{\theta}_{1}^{2}(0)\rangle=\langle\hat{\theta}_{1}(\xi_{c})\hat{\theta}_{1}(0)\rangle=\frac{c_{11}+c_{21}}{\pi c_{11}c_{21}}\left[-\gamma+\ln(\frac{L}{2\pi\xi_{c}})\right].
\end{align}
Then, we obtain the correlation function $\langle\hat{\Phi}_{1}^{\dagger}(x)\hat{\Phi}_{1}(0)\rangle$,
\begin{align}
\langle\hat{\Phi}_{1}^{\dagger}(x)\hat{\Phi}_{1}(0)\rangle= & \rho_{10}e^{\langle\hat{\theta}_{1}(x)\hat{\theta}_{1}(0)\rangle-\langle\hat{\theta}_{1}^{2}(0)\rangle}=\rho_{10}\left(\frac{\left|x\right|}{\xi_{c}}\right)^{-\frac{c_{11}+c_{21}}{\pi c_{11}c_{21}}}.
\end{align}
The correlation function $\langle\hat{\Phi}_{1}^{\dagger}(\mathbf{x})\hat{\Phi}_{1}(\mathbf{0})\rangle$ is a power-law decay.
The correlation function of $\hat{\theta}_{2}$ is 
\begin{align}
\langle\hat{\theta}_{2}(x)\hat{\theta}_{2}(0)\rangle= & \int\frac{dk}{2\pi}e^{ikx}\left(\frac{\left|k\right|}{c_{12}}+\frac{1}{c_{22}\left|k\right|^{3}}\right)=\frac{1}{2\pi c_{12}}\int dke^{ikx}\left|k\right|+\frac{1}{2\pi c_{22}}\int dk\frac{e^{ikx}}{\left|k\right|^{3}}.
\end{align}
Here we calculate the these integrals separately,
\begin{align}
\int_{-\frac{2\pi}{\xi_{c}}}^{\frac{2\pi}{\xi_{c}}}dke^{ikx}\left|k\right|=\frac{4\pi\left|x\right|\sin(\frac{2\pi\left|x\right|}{\xi_{c}})-4\xi_{c}\sin^{2}\left(\frac{\pi\left|x\right|}{\xi_{c}}\right)}{\left|x\right|^{2}\xi_{c}},
\end{align}
\begin{align}
& \int_{\frac{2\pi}{L}}^{\infty}dk\frac{e^{ikx}}{\left|k\right|^{3}}+\int_{-\infty}^{-\frac{2\pi}{L}}dk\frac{e^{ikx}}{\left|k\right|^{3}}\overset{L\rightarrow\infty}{=}\frac{L^{2}}{4\pi^{2}}+\frac{\left|x\right|^{2}}{2}\left[-3+2\gamma+2\ln(\frac{2\pi\left|x\right|}{L})\right].
\end{align}
Then, we obtain the correlation functions $\langle\hat{\theta}_{2}(x)\hat{\theta}_{2}(0)\rangle$
and $\langle\hat{\theta}_{2}^{2}(0)\rangle$, 
\begin{align}
\langle\hat{\theta}_{2}(x)\hat{\theta}_{2}(0)\rangle= & \int\frac{dk}{2\pi}e^{ikx}(\frac{\left|k\right|}{c_{12}}+\frac{1}{c_{22}\left|k\right|^{3}})=\frac{1}{2\pi c_{12}}\int dke^{ikx}\left|k\right|+\frac{1}{2\pi c_{22}}\int dk\frac{e^{ikx}}{\left|k\right|^{3}}\nonumber\\
= & \frac{1}{2\pi c_{12}}\left[\frac{4\pi\left|x\right|\sin(\frac{2\pi\left|x\right|}{\xi_{c}})-4\xi_{c}\sin^{2}(\frac{\pi\left|x\right|}{\xi_{c}})}{\left|x\right|^{2}\xi_{c}}\right]+\frac{1}{2\pi c_{22}}\left[\frac{L^{2}}{4\pi^{2}}+\frac{\left|x\right|^{2}}{2}(-3+2\gamma+2\ln(\frac{2\pi\left|x\right|}{L}))\right],
\end{align}
\begin{align}
\langle\hat{\theta}_{2}^{2}(0)\rangle= & \int\frac{dk}{2\pi}(\frac{\left|k\right|}{c_{12}}+\frac{1}{c_{22}\left|k\right|^{3}})=\int_{-\frac{2\pi}{\xi_{c}}}^{\frac{2\pi}{\xi_{c}}}dk\left|k\right|+\int_{\frac{2\pi}{L}}^{\infty}dk\frac{1}{\left|k\right|^{3}}+\int_{-\infty}^{-\frac{2\pi}{L}}dk\frac{1}{\left|k\right|^{3}}=\frac{1}{2\pi c_{12}}\frac{4\pi^{2}}{\xi_{c}^{2}}+\frac{1}{2\pi c_{22}}\frac{L^{2}}{4\pi^{2}}.
\end{align}
Then, we obtain the correlation function $\langle\hat{\Phi}_{2}^\dagger(\mathbf{x})\hat{\Phi}_{2}(\mathbf{0})\rangle$,
\begin{align}
\langle\hat{\Phi}_{2}^{\dagger}(x)\hat{\Phi}_{2}(0)\rangle= & \rho_{20}e^{\langle\hat{\theta}_{2}(x)\hat{\theta}_{2}(0)\rangle-\langle\hat{\theta}^{2}(0)\rangle}\rightarrow0.
\end{align}
The correlation function $\langle\hat{\Phi}_2^\dagger(x)\hat{\Phi}_2(0)\rangle$ decays exponentially with $\left|x\right|^2$.
The correlation function $\langle\hat{\Phi}_{1}^{\dagger}(\mathbf{x})\hat{\Phi}_{1}(\mathbf{0})\rangle$ is a power-law decay. Model Series B does not break the charge symmetry and is algebraically ordered in one spatial dimension. 

\subsection{Two spatial dimensions}

In two spatial dimensions, we set $k_{1}=\left|\mathbf{k}\right|\cos\theta$, and $k_{2}=\left|\mathbf{k}\right|\sin\theta$, for
$\left|\mathbf{k}\right|\geq0$, $0\leq\theta<2\pi$. Then, we can get the dispersion relations,
\begin{align}
\omega_{1}\approx & \sqrt{4g\Gamma\rho_{10}\rho_{20}^{2}}\left|\mathbf{k}\right|,
\end{align}
\begin{equation}
  \begin{split}\omega_{2}\approx\left\{ \begin{split} & \frac{1}{2}\rho_{20}\ell\left|\sin2\theta\right|\sqrt{g\rho_{10}\Gamma(1-\sin2\theta)}\left|\mathbf{k}\right|^{2}\quad(\theta\neq0,\frac{\pi}{4},\frac{\pi}{2},\frac{3\pi}{4},\pi,\frac{5\pi}{4},\frac{3\pi}{2},\frac{7\pi}{4})\\
   & \sqrt{\kappa_{2}g\rho_{20}^{3}+\frac{\kappa_{1}g\rho_{10}^{2}\ell^{2}}{4}}\left|\mathbf{k}\right|^{3}\quad(\theta=0,\frac{\pi}{2},\pi,\frac{3\pi}{2})\\
   & \sqrt{\kappa_{2}g\rho_{20}^{3}+\frac{\kappa_{1}g\rho_{10}^{2}\ell^{2}}{8}}\left|\mathbf{k}\right|^{3}\quad(\theta=\frac{\pi}{4},\frac{3\pi}{4},\frac{5\pi}{4},\frac{7\pi}{4})
  \end{split}
  \right.\end{split}
  \end{equation}
In two spatial dimensions, the dispersion relations are anisotropic.
The same is true for the two terms $\left|T_{21} (\mathbf{k}) \right|^{2}$
and $\left|T_{22} (\mathbf{k}) \right|^{2}$, 
\begin{align}
  \left|T_{11} (\mathbf{k}) \right|^{2}\approx & \frac{\sqrt{g\Gamma\rho_{10}}}{4\Gamma\rho_{10}\rho_{20}}\frac{1}{\left|\mathbf{k}\right|},\quad\left|T_{12} (\mathbf{k}) \right|^{2}\approx\frac{\sqrt{g\Gamma\rho_{10}}\ell^{2}(\cos^{3}\theta+\sin^{3}\theta)^{2}}{16\Gamma\rho_{10}\rho_{20}}\left|\mathbf{k}\right|,
  \end{align}
\begin{equation}
  \begin{split}\left|T_{21} (\mathbf{k}) \right|^{2} & \left\{ \begin{split}\approx & \frac{\sqrt{g\Gamma\rho_{10}}\ell(3\cos\theta+\cos3\theta+4\sin^{3}\theta)^{2}}{64\Gamma\rho_{10}\rho_{20}\left|\sin2\theta\right|\sqrt{1-\sin2\theta}}\quad(\theta\neq0,\frac{\pi}{4},\frac{\pi}{2},\frac{3\pi}{4},\pi,\frac{5\pi}{4},\frac{3\pi}{2},\frac{7\pi}{4})\\
  \approx & \frac{\sqrt{g}\ell^{2}}{4\sqrt{\kappa_{1}\rho_{10}^{2}\ell^{2}+4\kappa_{2}\rho_{20}^{3}}}\frac{1}{\left|\mathbf{k}\right|}\quad(\theta=0,\frac{\pi}{2},\pi,\frac{3\pi}{2})\\
  \approx & \frac{\sqrt{g}\ell^{2}}{4\sqrt{2}\sqrt{\kappa_{1}\rho_{10}^{2}\ell^{2}+8\kappa_{2}\rho_{20}^{3}}}\frac{1}{\left|\mathbf{k}\right|}\quad(\theta=\frac{\pi}{4},\frac{5\pi}{4})\\
  = & 0\quad(\theta=\frac{3\pi}{4},\frac{7\pi}{4})
  \end{split}
  \right.\end{split},
  \end{equation}
\begin{equation}
    \begin{split}\left|T_{22} (\mathbf{k}) \right|^{2} & \left\{ \begin{split}\approx & \frac{\sqrt{g\Gamma\rho_{10}}}{\Gamma \ell\rho_{10}\rho_{20}\sqrt{1-\sin2\theta}\left|\sin2\theta\right|\left|\mathbf{k}\right|^{2}}\quad(\theta\neq0,\frac{\pi}{4},\frac{\pi}{2},\frac{3\pi}{4},\pi,\frac{5\pi}{4},\frac{3\pi}{2},\frac{7\pi}{4})\\
    \approx & \frac{\sqrt{g}}{\sqrt{\kappa_{1}\rho_{10}^{2}\ell^{2}+4\kappa_{2}\rho_{20}^{3}}}\frac{1}{\left|\mathbf{k}\right|^{3}}\quad(\theta=0,\frac{\pi}{2},\pi,\frac{3\pi}{2})\\
    \approx & \frac{\sqrt{2g}}{\sqrt{\kappa_{1}\rho_{10}^{2}\ell^{2}+8\kappa_{2}\rho_{20}^{3}}}\frac{1}{\left|\mathbf{k}\right|^{3}}\quad(\theta=\frac{\pi}{4},\frac{5\pi}{4})\\
    = & 0\quad(\theta=\frac{3\pi}{4},\frac{7\pi}{4})
    \end{split}
    \right.\end{split}.
\end{equation}
In the long-wave length limit, to simplify the calculation, we can always choose the lowest order to give a rough approximation,
\begin{align}
\left|T_{11} (\mathbf{k}) \right|^{2}\approx\frac{1}{c_{11}\left|\mathbf{k}\right|},\quad\left|T_{12} (\mathbf{k}) \right|^{2}\lesssim\frac{\left|\mathbf{k}\right|}{c_{12}},\quad\left|T_{21} (\mathbf{k}) \right|^{2}\lesssim\frac{1}{c_{21}\left|\mathbf{k}\right|},\quad\left|T_{22} (\mathbf{k}) \right|^{2}\lesssim\frac{1}{c_{22}\left|\mathbf{k}\right|^{3}},
\end{align}
where $c_{11}=\frac{4\Gamma\rho_{10}\rho_{20}}{\sqrt{g\Gamma\rho_{10}}}$,
$c_{12}=\frac{16\Gamma\rho_{10}\rho_{20}}{\sqrt{g\Gamma\rho_{10}}\ell^{2}}$,
$c_{21}=\frac{4\sqrt{\kappa_{1}\rho_{10}^{2}\ell^{2}+4\kappa_{2}\rho_{20}^{3}}}{\sqrt{g}\ell^{2}}$,
$c_{22}=\frac{\sqrt{\kappa_{1}\rho_{10}^{2}\ell^{2}+8\kappa_{2}\rho_{20}^{3}}}{\sqrt{2g}}$.

Following this rough approximation, if the calculation results show that there is a spontaneous breaking of charge symmetry, there must be, but not the opposite.
With the above approximation, we can compute correlation functions,
\begin{align}
\langle\hat{\theta}_{1}(\mathbf{x})\hat{\theta}_{1}(\mathbf{0})\rangle= & \int\frac{kdkd\theta}{(2\pi)^{2}}e^{ik\left|\mathbf{x}\right|\cos\theta}\left(\frac{1}{c_{11}k}+\frac{1}{c_{21}k}\right)=\frac{c_{11}+c_{21}}{4\pi^{2}c_{11}c_{21}}\int dkd\theta e^{ik\left|\mathbf{x}\right|\cos\theta}=\frac{c_{11}+c_{21}}{2\pi c_{11}c_{21}\left|\mathbf{x}\right|}.
\end{align}
To get $\langle\hat{\theta}_{1}^{2}(\mathbf{0})\rangle=\lim_{\left|\mathbf{x}\right|\rightarrow0}\langle\theta_{1}(\mathbf{x})\theta_{a}(0)\rangle$,
we just need to replace $\left|x\right|$ by $\xi_{c}$. $\xi_{c}$ is the coherence length,
\begin{align}
\langle\hat{\theta}_{1}^{2}(\mathbf{0})\rangle=\langle\hat{\theta}_{1}(\xi_{c})\hat{\theta}_{1}(\mathbf{0})\rangle=\frac{c_{11}+c_{21}}{2\pi c_{11}c_{21}\xi_{c}}.
\end{align}
Then, we can calculate the correlation function $\langle\hat{\Phi}_{1}^{\dagger}(\mathbf{x})\hat{\Phi}_{1}(\mathbf{0})\rangle$,
\begin{align}
\langle\hat{\Phi}_{1}^{\dagger}(\mathbf{x})\hat{\Phi}_{1}(\mathbf{0})\rangle= & \rho_{10}e^{\langle\hat{\theta}_{1}(\mathbf{x})\hat{\theta}_{1}(\mathbf{0})\rangle-\langle\hat{\theta}_{1}^{2}(\mathbf{0})\rangle}\overset{\left|\mathbf{x}\right|\rightarrow\infty}{=}\rho_{10}e^{-\frac{c_{11}+c_{21}}{2\pi c_{11}c_{21}\xi_{c}}}.
\end{align}
The correlation function of $\hat{\theta}_{2}$ is 
\begin{align}
\langle\hat{\theta}_{2}(\mathbf{x})\hat{\theta}_{2}(\mathbf{0})\rangle= & \int\frac{kdkd\theta}{(2\pi)^{2}}e^{ik\left|\mathbf{x}\right|\cos\theta}\left(\frac{k}{c_{12}}+\frac{1}{c_{22}k^{3}}\right)=\frac{1}{4\pi^{2}c_{12}}\int dkd\theta e^{ik\left|\mathbf{x}\right|\cos\theta}k^{2}+\frac{1}{4\pi^{2}c_{22}}\int dkd\theta e^{ik\left|\mathbf{x}\right|\cos\theta}\frac{1}{k^{2}}.
\end{align}
Here we compute the two integrals separately. 
\begin{align}
& \frac{1}{4\pi^{2}c_{12}}\int dkd\theta e^{ik\left|\mathbf{x}\right|\cos\theta}k^{2}=\frac{1}{4\pi^{2}c_{12}}\int_{0}^{\frac{2\pi}{\xi_{c}}}2\pi J_{0}(k\left|\mathbf{x}\right|)k^{2}\nonumber\\
= & \frac{1}{4\pi^{2}c_{12}}\frac{2\pi^{3}\left[J_{1}(\frac{2\pi\left|\mathbf{x}\right|}{\xi_{c}})(4\left|\mathbf{x}\right|-\xi_{c}H_{0}(\frac{2\pi\left|\mathbf{x}\right|}{\xi_{c}}))+\xi_{c}J_{0}(\frac{2\pi\left|\mathbf{x}\right|}{\xi_{c}})H_{1}(\frac{2\pi\left|\mathbf{x}\right|}{\xi_{c}})\right]}{\left|\mathbf{x}\right|^{2}\xi_{c}^{2}}\overset{\left|\mathbf{x}\right|\rightarrow\infty}{=}0
\end{align}
where $J_{a}(z)$ is the Bessel function of the first kind,
and $H_{n}(z)$ is the Struve function. 
\begin{align}
\frac{1}{4\pi^{2}c_{22}}\int dkd\theta e^{ik\left|\mathbf{x}\right|\cos\theta}\frac{1}{k^{2}}= & \frac{1}{4\pi^{2}c_{22}}\int_{\frac{2\pi}{L}}^{\infty}\frac{2\pi J_{0}(k\left|\mathbf{k}\right|)}{k^{2}}\overset{L\rightarrow\infty}{=}\frac{1}{4\pi^{2}c_{22}}(L-2\pi\left|\mathbf{x}\right|)
\end{align}
Then, we can get 
\begin{align}
\langle\hat{\theta}_{2}(\mathbf{x})\hat{\theta}_{2}(\mathbf{0})\rangle\overset{L\rightarrow\infty,\quad\left|\mathbf{x}\right|\rightarrow\infty}{=}\frac{1}{4\pi^{2}c_{22}}(L-2\pi\left|\mathbf{x}\right|).
\end{align}
We can also calculate $\langle\hat{\theta}_{2}^{2}(\mathbf{0})\rangle$,
\begin{align}
\langle\hat{\theta}_{2}^{2}(\mathbf{0})\rangle=\frac{1}{4\pi^{2}c_{12}}\int_{0}^{\frac{2\pi}{\xi_{c}}}dk\int_{0}^{2\pi}d\theta k^{2}+\frac{1}{4\pi^{2}c_{22}}\int_{\frac{2\pi}{L}}^{\infty}dk\int_{0}^{2\pi}d\theta\frac{1}{k^{2}}=\frac{4\pi^{2}}{3c_{12}\xi_{c}^{3}}+\frac{L}{4\pi^{2}c_{22}}.
\end{align}
Then, we can calculate the correlation function $\langle\hat{\Phi}_{2}^{\dagger}(\mathbf{x})\hat{\Phi}_{2}(\mathbf{0})\rangle$,
\begin{align}
\langle\hat{\Phi}_{2}^{\dagger}(\mathbf{x})\hat{\Phi}_{2}(\mathbf{0})\rangle= & \rho_{20}e^{\langle\hat{\theta}_{2}(\mathbf{x})\hat{\theta}_{2}(\mathbf{0})\rangle-\langle\hat{\theta}_{2}^{2}(\mathbf{0})\rangle}\overset{\left|\mathbf{x}\right|\rightarrow\infty}{=}\rho_{20}e^{-\frac{4\pi^{2}}{3c_{12}\xi_{c}^{3}}-\frac{\left|\mathbf{x}\right|}{2\pi c_{22}}}\overset{\left|\mathbf{x}\right|\rightarrow\infty}{=}0.
\end{align}
According to the results of the above analytical calculation, the correlation function $\langle\hat{\Phi}_{1}^{\dagger}(\mathbf{x})\hat{\Phi}_{1}(\mathbf{0})\rangle$ saturates to a constant and $\langle\hat{\Phi}_{2}^{\dagger}(\mathbf{x})\hat{\Phi}_{2}(\mathbf{0})\rangle$ is a power-law decay when $\left|\mathbf{x}\right|\rightarrow\infty$.
We can test these results numerically. By setting $\kappa_{1}=1$,
$\kappa_{2}=1$, $\Gamma=1$, $g=1$, $\rho_{10}=1$, $\rho_{20}=1$,
$\ell=1$, $\xi_{c}=2\pi$, $\frac{2\pi}{\xi_{c}}=1$, $\left|\mathbf{x}\right|\sim10^{10}$,
$L\sim10^{20}$, we can calculate the correlation function by using
the numerical methods. Firstly, we can calculate $\langle\hat{\Phi}_{1}^{\dagger}(10^{10},0)\hat{\Phi}_{1}(\mathbf{0})\rangle$
numerically. For the integrals, we have 
\begin{align}
& \int_{0}^{1}dk\int_{0}^{2\pi}d\theta\left|T_{11} (\mathbf{k}) \right|^{2}ke^{i10^{10}k\cos\theta}-\int_{0}^{1}dk\int_{0}^{2\pi}d\theta\left|T_{11} (\mathbf{k}) \right|^{2}k\approx(5.90\times10^{-13}+8.41\times10^{-16}i)-1.42\approx-1.42,
\end{align}
\begin{align}
& \int_{0}^{1}dk\int_{0}^{2\pi}d\theta\left|T_{21} (\mathbf{k}) \right|^{2}ke^{i10^{10}k\cos\theta}-\int_{0}^{1}dk\int_{0}^{2\pi}d\theta\left|T_{21} (\mathbf{k}) \right|^{2}k\approx(-5.23\times10^{-4}-6.08\times10^{-4}i)-0.351\approx-0.351.
\end{align}
Then, we obtain the result,
\begin{align}
\langle\hat{\Phi}_{1}^{\dagger}(10^{10},0)\hat{\Phi}_{1}(\mathbf{0})\rangle\approx & e^{\frac{1}{(2\pi)^{2}}(-1.42-0.351)}\approx0.956.
\end{align}
From this we can believe that $\langle\hat{\Phi}_{1}^{\dagger}(\mathbf{x})\hat{\Phi}_{1}(\mathbf{0})\rangle$ saturates to a constant when $\left|\mathbf{x}\right|\rightarrow\infty$. We
can also calculate the correlation function $\hat{\Phi}_{2}^{\dagger}(10^{10},0)\hat{\Phi}_{2}(\mathbf{0})$.
For the integrals, we have 
\begin{align}
& \int_{0}^{1}dk\int_{0}^{2\pi}d\theta\left|T_{12} (\mathbf{k}) \right|^{2}ke^{i10^{10}k\cos\theta}-\int_{0}^{1}dk\int_{0}^{2\pi}d\theta\left|T_{12} (\mathbf{k}) \right|^{2}k\approx(2.34\times10^{-4}+3.56\times10^{-4}i)-0.0623\approx-0.0623,
\end{align}
\begin{align}
& \int_{10^{-20}}^{1}dk\int_{0}^{2\pi}d\theta\left|T_{22} (\mathbf{k}) \right|^{2}k(e^{i10^{10}k\cos\theta}-1)\approx-874-10.9i\approx-874,
\end{align}
\begin{align}
\langle\hat{\Phi}_{2}^{\dagger}(10^{10},0)\hat{\Phi}_{2}(\mathbf{0})\rangle\approx & e^{\frac{1}{(2\pi)^{2}}(-874)}\approx2.43\times10^{-10}.
\end{align}
From this we can know that $\langle\hat{\Phi}_{2}^{\dagger}(\mathbf{x})\hat{\Phi}_{2}(\mathbf{0})\rangle$
do not saturate to a constant when $\left|\mathbf{x}\right|\rightarrow\infty$.
The correlation function $\langle\hat{\Phi}_{2}^{\dagger}(\mathbf{x})\hat{\Phi}_{2}(\mathbf{0})\rangle$ is shown in Fig. \ref{fignumerical}(a) below. 
The figure shows that the correlation function $\langle\hat{\Phi}_{2}^{\dagger}(\mathbf{x})\hat{\Phi}_{2}(\mathbf{0})\rangle$ decays power-law.
Since the correlation function $\langle\hat{\Phi}_{1}^{\dagger}(\mathbf{x})\hat{\Phi}_{1}(\mathbf{0})\rangle$ decays in a power law when $\Gamma$ is 0, the value of $\Gamma$ affects the value of the correlation function. 
We can also give the dependence of the correlation function $\langle\hat{\Phi}_{1}^{\dagger}(\mathbf{x})\hat{\Phi}_{1}(\mathbf{0})\rangle$ on $\Gamma$, see the blue line of Fig. \ref{fignumerical}(b).
The constant behavior of the correlation function $\langle\hat{\Phi}_1^\dagger(\mathbf{x})\hat{\Phi}_1(\mathbf{0})\rangle$ shows the spontaneous breaking of the charge symmetry $U(1)_{1,C}$ of bosons of species 1. Model Series B has a true ODLRO in two spatial dimensions.

\subsection{Three spatial dimensions}

\begin{figure}[htbp]
  \centering
  \includegraphics[width=\textwidth]{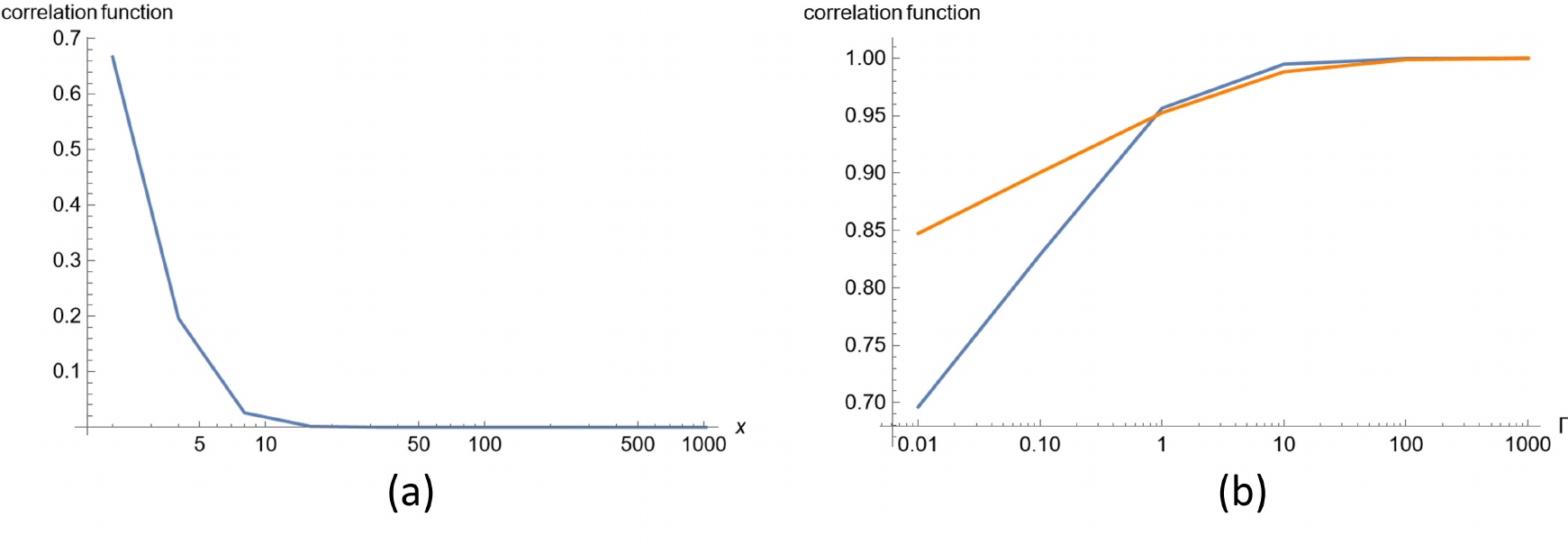}
  \caption{Results obtained by numerical calculations. Fig.~\ref{fignumerical}(a) shows the graph of the correlation function $\langle\hat{\Phi}_2^\dagger(x,0)\hat{\Phi}_2(\mathbf{0})\rangle$ decaying with increasing x in two spatial dimensions. In Fig.~\ref{fignumerical}(b), the blue line shows the variation of correlation function $\langle\hat{\Phi}_1^\dagger(10^{10},0)\hat{\Phi}_1(\mathbf{0})\rangle$ with increasing $\Gamma$ in two spatial dimensions, and the orange line shows the variation of correlation function $\langle\hat{\Phi}_2^\dagger(10^{10},0,0)\hat{\Phi}_2(\mathbf{0})\rangle$ with increasing $\Gamma$ in three spatial dimensions.}
  \label{fignumerical}
\end{figure}

In three spatial dimensions, we set $k_{1}=\left|\mathbf{k}\right|\cos\theta\sin\psi$, $k_{2}=\left|\mathbf{k}\right|\sin\theta\sin\psi$,
$k_{3}=\left|\mathbf{k}\right|\cos\psi$. Due to the large error in the analytical approximation of the 3D results, we directly use numerical methods to calculate the correlation functions,
\begin{align}
& \int_{0}^{1}dk\int_{0}^{2\pi}d\theta\int_{0}^{\pi}d\psi\left|T_{11} (\mathbf{k}) \right|^{2}k^{2}\sin\psi e^{i10^{10}k\cos\theta\sin\psi}-\int_{0}^{1}dk\int_{0}^{2\pi}d\theta\int_{0}^{\pi}d\psi\left|T_{11} (\mathbf{k}) \right|^{2}k^{2}\sin\psi\nonumber\\
\approx & (-0.0239+0.0113i)-1.39\approx-1.41,
\end{align}
\begin{align}
& \int_{0}^{1}dk\int_{0}^{2\pi}d\theta\int_{0}^{\pi}d\psi\left|T_{21} (\mathbf{k}) \right|^{2}k^{2}\sin\psi e^{i10^{10}k\cos\theta\sin\psi}-\int_{0}^{1}dk\int_{0}^{2\pi}d\theta\int_{0}^{\pi}d\psi\left|T_{21} (\mathbf{k}) \right|^{2}k^{2}\sin\psi\nonumber\\
\approx & (-9.34\times10^{-4}-9.71\times10^{-4})-0.258\approx-0.258.
\end{align}
From the above results, we can obtain the correlation function $\langle\hat{\Phi}_{1}^{\dagger}(\mathbf{x})\hat{\Phi}_{1}(\mathbf{0})\rangle$,
\begin{align}
\langle\hat{\Phi}_{1}^{\dagger}(10^{10},0,0)\hat{\Phi}_{1}(\mathbf{0})\rangle\approx & e^{\frac{1}{(2\pi)^{3}}(-1.41-0.258)}\approx0.993.
\end{align}
From this we can believe that $\langle\hat{\Phi}_{1}^{\dagger}(\mathbf{x})\hat{\Phi}_{1}(\mathbf{0})\rangle$
saturates to a constant when $\left|\mathbf{x}\right|\rightarrow\infty$.
\begin{align}
& \int_{0}^{1}dk\int_{0}^{2\pi}d\theta\int_{0}^{\pi}d\psi\left|T_{12} (\mathbf{k}) \right|^{2}k^{2}\sin\psi e^{i10^{10}k\cos\theta\sin\psi}-\int_{0}^{1}dk\int_{0}^{2\pi}d\theta\int_{0}^{\pi}d\psi\left|T_{12} (\mathbf{k}) \right|^{2}k^{2}\sin\psi\nonumber\\
\approx & (1.05\times10^{-3}+6.24\times10^{-5}i)-0.0654\approx-0.064,
\end{align}
\begin{align}
& \int_{10^{-20}}^{1}dk\int_{0}^{2\pi}d\theta\int_{0}^{\pi}d\psi\left|T_{22} (\mathbf{k}) \right|^{2}k^{2}\sin\psi(e^{i10^{10}k\cos\theta\sin\psi}-1)\approx-11.6,
\end{align}
\begin{align}
\langle\hat{\Phi}_{2}^{\dagger}(10^{10},0,0)\hat{\Phi}_{2}(\mathbf{0})\rangle\approx & e^{\frac{1}{(2\pi)^{3}}(-0.064-11.6)}\approx0.954.
\end{align}
From this we can believe that $\langle\hat{\Phi}_{2}^{\dagger}(\mathbf{x})\hat{\Phi}_{2}(\mathbf{0})\rangle$
saturates to a constant when $\left|\mathbf{x}\right|\rightarrow\infty$. We
can also give the dependence of the correlation function $\langle\hat{\Phi}_{2}^{\dagger}(\mathbf{x})\hat{\Phi}_{2}(\mathbf{0})\rangle$
on $\Gamma$, see the orange line of Fig. \ref{fignumerical}(b).
The constant behaviors of the correlation functions $\langle\hat{\Phi}_1^\dagger(\mathbf{x})\hat{\Phi}_1(\mathbf{0})\rangle$ and $\langle\hat{\Phi}_2^\dagger(\mathbf{x})\hat{\Phi}_2(\mathbf{0})\rangle$ shows the spontaneous breaking of charge symmetry $U(1)_{1,C}$ and $U(1)_{2,C}$. This system has a true ODLRO in three spatial dimensions.

\section{Mean-field theory of Lattice Model A}
\label{meanfieldapp}

To determine the phase boundary between the MI phase and the MDC phase in Lattice Model Series A, we present a simple mean-field approximation. 

We define the local operators of dipoles: $\hat{d}_{a,\mathbf{j}}^{i}=\hat{b}_{a,\mathbf{j}+\hat{x}_i}^{\dagger}\hat{b}_{a,\mathbf{j}}$, and set $t_{a}^{(ij)}=t_{1}$, $t_{ab}^{(i)}=t_{2}$,
which mean the coefficients of the hopping terms for the same species and different species respectively.
Then, the hopping terms in the Hamiltonian [Eq.~(\ref{LM1})] can be written as: 
\begin{align}
H_{\text{hop}}= & \sum_{a}^{2}\Bigg[\sum_{\mathbf{i},i,j}-t_1(\hat{d}_{a,\mathbf{i}}^{i})^{\dagger}\hat{d}_{a,\mathbf{i}+\hat{x}_j}^{i}+\sum_{b\neq a}^{2}\sum_{\mathbf{i},i}-t_2(\hat{d}_{a,\mathbf{i}}^{i})^{\dagger}\hat{d}_{b,\mathbf{i}}^{i}\Bigg]+\text{h.c.}
\end{align}
We introduce the mean dipole fields to decompose the hopping terms.  The mean dipole fields $\varPsi_{a,i}$ are given by 
\begin{align}
\varPsi_{a,i}=&\sum_{j}t_1\langle\hat{d}_{a,\mathbf{i}-\hat{x}_j}^{i}+\hat{d}_{a,\mathbf{i}+\hat{x}_j}^{i}\rangle +\sum_{b\neq a}^{2}t_2\langle\hat{d}_{b,\mathbf{i}}^{i}\rangle.
\end{align}
By the mean-field approximation, the hopping terms become 
\begin{align}
H_{\text{hop}}\approx & -\sum_{a}^{2}\sum_{\mathbf{i},i}\Bigg[\varPsi_{a,i}(\hat{d}_{a,\mathbf{i}}^{i})^{\dagger}+\varPsi_{a,i}^{\ast}\hat{d}_{a,\mathbf{i}}^{i}-\frac{\left|\varPsi_{a,i}\right|^{2}}{2dt_{1}+t_{2}}\Bigg]\,,
\end{align}
and the total Hamiltonian is written as a local piece $h_{\mathbf{i}}^{(0)}$ plus a perturbation:
\begin{align}
H=\sum_{\mathbf{i}}\left[h_{\mathbf{i}}^{(0)}-\sum_{a}^{2}\sum_{i}(\varPsi_{a,i}(\hat{d}_{a,\mathbf{i}}^{i})^{\dagger}+\varPsi_{a,i}^{\ast}\hat{d}_{a,\mathbf{i}}^{i})\right],
\end{align}
where 
\begin{align}
h_{\mathbf{i}}^{(0)}= & \sum_{a}^{2}\left[-\mu\hat{n}_{a,\mathbf{i}}+\frac{U}{2}\hat{n}_{a,\mathbf{i}}(\hat{n}_{a,\mathbf{i}}-1)+\sum_{i}\frac{\left|\varPsi_{a,i}\right|^{2}}{2dt_{1}+t_{2}}\right].
\end{align}
The terms $-\sum_{a}^{2}\sum_{i}(\varPsi_{a,i}(\hat{d}_{a,\mathbf{i}}^{i})^{\dagger}+\varPsi_{a,i}^{\ast}\hat{d}_{a,\mathbf{i}}^{i})$ are treated as the perturbation.
The energy density can be expressed as a series expansion in even powers of the parameter $\varPsi_{a,i}$: 
$ E[\varPsi_{a,i}]=\sum_{a}^{2}\sum_{i}\left[\text{constant}+R\left|\varPsi_{a,i}\right|^{2}+W\left|\varPsi_{a,i}\right|^{4}+\dots\right]\,.$ 
We assume the coefficient $W>0$, and the phase transition between the MI phase and the MDC phase occurs at the point $R=0$.
We calculate the coefficient $R$ by using second-order perturbation theory.
The zeroth-order energy from $h_{\mathbf{i}}^{(0)}$ is given by
\begin{align}
E_{a,n}^{(0)}= & \langle n|h_{a,\mathbf{i}}^{(0)}|n\rangle=-\mu n+\frac{U}{2}n(n-1)+\sum_{i}\frac{\left|\varPsi_{a,i}\right|^{2}}{2dt_{1}+t_{2}}\,,
\end{align}
where $n$ is the filling numbers of bosons of species $1$ and $2$.
To calculate second-order energy contribution, we consider a site and one of its nearest neighbors in the direction labeled by $i$,

\begin{align}
  E_{a,(n,n)}^{(2)}= & \!\!\sum_{i}\left|\varPsi_{a,i}\right|^{2}\!\!\sum_{(m,m^{\prime})\neq(n,n)}\!\!\frac{\left|\langle n,n|(\hat{d}_{a,\mathbf{i}}^{i})^{\dagger}+\hat{d}_{a,\mathbf{i}}^{i}|m,m^{\prime}\rangle\right|^{2}}{E_{a,(n,n)}^{(0)}-E_{a,(m,m^{\prime})}^{(0)}}\nonumber\\
  =& \!\!\sum_{i}\left|\varPsi_{a,i}\right|^{2}\!\!\Bigg[\frac{\left|\langle n,n|(\hat{d}_{a,\mathbf{i}}^{i})^{\dagger}+\hat{d}_{a,\mathbf{i}}^{i}|n-1,n+1\rangle\right|^{2}}{E_{a,(n,n)}^{(0)}-E_{a,(n-1,n+1)}^{(0)}}+\frac{\left|\langle n,n|(\hat{d}_{a,\mathbf{i}}^{i})^{\dagger}+\hat{d}_{a,\mathbf{i}}^{i}|n+1,n-1\rangle\right|^{2}}{E_{a,(n,n)}^{(0)}-E_{a,(n+1,n-1)}^{(0)}}\Bigg]\,,
\end{align}
where $E_{a,(n-1,n+1)}^{(0)}=E_{a,(n+1,n-1)}^{(0)}=E_{a,(n,n)}^{(0)}+U$.
Then, we obtain the second-order energy, 
\begin{align}
E_{a,(n,n)}^{(2)}
= & -\sum_{i}\left|\varPsi_{a,i}\right|^{2}\frac{2n(n+1)}{U}.\label{secondorder}
\end{align}
The energy $E_{a,(n,n)}^{(2)}$ is shared by the two sites, and each site has two nearest neighbors in the direction labeled by $i$.
Thus, the energy in Eq. (\ref{secondorder}) is exactly the second-order energy of a single site.
With the contribution of the zeroth-order and second-order energy,
the quadratic coefficient $R$ is 
\begin{align}
R=\frac{1}{2dt_{1}+t_{2}}-\frac{2n(n+1)}{U}.
\end{align}
At $R>0$, this system is under the Mott insulator phase, while at $R<0$, the dipole condensation occurs. 
In the coefficient $R$, the coupling coefficient $t_{2}$ plays a similar role as $t_{1}$. It is worth mentioning that at $t_{1}>0$, $t_{2}>0$, we have $\langle \hat{b}_{a,\mathbf{i}+\hat{x}_i}^\dagger \hat{b}_{a,\mathbf{i}}\rangle\neq 0$, $\langle \hat{b}_{1,\mathbf{i}}^\dagger \hat{b}_{2,\mathbf{i}}\rangle\neq 0$, which implies condensates with a variety of particle-hole bound states in this system as long as $t_{2}>0$. This phase is called multi-dipole condensate phase (MDC).

According to the result of mean-field theory, we can draw schematic phase diagrams of this model, as shown in Fig.~\ref{phase_t} and Fig.~\ref{phase_mu}.

\section{Derivation of Lattice Model A in a tilted lattice}\label{detailtilted}
In a microscopic Hamiltonian with two species of
bosons in a tilted optical lattice chain, a $U^\prime$ term Hubbard term between different species is included:
\begin{align}
  H_{\text{tilted}}=&-t\sum_{a,\mathbf{i}}(\hat{b}_{a,\mathbf{i}}^\dagger\hat{b}_{a,\mathbf{i}+1}+\text{h.c.})-\mu\sum_{a,\mathbf{i}}\hat{n}_{a,\mathbf{i}}+\frac{U}{2}\sum_{a,\mathbf{i}}\hat{n}_{a,\mathbf{i}}(\hat{n}_{a,\mathbf{i}}-1)+U^\prime\sum_{i}\hat{n}_{1,\mathbf{i}}\hat{n}_{2,\mathbf{i}}+V\sum_{a,\mathbf{i}}\mathbf{i}\hat{n}_{a,\mathbf{i}}\nonumber\\
  \equiv&H_t+H_U+H_V\,.
\end{align}
In the limit $V\gg t\,,U\,,U^\prime$, we use a Nakajima transformation $
  e^{S}H e^{-S}=H+[S,H]+\frac{1}{2}[S,[S,H]]+\cdots\,,$ where $S$ is anti-Hermitian.
We take $S=\sum_{n=1}^{\infty}S_n$, where $S_n$ is order $n$ in $t/V$, $U/V$, and $U^\prime/V$. 
In this derivation, we need a effective Hamiltonian to the third-order, and thus we can set $S_n$ to 0 for $n\geq4$. 
Then, the effective Hamiltonian becomes
\begin{align}
  &H_{\text{eff}}=H_V+H_t+H_U+[S_1,H_V]+[S_1,H_U]+[S_1,H_t]+[S_2,H_V]+[S_2,H_U]+[S_2,H_t]+[S_3,H_V]\nonumber\\
  &+\frac{1}{2}[S_1,[S_1,H_V]]+\frac{1}{2}[S_1,[S_1,H_U]]+\frac{1}{2}[S_1,[S_1,H_t]]+\frac{1}{2}[S_1,[S_2,H_V]]+\frac{1}{2}[S_2,[S_1,H_V]]+\frac{1}{6}[S_1,[S_1,[S_1,H_V]]]
\end{align}
Firstly, we need $[S_1,H_V]+H_t=0$, and we can get:
 $  S_1=\frac{t}{V}\sum_{a,\mathbf{i}}(\hat{b}_{a,\mathbf{i}}^\dagger\hat{b}_{a,\mathbf{i}+1}-\text{h.c.})\,,
 $ and we have
\begin{align}
  &[S_1,H_V]=-H_t\,,\quad
  [S_1,H_t]=0\label{S1cr}\\
  &[S_1,H_U]=\frac{t_0U}{2V}\sum_{a,\mathbf{i}}\{\hat{n}_{a,\mathbf{i}},(\hat{b}_{a,\mathbf{i}-1}^\dagger \hat{b}_{a,\mathbf{i}}-\hat{b}_{a,\mathbf{i}}^\dagger \hat{b}_{a,\mathbf{i}+1}+\text{h.c.})\}+\frac{t_0U^\prime}{V}\sum_{a\neq b\,,\mathbf{i}}(\hat{b}_{a,\mathbf{i}-1}^\dagger \hat{b}_{a,\mathbf{i}}-\hat{b}_{a,\mathbf{i}}^\dagger \hat{b}_{a,\mathbf{i}+1}+\text{h.c.})\hat{n}_{b,\mathbf{i}}\,.
\end{align}
Plugging Eq.~(\ref{S1cr}) into the effective Hamiltonian, we obtain
\begin{align}
  H_{\text{eff}}=&H_V+H_U+[S_1,H_U]+[S_2,H_V]+[S_2,H_U]+\frac{1}{2}[S_2,H_t]+[S_3,H_V]+\frac{1}{2}[S_1,[S_1,H_U]]+\frac{1}{2}[S_1,[S_2,H_V]]\,.
\end{align}
Then we require $S_2$ to satisfy $[S_2,H_V]+[S_1,H_U]=0$. $S_2$ is given by:
\begin{align}
  S_2=&-\frac{t_0U}{2V^2}\sum_{a,\mathbf{i}}\{\hat{n}_{a,\mathbf{i}},(\hat{b}_{a,\mathbf{i}-1}^\dagger \hat{b}_{a,\mathbf{i}}-\hat{b}_{a,\mathbf{i}}^\dagger \hat{b}_{a,\mathbf{i}+1}-\text{h.c.})\}-\frac{t_0U^\prime}{V^2}\sum_{a\neq b\,,\mathbf{i}}(\hat{b}_{a,\mathbf{i}-1}^\dagger \hat{b}_{a,\mathbf{i}}-\hat{b}_{a,\mathbf{i}}^\dagger \hat{b}_{a,\mathbf{i}+1}-\text{h.c.})\hat{n}_{b,\mathbf{i}}\,.
\end{align}
And thus, the effective Hamiltonian becomes
 $  H_{\text{eff}}=H_V+H_U+[S_2,H_U]+\frac{1}{2}[S_2,H_t]+[S_3,H_V]\,.
 $ To the third-order, we need $[S_3,H_V]$ to cancel the terms in $[S_2,H_U]+\frac{1}{2}[S_2,H_t]$ which do not conserve the total dipole moment. Here we define a projection operator $P$ which projects onto total dipole-conserving terms, and we have
\begin{align}
  &[S_2,H_U]+\frac{1}{2}[S_2,H_t]+[S_3,H_V]=P([S_2,H_U]+\frac{1}{2}[S_2,H_t])P=\frac{1}{2}P[S_2,H_t]P\nonumber\\
  =&\sum_{a,\mathbf{i}}\left(-\frac{2t_0^2U}{V^2}\hat{n}_{a,\mathbf{i}}^2+\frac{4t_0^2U}{V^2}\hat{n}_{a,\mathbf{i}}\hat{n}_{a,\mathbf{i}+1}\right)+\left(-\frac{t^2U}{V^2}\hat{b}_{a,\mathbf{i}}(\hat{b}_{a,\mathbf{i}+1}^{\dagger})^2\hat{b}_{a,\mathbf{i}+2}+\text{h.c.}\right)\nonumber\\
  &+\sum_{a\neq b\,,i}\left(-\frac{2t_0^2U^\prime}{V^2}\hat{n}_{a,\mathbf{i}}\hat{n}_{b,\mathbf{i}}+\frac{2t_0^2U^\prime}{V^2}\hat{n}_{a,\mathbf{i}}\hat{n}_{b,\mathbf{i}+1}\right)+\left(\frac{t^2U^\prime}{V^2}\hat{b}_{a,\mathbf{i}}^\dagger \hat{b}_{a,\mathbf{i}+1}\hat{b}_{b,\mathbf{i}+1}\hat{b}_{b,\mathbf{i}+2}^\dagger-\frac{t^2U^\prime}{V^2} \hat{b}_{a,\mathbf{i}}^\dagger \hat{b}_{a,\mathbf{i}+1}\hat{b}_{b,\mathbf{i}}\hat{b}_{b,\mathbf{i}+1}^\dagger+\text{h.c.}\right)
\end{align}
At last, the effective Hamiltonian is 
\begin{align}
  H_{\text{eff}}=&\sum_{a,\mathbf{i}}\left(-\frac{t^2U}{V^2}\hat{b}_{a,\mathbf{i}}(\hat{b}_{a,\mathbf{i}+1}^{\dagger})^2\hat{b}_{a,\mathbf{i}+2}+\text{h.c.}\right)+\sum_{a\neq b\,,\mathbf{i}}\left(\frac{t^2U^\prime}{V^2}\hat{b}_{a,\mathbf{i}}^\dagger \hat{b}_{a,\mathbf{i}+1}\hat{b}_{b,\mathbf{i}+1}\hat{b}_{b,\mathbf{i}+2}^\dagger-\frac{t^2U^\prime}{V^2} \hat{b}_{a,\mathbf{i}}^\dagger \hat{b}_{a,\mathbf{i}+1}\hat{b}_{b,\mathbf{i}}\hat{b}_{b,\mathbf{i}+1}^\dagger+\text{h.c.}\right)\nonumber\\
  &+\sum_{a,i}\left(-\mu-\frac{U}{2}+V\mathbf{i}\right)\hat{n}_{a,\mathbf{i}}+\sum_{a,\mathbf{i}}\left(\frac{U}{2}-\frac{2t_0^2U}{V^2}\right)\hat{n}_{a,\mathbf{i}}^2+\frac{4t_0^2U}{V^2}\hat{n}_{a,\mathbf{i}}\hat{n}_{a,\mathbf{i}+1}\nonumber\\
  &+\sum_{a\neq b\,,\mathbf{i}}\left[\left(\frac{U^\prime}{2}-\frac{2t_0^2U^\prime}{V^2}\right)\hat{n}_{a,\mathbf{i}}\hat{n}_{b,\mathbf{i}}+\frac{2t_0^2U^\prime}{V^2}\hat{n}_{a,\mathbf{i}}\hat{n}_{b,\mathbf{i}+1}\right]
\end{align}
After the above derivation, we obtain effective hopping terms with the conservation of the total dipole moments on the tilted optical lattice.

\end{document}